\documentclass[&letterpaper,11pt]{article}
\pdfoutput=1
\pdfoutput=1

\usepackage{jheppub}
\usepackage{multirow}

\usepackage{subfig}
\usepackage{xspace}
\usepackage{xcolor}
\usepackage[countmax]{subfloat}
\usepackage{enumitem}
\usepackage{amssymb}
\usepackage{amsmath}
\usepackage{cancel}
\usepackage{color}
\usepackage{nccmath}
\usepackage{braket}
\usepackage{graphicx}
\usepackage{multirow}
\usepackage{verbatim}
\usepackage{amsthm}
\usepackage{slashed}
\usepackage{wasysym}
\usepackage{simplewick}
\usepackage{mathtools}
\usepackage{soul}
\usepackage{xspace}

\definecolor{dancomment}{RGB}{0,159,0}
\newcommand{\fd}[2]{\parbox{#1}{\includegraphics[width=#1]{#2}}}

\def\cA{\mathcal{A}}
\def\cB{\mathcal{B}}
\def\cC{\mathcal{C}} 
\def\cD{\mathcal{D}}

\def\cJ{\mathcal{J}}

\def\cL{\mathcal{L}}
\def\cM{\mathcal{M}}

\def\cO{\mathcal{O}}
\def\cS{\mathcal{S}}

\def\cP{\mathcal{P}}
\def\cJ{\mathcal{J}}
\def\cW{\mathcal{W}}

\def\cY{\mathcal{Y}}

\def\tr{{\rm tr}}
\def\cPbar{\bar{\mathcal{P}}}

\def\TO{{\bf{T}}}
\def\ATO{{\bar{\TO}}}
\def\tO{\tilde{O}}

\def\nn{{\nonumber}}

\newcommand{\hard}{\mathrm{hard}}
\newcommand{\dyn}{\mathrm{dyn}}

\newcommand{\BPS}{\mathrm{BPS}}

\def\dg{\dagger}

\newcommand{\Eq}[1]{Equation~\eqref{#1}}

\DeclareRobustCommand{\Sec}[1]{Sec.~\ref{#1}}
\DeclareRobustCommand{\Secs}[2]{Secs.~\ref{#1} and \ref{#2}}
\DeclareRobustCommand{\App}[1]{App.~\ref{#1}}
\DeclareRobustCommand{\Tab}[1]{Table~\ref{#1}}

\DeclareRobustCommand{\Fig}[1]{Fig.~\ref{#1}}

\DeclareRobustCommand{\Eq}[1]{Eq.~(\ref{#1})}
\DeclareRobustCommand{\Eqs}[2]{Eqs.~(\ref{#1}) and (\ref{#2})}
\DeclareRobustCommand{\Ref}[1]{Ref.~\cite{#1}}
\DeclareRobustCommand{\Refs}[1]{Refs.~\cite{#1}}

\def\be{\begin{equation}}
\def\ee{\end{equation}}

\newcommand{\nbar}{{\bar n}}

\newcommand{\SCETi}{\mbox{${\rm SCET}_{\rm I}$}\xspace}

\def\l{\langle}
\def\r{\rangle}
\def\a{\alpha}

\def\b{\beta}

\newcommand{\Sl}[1]{\slashed{#1}}

\renewcommand{\arraystretch}{1.05}
\allowdisplaybreaks[3]

\renewcommand{\eq}[1]{Eq.~\eqref{eq:#1}}
\renewcommand{\eqs}[2]{Eqs.~\eqref{eq:#1} and \eqref{eq:#2}}
\renewcommand{\sec}[1]{Sec.~\ref{sec:#1}}

\newcommand{\ord}[1]{\mathcal{O}(#1)}

\newcommand{\df}{\mathrm{d}}

\newcommand{\sdt}{\!\cdot\!}

\newcommand{\Tau}{\mathcal{T}}
\newcommand\bn{{\bar n}}

\newcommand{\la}{\lambda}

\newcommand{\w}{\omega}
\newcommand{\balpha}{{\bar \alpha}}
\newcommand{\bbeta}{{\bar \beta}}

\newcommand{\nslash}{\slashed{n}}
\newcommand{\bnslash}{\slashed{\bar{n}}}

\newcommand{\lp}{\tilde p}        

\newcommand{\bnP}{\overline {\mathcal P}}

\newcommand{\id}{\mathbf{1}}

  \newcount\hour \newcount\minute
  \hour=\time \divide \hour by 60 \minute=\time
  \newcommand{\todaytime}{\today \ -- \number\hour :\ifnum \minute<10 0\fi\number\minute}

\preprint{MIT-CTP 4948}

\title{Subleading Power Factorization with Radiative Functions}

\author[1,2]{Ian Moult,}
\author[3]{Iain W. Stewart,}
\author[3]{and Gherardo Vita}

\affiliation[1]{Berkeley Center for Theoretical Physics, University of California, Berkeley, CA 94720, USA}
\affiliation[2]{Theoretical Physics Group, Lawrence Berkeley National Laboratory, Berkeley, CA 94720, USA}
\affiliation[3]{Center for Theoretical Physics, Massachusetts Institute of Technology, Cambridge, MA 02139, USA}

\emailAdd{ianmoult@lbl.gov}
\emailAdd{iains@mit.edu}
\emailAdd{vita@mit.edu}

\abstract{The study of amplitudes and cross sections in the soft and collinear limits allows for an understanding of their all orders behavior, and the identification of universal structures. At leading power soft emissions are eikonal, and described by Wilson lines. Beyond leading power the eikonal approximation breaks down, soft fermions must be added, and soft radiation resolves the nature of the energetic partons from which they were emitted. For both subleading power soft gluon and quark emissions, we use the soft collinear effective theory (SCET) to derive an all orders gauge invariant bare factorization, at both amplitude and cross section level. This yields universal multilocal matrix elements, which we refer to as radiative functions.  These appear from subleading power Lagrangians inserted along the lightcone which dress the leading power Wilson lines. The use of SCET enables us to determine the complete set of radiative functions that appear to $\mathcal{O}(\lambda^2)$ in the power expansion, to all orders in $\alpha_s$. 
For the particular case of event shape observables in $e^+e^-\to$ dijets we derive how the radiative functions contribute to the factorized cross section to $\mathcal{O}(\lambda^2)$. 
}

\begin{document} 

\maketitle

\section{Introduction}\label{sec:intro}

The simplicity of the soft and collinear limits of gauge theories allows for an all orders understanding of the behavior of amplitudes and cross sections, typically formulated in terms of factorization theorems~\cite{Collins:1985ue,Collins:1988ig,Collins:1989gx}. Unlike for observables which are amenable to a local operator product expansion (OPE)~\cite{Wilson:1969zs}, these general factorization theorems typically involve non-local matrix elements with Wilson lines. While the structure of these matrix elements is well understood at leading power, the structure of power corrections is much less well understood. In general, complicated non-local matrix elements, typically involving operators strung along the light cone dressing the leading power Wilson line structure, are required \cite{Balitsky:1987bk,Balitsky:1990ck,Bauer:2001mh,Bosch:2004cb,Lee:2004ja}.

The soft collinear effective theory (SCET) \cite{Bauer:2000ew, Bauer:2000yr, Bauer:2001ct, Bauer:2001yt}, an effective field theory describing the soft and collinear limits of QCD, provides an operator and Lagrangian based formalism for deriving factorization theorems at subleading power. As an example, SCET has been used to systematically study power corrections to the leading power factorization for $B\to X_s \gamma$, $B\to X_u l \bar \nu$  \cite{Korchemsky:1994jb,Bauer:2001yt} in the shape function region  \cite{Bigi:1993ex,Neubert:1993um,Mannel:1994pm}, and derive subleading factorization theorems in terms of universal non-local operators \cite{Beneke:2002ph,Leibovich:2002ys,Kraetz:2002rv,Neubert:2002yx,Burrell:2003cf,Bosch:2004cb,Lee:2004ja,Beneke:2004in}. In this case, the power corrections take the form of non-local operators describing both soft fluctuations at the scale  $\Lambda_{\text{QCD}}$, in terms of matrix elements of the $B$ meson, as well as the coupling of soft and collinear modes. 

Recently, there has been significant interest in understanding the subleading power soft and collinear limits of perturbative scattering amplitudes and event shape observables. 
This has been motivated at the amplitude level both by their relation to asymptotic symmetries (see e.g. \cite{Strominger:2013jfa,Cachazo:2014fwa,Casali:2014xpa,Cheung:2016iub,Bern:2014oka,He:2014bga,Larkoski:2014bxa,He:2017fsb}), as well as to better understand the structure of amplitudes by studying their limits (see e.g. \cite{Dixon:2011pw,Dixon:2014iba,Dixon:2015iva,Caron-Huot:2016owq,Dixon:2016nkn}).  
At the cross section level an understanding of subleading power corrections will allow for the improved accuracy of perturbative predictions involving resummation, and improvements to next-to-next-to-leading order subtraction schemes \cite{Boughezal:2015dva,Boughezal:2015aha,Gaunt:2015pea} by analytically calculating subleading power corrections \cite{Moult:2016fqy,Boughezal:2016zws,Moult:2017jsg,Boughezal:2018mvf,Ebert:2018lzn,Ebert:2018gsn,Bhattacharya:2018vph}, amongst many other applications. 
From explicit calculations, there are hints for the simplicity of power corrections at higher loop order, for example in splitting functions \cite{Dokshitzer:2005bf},  in the threshold limit \cite{Matsuura:1987wt,Matsuura:1988sm,Hamberg:1990np,DelDuca:2017twk,Dulat:2017prg,Bahjat-Abbas:2018hpv}, for event shape observables \cite{Moult:2016fqy,Boughezal:2016zws,Moult:2017jsg,Balitsky:2017flc,Dixon:2018qgp}, for power corrections in quark masses \cite{Liu:2017vkm,Liu:2018czl}, and in the Regge limit \cite{Bruser:2018jnc}. 
To obtain an all loop understanding, and identify universal structures which persist at subleading powers, it is desirable to formulate subleading power factorization theorems whose renormalization group structure allows the prediction of higher loop results from lower loop data, as has been successful at leading power. 
Recently this was used to derive the first resummation at subleading power for the thrust event shape observable in $H\to gg$ \cite{Moult:2018jjd} and for threshold in \cite{Beneke:2018gvs}.

In this paper we use SCET to derive an all orders gauge invariant factorization for subleading power soft emissions, focusing in particular on non-local corrections described by so called radiative functions. 
We use the SCET Lagrangian, formulated in terms of non-local gauge invariant quark and gluon fields to provide gauge invariant definitions of the radiative functions for the emission of both soft quarks and gluons.
Gauge invariance is guaranteed by an intricate Wilson line structure, dictated by the symmetries of the effective theory. 
We show how these radiative functions appear in factorization formulas at subleading power, both at the level of the amplitude and the cross section, as multilocal matrix elements with convolutions of operators along the lightcone. 
These operator insertions correct the leading power Wilson line structure.
This completes our derivation of all the required components for subleading power factorization initiated in \cite{Feige:2017zci,Moult:2017rpl}, and we review in detail the complete factorization structure at subleading power, highlighting the role that radiative functions play.

\begin{figure}[t!]
\begin{center}
\includegraphics[width=0.23\columnwidth]{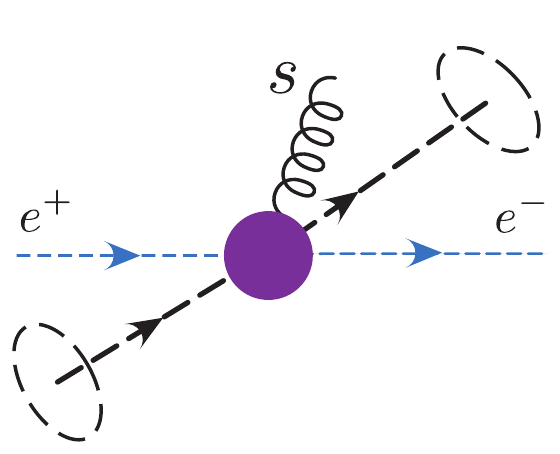} 
\hspace{0.1cm}
\includegraphics[width=0.23\columnwidth]{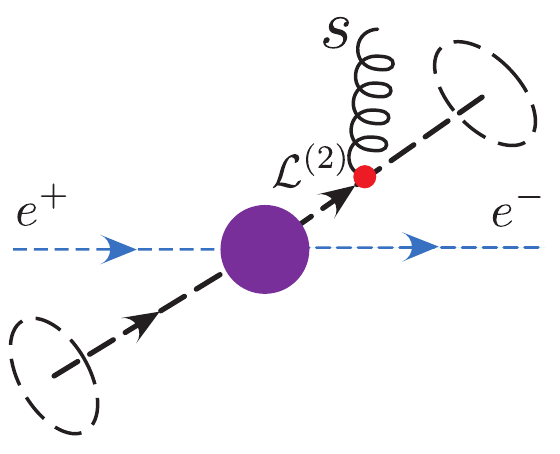} 
\hspace{0.1cm}
\includegraphics[width=0.23\columnwidth]{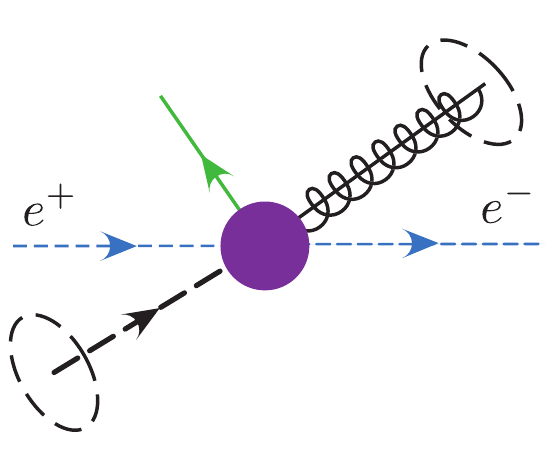} 
\hspace{0.1cm}
\includegraphics[width=0.23\columnwidth]{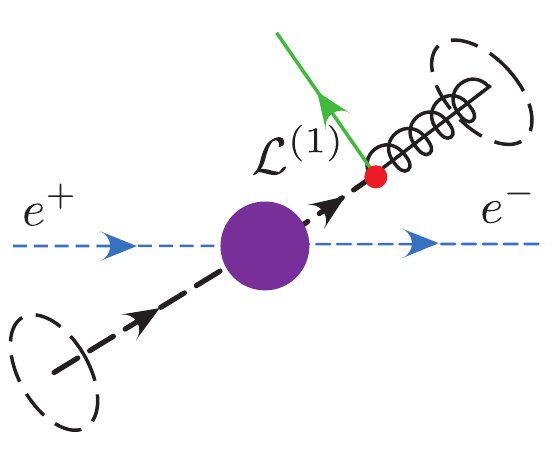} 
\raisebox{0cm}{ \hspace{-0.2cm} 
  $a$)\hspace{3.4cm}
  $b$)\hspace{3.4cm} 
  $c$)\hspace{3.4cm}
  $d$)\hspace{4cm} } 
\\[-25pt]
\end{center}
\vspace{-0.4cm}
\caption{ 
Subleading power contributions from the emission of a soft quark or gluon in a hard scattering. The subleading emission can either be from a local hard scattering operator, as shown in a), c), or from a radiative contribution from the energetic partons, as in b), d). } 
\label{fig:subleadingamp}
\end{figure}

If we consider the subleading power emission of a soft quark or gluon from a hard scattering vertex, there are two potential classes of contributions, as shown in \Fig{fig:subleadingamp}. First, there are contributions where the soft emission localizes to the hard scattering vertex, as shown in \Fig{fig:subleadingamp}. Here the power suppression arises due the lack of a nearly on-shell propagator, and these contributions are described by local hard scattering operators, complete bases of which are known for  $\bar q \Gamma q$  \cite{Feige:2017zci,Chang:2017atu} and $gg$ currents \cite{Moult:2017rpl}, as well as recently for $N$-jet configurations \cite{Beneke:2017ztn,Beneke:2018rbh}. Second, there are contributions from a non-local emission from the energetic parton, which arise from corrections beyond the eikonal limit to the dynamics of the interaction of the soft and collinear particles. Such contributions were studied in the abelian case in the work of Del Duca  \cite{DelDuca:1990gz}, extending the work of Low, Burnett and Kroll (LBK) \cite{Low:1958sn,Burnett:1967km}, and were referred to as radiative jet functions. For the emission of a single soft gluon from an energetic quark, they have been extended to the non-abelian case in \cite{Bonocore:2015esa, Bonocore:2016awd}. They were also studied in \cite{Larkoski:2014bxa} using SCET, where a one-loop expression for soft emission was derived. Our work goes beyond this, by providing explicit all orders factorization in terms of gauge invariant soft and collinear matrix elements. Here we will refer to the general class of such objects as radiative functions. We reserve ``radiative jet function" for the analogous objects at cross section level.

\begin{figure}[t!]
\begin{center}
\includegraphics[width=0.43\columnwidth]{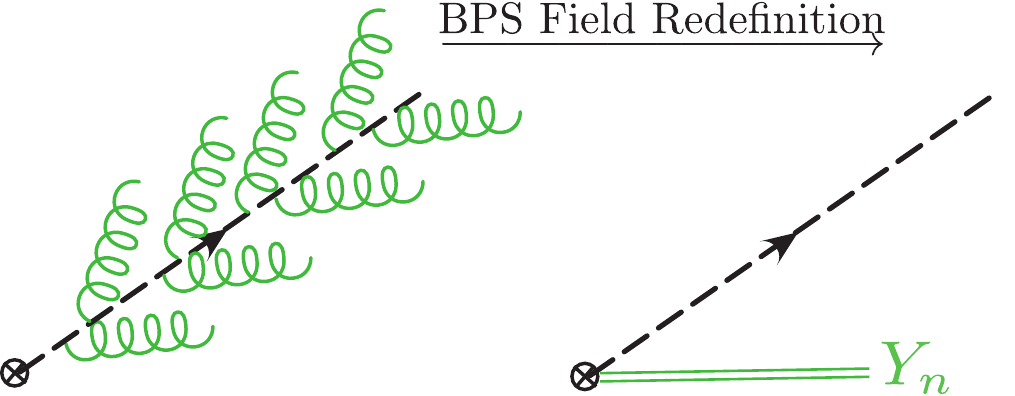} 
\hspace{0.1cm}
\includegraphics[width=0.43\columnwidth]{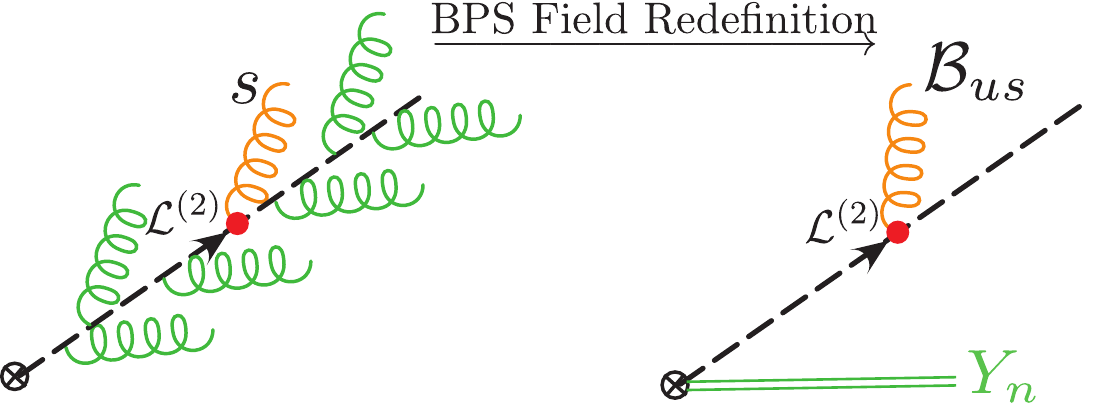} 
\raisebox{0cm}{ \hspace{-0.2cm} 
  $a$)\hspace{7cm}
  $b$)\hspace{3.4cm} 
} 
\\[-25pt]
\end{center}
\vspace{-0.4cm}
\caption{ 
The all orders factorization of soft gluons at leading power a) and next to leading power b). In the leading power case, all soft emissions are absorbed into a Wilson line $Y_n$. In the presence of a next to leading power emission, they are replaced by a Wilson line, and a gauge invariant soft gluon field emitted from the light cone.
 } 
\label{fig:sprig_NLP}
\end{figure}

To provide an all orders description, one must consider a subleading power soft emission in the presence of an arbitrary number of leading power soft, or collinear emissions.  At leading power, the energetic partons emitted from the hard scattering eikonalize, and act as a source for the long wavelength soft radiation. In this limit, the dynamics of the energetic partons can be integrated out, and replaced with a Wilson line along their path. This is shown schematically in \Fig{fig:sprig_NLP}. This leads to the ubiquitous appearance of cusped light-like Wilson lines in the description of the soft and collinear limits of gauge theory amplitudes and cross sections, whose renormalization is controlled by the universal $\Gamma_{\text{cusp}}$ \cite{Korchemsky:1987wg,Korchemsky:1991zp}. Beyond leading power we expect corrections to this picture associated with the breakdown of eikonalization, namely we expect the Wilson lines to be decorated with operators, which we will associate with radiative functions. To achieve subleading power factorization of amplitudes and cross sections, and to understand the universality of these factorizations, we would like to have a systematic approach to the construction of gauge invariant radiative functions in terms of well defined field theoretic objects. This is more difficult due to the nature of the operators, which possess intricate Wilson line structure to ensure gauge invariance in a non-abelian theory.  In QED, this is not an issue since the gauge group is abelian.

To see how radiative functions naturally emerge from the effective theory, we consider the SCET Lagrangian (here we restrict ourselves to the case of \SCETi), which consists of both hard scattering operators, and a dynamical Lagrangian
\begin{align} \label{eq:SCETLagExpand_intro}
\cL_{\text{SCET}}=\cL_\hard+\cL_\dyn= \sum_{i\geq0} \cL_\hard^{(i)}+\sum_{i\geq0} \cL^{(i)}+\cL_G^{(0)} \,,
\end{align}
each of which is a power expansion in $\lambda$. The hard scattering operators, included in $\cL_\hard$, describe all the localized contributions in \Fig{fig:subleadingamp}, while the non-local contributions are described by the dynamical Lagrangians, $\cL^{(i)}$. The dynamical Lagrangian is universal, and known up to $\cO(\lambda^2)$ \cite{Beneke:2002ni,Chay:2002vy,Manohar:2002fd,Pirjol:2002km,Beneke:2002ph,Bauer:2003mga}. Finally, $\cL_G^{(0)}$ is the leading power Glauber Lagrangian \cite{Rothstein:2016bsq}. The SCET Lagrangian is fixed by the symmetries of the theory, namely soft and collinear gauge symmetries and reparametrization invariance \cite{Manohar:2002fd,Chay:2002vy}, and is known to not be renormalized to all orders in $\alpha_s$ \cite{Beneke:2002ph}, which will allow us to prove the universality of the radiative functions.

In the SCET framework, the leading power eikonalization of \Fig{fig:sprig_NLP} is achieved at the level of the Lagrangian through the BPS field redefinition \cite{Bauer:2002nz}, 
\be \label{eq:BPSfieldredefinition}
\cB^{a\mu}_{n\perp}\to \cY_n^{ab} \cB^{b\mu}_{n\perp} , \qquad \chi_n^\alpha \to Y_n^{\alpha \bbeta} \chi_n^\beta,
\ee
which is performed in each collinear sector. Here $Y_n$, $\cY_n$ are fundamental and adjoint ultrasoft Wilson lines, where for a general representation, r, the ultrasoft Wilson line is defined by
\be
Y^{(r)}_n(x)=\bold{P} \exp \left [ ig \int\limits_{0}^\infty ds\, n\cdot A^a_{us}(x+sn)  T_{(r)}^{a}\right]\,,
\ee
where $\bold P$ denotes path ordering.  The BPS field redefinition decouples the ultrasoft degrees of freedom from the leading power collinear Lagrangian \cite{Bauer:2002nz}, so that they appear only in the hard scattering vertex. Beyond leading power, the BPS field redefinition does not decouple the ultrasoft and collinear interactions. However, since subleading power interactions can only occur a finite number of times at a given power, they can be viewed as decorating the Wilson line, and giving rise to radiative functions. 

Consider a subleading power emission, in the presence of an arbitrary number of additional soft emissions, as shown in \Fig{fig:sprig_NLP}. The insertion of the subleading power Lagrangian implies that the soft emissions cannot simply be pulled back into a Wilson line at the hard scattering vertex, since they become trapped at the subleading power Lagrangian insertion. The Wilson lines appearing in the BPS field redefinition can be used to sandwich the covariant derivative describing the soft gluon emission. For a derivative in an arbitrary representation, $r$, we have
\begin{align}\label{eq:soft_gluon_intro}
Y^{(r)\,\dagger}_{n_i} i D^{(r)\,\mu}_{us} Y^{(r)}_{n_i }=i \partial^\mu_{us} + [Y_{n_i}^{(r)\,\dagger} i D^{(r)\,\mu}_{us} Y^{(r)}_{n_i}]=i\partial^\mu_{us}+T_{(r)}^{a} g \cB^{a\mu}_{us(i)}\,,
\end{align}
which allows us to define the ultrasoft gauge invariant gluon building block field by
\begin{align} \label{eq:softgluondef_intro}
g \cB^{a\mu}_{us(i)}= \left [   \frac{1}{in_i\cdot \partial_{us}} n_{i\nu} i G_{us}^{b\nu \mu} \cY^{ba}_{n_i}  \right] \,.
\end{align}
Furthermore, there will be a single remaining Wilson line at the hard scattering vertex. This is shown schematically in \Fig{fig:sprig_NLP}. After applying the BPS field redefinition, there are no further interactions between soft and collinear partons. The finite number of subleading power interactions between soft and collinear fields at a given power are represented by gauge invariant operator insertions along the lightcone, which dress the leading power Wilson lines. These give rise to universal non-local string operators appearing in subleading power factorization theorems. A similar picture also applies to soft quark emission. This provides a systematic way to provide gauge invariant operator definitions of radiative functions in the effective theory, which is the goal of this paper. The use of non-local gauge invariant fields is crucial to achieve factorization for these non-local operators, since it enables gauge invariant definitions of soft and collinear matrix elements tied together by  convolution variables, that can be separately renormalized. This is non-trivial in a non-abelian gauge theory, where the soft emission carries a color charge.

An important result of this paper is that we will show how to achieve an all orders gauge invariant factorization at the cross section level for the radiative contributions to event shape observables. This will allow us to express the cross section as a convolution of gauge invariant collinear and soft factors, each of which can in principle be separately renormalized. Unlike at leading power, these subleading power factorizations will involve an additional convolution over the gauge invariant momentum (or equivalently position along the light cone) of the insertion of the gauge invariant field. 

As an example, we can consider the factorization involving a radiative contribution at cross section level involving a soft quark field. We will show that such a contribution can be writen as a convolution which can be shown schematically as
\begin{align}
\fd{3cm}{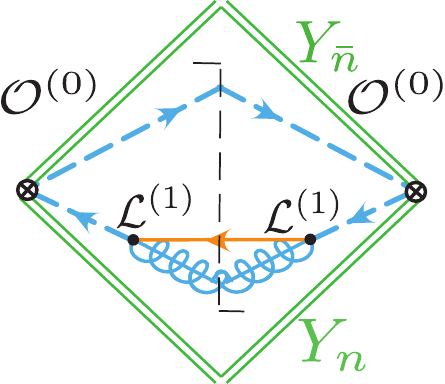} =  \int dr_2^+ dr_3^+ \fd{3cm}{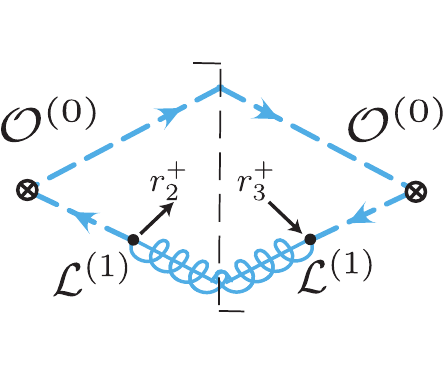} \otimes  \fd{3cm}{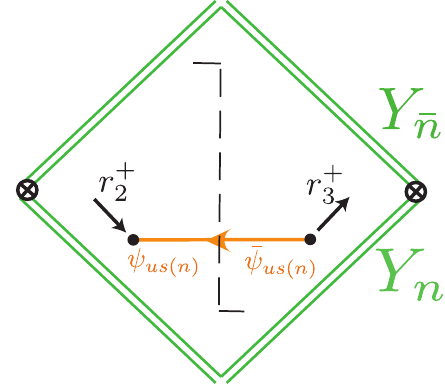}\,.
\end{align}
This involves a factorization into a convolution over a collinear radiative function which emits the soft quark, and a soft function involving a gauge invariant soft quark field, in addition to the standard Wilson lines. This form of factorization allows us to describe systematically, and in a gauge invariant manner, all subleading power corrections to event shape observables.

As noted earlier, non-local gauge invariant  operators describing the coupling of soft and collinear particles have appeared in the SCET literature on subleading power corrections in $B$-physics \cite{Bigi:1993ex,Neubert:1993um,Mannel:1994pm,Korchemsky:1994jb,Bauer:2001yt,Beneke:2002ph,Leibovich:2002ys,Kraetz:2002rv,Neubert:2002yx,Burrell:2003cf,Bosch:2004cb,Lee:2004ja,Beneke:2004in}, many of which have a similar structure to those discussed in this paper.
Here we provide a simplified approach by determining the subleading power Lagrangians in terms of gauge invariant building blocks, and then we focus on the application to perturbative scattering amplitudes and collider event shape observables.  
We will also emphasize the connection to the radiative functions approach which has been studied in the literature \cite{Bonocore:2015esa, Bonocore:2016awd}.

An outline of this paper is as follows. In \Sec{sec:scet} we review the key features of SCET that will be important for our analysis, in particular,  the definition of gauge invariant collinear and soft fields and the BPS field redefinition. In \Sec{sec:sub-fact} we discuss the different contributions to factorization at subleading power, and how they appear in SCET, placing in context the contributions from radiative functions. In \Sec{sec:subleading_lagrangians} we derive the form of the subleading power SCET Lagrangians describing the interactions of the non-local gauge invariant quark and gluon fields, using the BPS field redefinition and the equations of motion. In \Sec{sec:RadiativeFunction_intro} we use these Lagrangians to study subleading power factorization at the amplitude level, showing how radiative functions naturally emerge, and deriving in detail their structure. We also compare our radiative functions to those previously discussed in the literature. In \Sec{sec:fact_RadiativeFunction} we consider radiative functions at the level of the cross section for event shape observables, and derive the structure of convolutions between the radiative functions. In \Sec{sec:RF_thrust} we classify all radiative functions contributing to $\SCETi$ type observables in $e^+e^-\to$ dijets. We conclude in \Sec{sec:conclusions}.

\section{Review of SCET}\label{sec:scet}

In this section we briefly review the aspects of SCET that will be needed in this paper, in particular the use of gauge invariant quark and gluon fields. Reviews of SCET can be found in \Refs{iain_notes,Becher:2014oda}, and \cite{iain_notes} in particular contains a discussion of the formalism required for SCET at subleading power.

\subsection{General Formalism}\label{sec:scet_gen}

SCET is an effective field theory of QCD describing the interactions of collinear and soft particles in the presence of a hard interaction \cite{Bauer:2000ew, Bauer:2000yr, Bauer:2001ct, Bauer:2001yt, Bauer:2002nz}. Since the dynamics of collinear particles is concentrated along the light cone, it is convenient to use light cone coordinates. For each relevant light-like direction (jet direction) we define two reference vectors $n_i^\mu$ and $\bn_i^\mu$ 
such that $n_i^2 = \bn_i^2 = 0$ and $n_i\sdt\bn_i = 2$. Any momentum $p$ can then be written as
\begin{equation} \label{eq:lightcone_dec}
p^\mu = \bn_i\sdt p\,\frac{n_i^\mu}{2} + n_i\sdt p\,\frac{\bn_i^\mu}{2} + p^\mu_{n_i\perp}\
\,.\end{equation}
A particle is referred to as ``$n_i$-collinear'' if it has momentum $p$ close to the $\vec{n}_i$ direction. More precisely, the components of its momentum must scale as $(n_i\!\cdot\! p, \bn_i \!\cdot\! p, p_{n_i\perp}) \sim \bn_i\!\cdot\! p$ $\,(\la^2,1,\la)$. Here $\la \ll 1$ is a formal power counting parameter defining the expansion. Different reference vectors $n_i$ and $n_i'$, with $n_i\cdot n_i' \sim \ord{\lambda^2}$ provide equivalent descriptions. This enforces a symmetry on the effective theory known as reparametrization invariance (RPI) \cite{Manohar:2002fd,Chay:2002vy}. We will often use this symmetry to simplify our description, for example by choosing that the total $\perp$ momentum of a particular collinear sector vanishes.
Extensions of SCET to describe more complicated situations with collinear particles with soft momentum $\bar n_i \cdot p \ll 1$ are referred to as SCET$_+$ \cite{Bauer:2011uc,Procura:2014cba,Larkoski:2015zka,Pietrulewicz:2016nwo}.

\begin{table}
\begin{center}
\begin{tabular}{| l | c | c |c |c|c|c |c|r| }
  \hline                       
  Operator & $\cB_{n\perp}^\mu$ & $\bar\cP^{}$ & $\cP_\perp^\mu$& $\bar n\cdot\cB_{us}$ &$\cB_{us\,\perp}^\mu$ & $\bar n\cdot \partial_{us}$ & $ n\cdot \partial_{us}$ & $ \partial_{us}^\perp$ \\
  Power Counting & $\lambda$  &  $\lambda^0$ & $\lambda$& $\lambda^2$& $\lambda^2$& $\lambda^2$& $\lambda^2$& $\lambda^2$ \\
  \hline  
\end{tabular}
\end{center}
\caption{
Power counting for building block operators in $\text{SCET}_\text{I}$.
}
\label{tab:PC}
\end{table}

SCET is constructed as an expansion about the light cone in powers of $\lambda$. Explicitly, momenta are expanded into label and residual components
\begin{equation} \label{eq:label_dec}
p^\mu = \lp^\mu + k^\mu = \bn_i \sdt\lp\, \frac{n_i^\mu}{2} + \lp_{n_i\perp}^\mu + k^\mu\,,
\,\end{equation}
where, $\bn_i \cdot\lp \sim Q$ and $\lp_{n_i\perp} \sim \la Q$ are the large label momentum components, with $Q$ a characteristic scale of the hard interaction, while $k\sim \la^2 Q$ is a small residual momentum. A multipole expansion is then performed to obtain fields with momenta of definite scaling, namely collinear quark and gluon fields for each collinear direction, as well as soft quark and gluon fields. Independent gauge symmetries are enforced for each set of collinear or soft fields.  As a consequence of the multipole expansion all fields and their derivatives acquire a definite power counting \cite{Bauer:2001ct}, shown in \Tab{tab:PC}. The detailed structure of the fields will be described shortly. Similarly, the SCET Lagrangian is expanded as
\begin{align} \label{eq:SCETLagExpand}
\cL_{\text{SCET}}=\cL_\hard+\cL_\dyn= \sum_{i\geq0} \cL_\hard^{(i)}+\sum_{i\geq0} \cL^{(i)} +\cL_G^{(0)} \,,
\end{align}
with each term having a definite power counting ${\cal O}(\lambda^i)$. As written, the SCET Lagrangian is divided into three different contributions. The $ \cL_\hard^{(i)}$ contains local hard scattering operators, and is derived by a matching calculation. The $\cL^{(i)}$ describe the long wavelength dynamics of ultrasoft and collinear modes in the effective theory, and can be found in \cite{iain_notes} to $\cO(\lambda^2)$. The Glauber Lagrangian~\cite{Rothstein:2016bsq}, $\cL_G^{(0)}$, describes interactions between soft and collinear modes in the form of potentials, which break factorization unless they can be shown to cancel.  In this paper, we will considerably simplify the structure of the subleading power SCET Lagrangians not involving Glauber exchange, re-organizing them using the equations of motion of the effective theory, and writing them in terms of gauge invariant operators.  Subleading power contributions to the Glauber Lagrangian that describe quark reggeization are also known \cite{Moult:2017xpp}.

We will write the SCET fields for $n_i$-collinear quarks and gluons, $\xi_{n_i,\lp}(x)$ and $A_{n_i,\lp}(x)$,  in position space with respect to the residual momentum and in momentum space with respect to the large momentum components. The large momentum $\lp$, and the collinear direction then act as labels for the fields. Derivatives acting on the fields give the residual momentum dependence, $i \partial^\mu \sim k \sim \la^2 Q$, while label momentum operators  $\cP_{n_i}^\mu$ give the label momentum $\cP_{n_i}^\mu\, \xi_{n_i,\lp} = \lp^\mu\, \xi_{n_i,\lp}$.  An important feature of the multipole expansion is that the propagator for collinear fields 
\begin{align}\label{eq:prop}
\frac{1}{\bar n \cdot \tilde p~ n \cdot p_r + \tilde p_{\perp} ^2}\,,
\end{align}
is independent of $\bar n \cdot p_r$ and $p_{r\perp}^\mu$ and hence is local in the residual $x^-$, $x^\perp$ components. This will allow for factorized expressions involving convolutions to be reduced to a single variable convolution in the position along the lightcone, and will play an important role in our definitions of the radiative functions.

Soft degrees of freedom are described in the effective theory by separate quark and gluon fields, $q_{us}(x)$ and $A_{us}(x)$. We will assume that we are working in the SCET$_\text{I}$ theory where these soft degrees of freedom are referred to as ultrasoft. These fields do not carry label momentum, and have $i \partial^\mu \sim \la^2Q$.

\subsection{Gauge Invariant Fields and the BPS Field Redefinition}\label{sec:BPS}

In our study of radiative functions the use of gauge invariant soft and collinear fields will play a central role, allowing us to systematically construct gauge invariant non-local operators. Collinearly gauge invariant quark and gluon fields are defined as \cite{Bauer:2000yr,Bauer:2001ct}
\begin{align} \label{eq:chiB}
\chi_{{n_i},\w}(x) &= \Bigl[\delta(\w - \bnP_{n_i})\, W_{n_i}^\dagger(x)\, \xi_{n_i}(x) \Bigr]
\,,\\
\cB_{{n_i}\perp,\w}^\mu(x)
&= \frac{1}{g}\Bigl[\delta(\w + \bnP_{n_i})\, W_{n_i}^\dagger(x)\,i  D_{{n_i}\perp}^\mu W_{n_i}(x)\Bigr]
 \,, \nn
\end{align}
where
\begin{equation}
i  D_{{n_i}\perp}^\mu = \cP^\mu_{{n_i}\perp} + g A^\mu_{{n_i}\perp}\,,
\end{equation}
is the collinear covariant derivative and
\begin{equation} \label{eq:Wn}
W_{n_i}(x) = \biggl[~\sum_\text{perms} \exp\Bigl(-\frac{g}{\bnP_{n_i}}\,\bn\sdt A_{n_i}(x)\Bigr)~\biggr]\,,
\end{equation}
is a Wilson line of ${n_i}$-collinear gluons in label momentum space, and the label operators in \eqs{chiB}{Wn} only act inside the square brackets. The collinear Wilson line $W_{n_i}(x)$ is localized with respect to the residual position $x$, and we can therefore treat
$\chi_{{n_i},\w}(x)$ and $\cB_{{n_i},\w}^\mu(x)$ as local quark and gluon fields from the perspective of ultrasoft derivatives $\partial^\mu$ that act on $x$.

To formulate gauge invariant radiative functions, which describe the subleading power emission of a ultrasoft quark or gluon, it will be essential to define gauge invariant ultrasoft quark and gluon fields. This can be done using the BPS field redefinition of \Eq{eq:BPSfieldredefinition}, which we repeat here for convenience
\be \label{eq:BPSfieldredefinition_v2}
\cB^{a\mu}_{n\perp}\to \cY_n^{ab} \cB^{b\mu}_{n\perp} , \qquad \chi_n^\alpha \to Y_n^{\alpha \bbeta} \chi_n^\beta\,.
\ee
Gauge invariant ultrasoft operators will be necessarily non-local at the ultrasoft scale, involving ultrasoft Wilson lines. 
However, the form of this non-locality is completely determined by the BPS field redefinition. 
Matching calculations from QCD to the effective theory are performed prior to the BPS field redefinition, when the theory is local at the hard scale, and then the non-locality arises only from the BPS field redefinition. 
However, one can take a bottom-up approach and consider gauge invariant soft gluon fields as building blocks of the theory. This approach has been used in \cite{Feige:2017zci,Moult:2017rpl,Chang:2017atu} to construct the operator bases at subleading power. Here we will show that the subleading power Lagrangians of SCET can be rewritten after BPS field redefinition purely in terms of gauge invariant ultrasoft quark and gluon building blocks and collinear fields. This will play an important role, allowing for the definition of gauge invariant radiative functions.

To define gauge invariant ultrasoft fields, we can group all Wilson lines arising from the BPS field redefinition with fields to form gauge invariant combinations. In particular, we can define an ultrasoft gauge invariant quark field as
\begin{align} \label{eq:usgaugeinvdef}
\psi_{us(i)}=Y^\dagger_{n_i} q_{us}\,,
\end{align}
Similarly, we can group Wilson lines with gauge covariant derivatives in an arbitrary representation, $r$,
\begin{align}\label{eq:soft_gluon}
Y^{(r)\,\dagger}_{n_i} i D^{(r)\,\mu}_{us} Y^{(r)}_{n_i }&=i \partial^\mu_{us} + [Y_{n_i}^{(r)\,\dagger} i D^{(r)\,\mu}_{us} Y^{(r)}_{n_i}]=i\partial^\mu_{us}+T_{(r)}^{a} g \cB^{a\mu}_{us(i)}\,, \nn \\
Y^{(r)\,\dagger}_{n_i} i \overleftarrow{D}^{(r)\,\mu}_{us} Y^{(r)}_{n_i }&=i \overleftarrow{\partial}^\mu_{us} + [Y_{n_i}^{(r)\,\dagger} i \overleftarrow{D}^{(r)\,\mu}_{us} Y^{(r)}_{n_i}]=i \overleftarrow{\partial}^\mu_{us}-T_{(r)}^{a} g \cB^{a\mu}_{us(i)}\,,
\end{align}
to define gauge invariant ultrasoft derivatives, and gauge invariant ultrasoft gluon fields
\begin{align} \label{eq:softgluondef}
g \cB^{a\mu}_{us(i)}= \left [   \frac{1}{in_i\cdot \partial_{us}} n_{i\nu} i G_{us}^{b\nu \mu} \cY^{ba}_{n_i}  \right] \,.
\end{align}
Here the square brackets indicate that the covariant derivative acts only on the Wilson lines within the brackets. In both \Eqs{eq:usgaugeinvdef}{eq:soft_gluon}, the fields are ultrasoft gauge invariant for an arbitrary lightlike direction, $n_i$, and the soft fields themselves are not naturally associated with a given direction. Due to \Eq{eq:BPSfieldredefinition_v2}, this direction is usually naturally taken to coincide with that of a collinear direction, for example the direction of the collinear particles emitting the soft quark or gluon. Note that the ultrasoft gauge invariant gluon field is the analogue of the gauge invariant collinear gluon field of \Eq{eq:chiB}, which can also be written
\begin{align}  
g\cB_{n_i\perp}^{A\mu} =\left [ \frac{1}{\bar \cP}    \bar n_{i\nu} i G_{n_i}^{B\nu \mu \perp} \cW^{BA}_{n_i}         \right]\,.
\end{align}

The gauge invariance of the ultrasoft fields in \Eqs{eq:usgaugeinvdef}{eq:soft_gluon}, is enforced by the presence of the soft Wilson line. This also implies that it has Feynman rules describing an arbitrary number of soft emissions. For example, expanded up to two emissions, we have
\begin{align}\label{eq:twogluonBus}
g\cB^\mu_{us(n)}&=g\left(   A^{\mu a}_{us}(k)\, T^a -k^\mu \frac{n \cdot A^a_{us}(k)\, T^a}{n \cdot k} \right)+g^2 [T^a, T^b] \frac{n \cdot A^a_{us}(k_1)}{n \cdot k_1} A^{b\,\mu }_{us}(k_2) + \nn \\[0.2cm]
&+\frac{g^2}{2}\left( \frac{k^\mu_{1} +k^\mu_{2}}{n \cdot k_1+n \cdot k_2}\right) \left(   \frac{T^a T^b}{n \cdot k_1}+\frac{T^b T^a}{n \cdot k_2 } \right)n \cdot A_{us}^a(k_1)\, n \cdot A^b_{us}(k_2) +\cdots\,.
\end{align}
As was described in the \Sec{sec:intro}, and as illustrated in \Fig{fig:sprig_NLP}, this particular combination sandwiching an ultrasoft emission will naturally appear when considering subleading power emission from a collinear sector, giving rise to the emission of a gauge invariant gluon field $\cB_{us(i)}$. The ability to formulate the emission of a soft parton in a gauge invariant manner in a non-abelian theory relies crucially on the use of non-local gauge invariant fields of \Eq{eq:softgluondef}, as it will allow the definition of separately gauge invariant soft and collinear matrix elements tied together by a convolution variable which represents the momentum flowing into the soft emission.

\section{Components of Subleading Power Factorization}
\label{sec:sub-fact}

In this section we discuss the different components contributing to the formulation of a subleading power factorization theorem in SCET, extending the brief discussion provided in \Sec{sec:intro}. Although some of the different components of the factorization, such as the construction of operator bases \cite{Feige:2017zci,Moult:2017rpl,Chang:2017atu} and the renormalization of subleading power operators \cite{Freedman:2014uta,Beneke:2017ztn,Goerke:2017lei,Moult:2018jjd,Beneke:2018rbh,Beneke:2018gvs}, have been studied extensively in the literature, we wish to provide here a self contained discussion, and we hope that it puts into context the role of radiative functions in subleading power factorization more generally.  We will take as a concrete example an SCET$_{\text{I}}$ event shape in $e^+e^-\to $ dijets, however, the considerations are more general. Throughout this section, we will not consider possible contributions from leading power Glauber modes, which we assume decouple from the soft and collinear modes. For the case of $e^+e^-$ this is reasonable, since all QCD particles are in the final state, where we expect leading Glauber effects can be absorbed into the direction of soft Wilson lines. Under this assumption, subleading power corrections to the Glauber Lagrangian of \cite{Rothstein:2016bsq} could be constructed and analyzed in the same manner discussed here, although this is beyond the scope of this paper.\footnote{A subset of power corrections to the Glauber Lagrangian, namely those giving rise to the Reggeization of the quark, were studied in \cite{Moult:2017xpp}.}

Consider a dimensionless SCET$_\text{I}$ event shape observable, $\tau$, in $e^+e^-\to $ dijets, which is chosen to vanish in the dijet limit. As a concrete example, one can consider $\tau=1-T$, where $T$ is the thrust observable \cite{Farhi:1977sg}. We can write the cross section as a power expansion
\begin{align}\label{eq:xsec_homog}
\frac{\df\sigma}{\df \tau} &=\frac{\df\sigma^{(0)}}{\df \tau} +\frac{\df\sigma^{(1)}}{\df \tau} +\frac{\df\sigma^{(2)}}{\df \tau}+\frac{\df\sigma^{(3)}}{\df \tau} +{\cal O}(\tau)\,,
\end{align}
where $\df\sigma^{(n)}/\df \tau\sim \tau^{n/2-1}$ due to the scaling relation $\lambda \sim \sqrt{\tau}$. We wish to find factorized expressions for the non-zero contributions in this series, in terms of hard, jet and soft functions of the schematic form
\begin{align} \label{eq:sec5facsigma}
&\hspace{-0.25cm}\frac{\df\sigma^{(n)}}{\df \tau} =
 Q\sigma_0 
\sum_{j}  H^{(n_{Hj})}_{j} \otimes J^{(n_{Jj})}_{j} \otimes S_j^{(n_{Sj})}  
\,.\end{align}
Here $H^{(n_{Hj})}_{j}$ describes matching coefficients, while $J^{(n_{Jj})}_{j}$ and $S_j^{(n_{Sj})}$ are field theoretic matrix elements involving only collinear or soft fields, and $Q$ is the center of mass energy. Here the superscipts denote the power suppression, and we have
\begin{align}
n=n_{Hj}+n_{Jj}+n_{Sj}\,.
\end{align}
In general, there will be multiple distinct hard, jet and soft functions at each power.

In the effective theory approach, the first step towards this goal is to write the expression for the differential cross section in terms of full theory QCD matrix elements,
\begin{align}
\frac{d\sigma}{\df \tau} = \frac{1}{2Q^2} \sum_X \tilde \delta^{(4)}_q  
    \bra{L} O (0) \ket{X}\bra{X} O(0) \ket{L} \delta\big( \tau - \tau(X) \big)\,,
\end{align}
where for  $e^+e^-\to$ dijets through a virtual photon, $O=\mathcal{J}^\mu L_\mu$, where $L_\mu$ is the leptonic current which includes the photon propagator and couplings, and $\mathcal{J}^\mu=\bar q \gamma^\mu q$.  Here we use the shorthand notation $\tilde \delta^{(4)}_q=(2\pi)^4\delta^4(q-p_X)$ for the momentum conserving delta function. The summation over all final states, $X$, includes phase space integrations. Here $|L\rangle$ denotes the $e^+e^-$ leptonic initial state. The measurement of the observable is enforced by $ \delta\big( \tau - \tau(X) \big) $, where $\tau(X)$, returns the value of the observable $\tau$ as measured on the final state $X$.

For $\tau \ll 1$, we are in the dijet limit and  can match onto SCET hard scattering operators with two collinear sectors
\begin{align}\label{eq:match_intro}
O=\sum_{\lambda_l,j,k} \cC_{\lambda_l\,j}^{(k)} O_{\lambda_l\,j}^{(k)}\,.
\end{align}
Here the sum is over powers in $\lambda$ indicated by the superscript $(k)$, and at each power distinct operators are labeled by $\lambda_l$ and $j$ which include helicity and color labels.  Our labels are split such that $\lambda_l = \pm$ indicates the helicity of the lepton current and the index $j$ denotes all helicity and color labels of the QCD component of the current. The $\cC_{\lambda_l j}^{(k)}$ coefficients include the electromagnetic coupling and charges.

We work to all orders in the strong coupling, $\alpha_s$, but to leading order in the electroweak couplings. We can therefore factorize out the leptonic component, $J_{\lambda l}$ of the hard scattering operators
\begin{align}
O_{\lambda_l\,j}^{(k)}=J_{\lambda_l}\tilde O_j^{(k)}\,.
\end{align}
Evaluating the tree level matrix element involving the external electron states, the expression for the cross section can be written in terms of matrix elements in the effective theory  as
\begin{align}\label{eq:pre_expand}
&\frac{d\sigma}{d \tau} = N \sum_X    \tilde \delta^{(4)}_q  \sum_{\lambda_l}
 \Big \l 0 \Big| \sum_{j,k} \cC_{\lambda_l\,j}^{(k)} \tO_j^{(k)}  \Big| X \Big \r_{\cL_{\text{dyn}} }     
 \Big \l X \Big |\sum_{j,k} \cC_{\lambda_l\,j}^{(k)} \tO_j^{(k)} \Big |0\Big \r_{\cL_{\dyn}}   
 \delta\Big( \tau - \sum_l \tau^{(l)}(X) \Big)
 \,.
\end{align}
After having calculated the leptonic matrix element we are left with a normalization factor $N$, whose explicit form is not relevant for the current discussion.

To achieve an expression with homogeneous power counting, as in \Eq{eq:xsec_homog}, we must systematically expand \Eq{eq:pre_expand} in $\lambda$, working to all orders in $\alpha_s$. At leading power, assuming that the action of the measurement function factorizes, this is simple. The BPS field redefinition decouples leading power soft and collinear interactions so that the Hilbert space factorizes, and the state $|X\r$ can be written
\begin{align}
|X\r =|X_n \r |X_\bn \r |X_{us} \r\,.
\end{align}
Algebraic manipulations can then be used to organize  \Eq{eq:pre_expand} into a form involving separate matrix elements of soft and collinear fields, and hence derive a bare factorization formula.

In the effective field theory organization it is then evident from \Eq{eq:pre_expand} that there are three sources of power corrections
\begin{enumerate}
\item Subleading power hard scattering operators.
\item Subleading power corrections to the measurement function.
\item Subleading power Lagrangian insertions.
\end{enumerate}
We will briefly discuss the structure of each of these sources of power corrections in turn, but our primary focus will be on the factorization of subleading power Lagrangian insertions, as these give rise to the subleading power radiative functions which are the focus of this paper.

\subsection{Hard Scattering Operators}\label{sec:hard_fact}

The matching of QCD onto SCET gives rise to hard scattering operators, see \Eq{eq:match_intro}. These operators are local at the scale of the matching, as shown schematically in \Fig{fig:subleadingamp}. Subleading power hard scattering operators with two collinear directions were recently discussed in detail in  \Refs{Feige:2017zci,Moult:2017rpl,Chang:2017atu} where complete bases were derived for $\bar q \Gamma q$ and $gg$ currents using the approach of helicity operators \cite{Moult:2015aoa,Kolodrubetz:2016uim}. The leading order matching was also performed. In addition, operator bases for $N$-jet configurations were studied in \cite{Beneke:2017ztn}.

Hard scattering operators at subleading power are similar to those at leading power in that they are formed from products of the SCET operator building blocks of \Tab{tab:PC}. These building blocks provide a complete basis of building blocks to all powers, as can be proven by the use of equations of motion and operator relations~\cite{Marcantonini:2008qn}. The difference between leading and subleading power comes from additional collinear or ultrasoft fields, or $\cP_\perp$ operators which are inserted into the hard scattering operator to give the power suppression. For example, leading power hard scattering operators for more inclusive processes typically have a single collinear field in each collinear sector, but at subleading power can have multiple collinear fields in a single sector.

If the entire power suppression comes from the hard scattering operator then the factorization proceeds similar to at leading power\footnote{Although the final formulas typically have a richer convolution structure since subleading power hard scattering operators can have multiple fields per collinear sector.}. In particular, in this case only the leading power Lagrangian is required since subleading power Lagrangian insertions would induce additional power suppression, and therefore factorization formulae can be derived through the BPS field redefinition.  We include the case that the power suppression comes from a mixture of suppression in the hard scattering operator and suppression from the subleading power Lagrangian in \Sec{sec:subl_insert}, where the treatment is more complicated due to the subleading power Lagrangian.

\subsection{Measurement Function Factorization}\label{sec:obs_fact}

The action of the measurement function $\tau(X)$, which is a function of the soft and collinear momenta, must also be expanded homogeneously in the soft and collinear limits.  As shown in \Eq{eq:pre_expand}, we expand the measurement function as
\begin{align} \label{eq:measurement_exp}
 \tau(X) &= \tau^{(0)}(X) + \tau^{(2)}(X) + \ldots \,.
\end{align}
Here we have assumed that any $\mathcal{O}(\lambda) = \mathcal{O}(\sqrt{\tau})$ corrections to the measurement function vanish.\footnote{This is true for most observables in the SCET formulation of \Ref{Bauer:2000ew, Bauer:2000yr, Bauer:2001ct, Bauer:2001yt} in which label and residual momentum are exactly conserved. In the approach of \Ref{Freedman:2011kj}, where momentum is not strictly conserved, an $\mathcal{O}(\lambda)$ contribution to the measurement function does appear. As shown in \Ref{Freedman:2013vya} this $\mathcal{O}(\lambda)$ contribution to the measurement function contributes as a product with an $\cO(\lambda)$ operator arising from the expansion of momentum conserving delta functions, and contributes at $\lambda^2$.} This was shown explicitly for the case of thrust in \cite{Feige:2017zci}. 

The measurement function enters  \Eq{eq:pre_expand} as a delta function constraint on the final state. This constraint can be expanded as
\begin{align}\label{eq:meas_expand}
\delta\big( \tau - \tau(X) \big) &=
\delta\big( \tau - \tau^{(0)}(X) - \tau^{(2)}(X) - \ldots \big) 
\\&= \delta\big( \tau - \tau^{(0)}(X) \big) - \tau^{(2)}(X)\: \delta' \big( \tau - \tau^{(0)}(X) \big) + \ldots
 \,,\nn
\end{align}
where the dots represent higher derivatives of delta functions.

To achieve a factorization, one must show that the measurement operators at each power can be factorized into contributions from soft or collinear degrees of freedom. To be specific, we restrict ourselves to what we have referred to as ``pseudo-additive observables'' \cite{Feige:2017zci} which we defined as those observables with measurement functions that can be factorized into contributions from collinear and ultrasoft modes at each order in the power expansion in the form
\begin{align}\label{eq:fact_meas}
 \tau^{(i)}(X)= \tau_n^{(i)}(X_n,G_\bn,G_s)+ \tau_\bn^{(i)}(X_\bn,G_n,G_s) + \tau_{us}^{(i)}(X_{us},G_n,G_\bn)\,.
\end{align}
The factors $G_n$, $G_\bn$, $G_{us}$, which can enter the measurement function for any sector, are global properties of a sector, and must be defined independent of the order in perturbation theory.\footnote{While the factors $G_n, G_\bn, G_{us}$ are often trivial, an example where they are not is the factorization for the ``soft haze'' region of \Ref{Larkoski:2015kga,Larkoski:2017iuy,Larkoski:2017cqq}, describing the factorization in endpoint region of energy correlation function based jet substructure observables \cite{Larkoski:2014gra,Moult:2016cvt}.}  In this case one can define field theoretic measurement functions $\widehat \cM_n^{(i)}$, $\widehat \cM_\bn^{(i)}$, and $\widehat \cM_{us}^{(i)}$ from the energy momentum tensor of the theory \cite{Lee:2006nr,Sveshnikov:1995vi,Korchemsky:1997sy,Bauer:2008dt,Belitsky:2001ij}. The measurement functions act as 
\begin{align}
\widehat \cM_n^{(i)} |X_n\r &= \delta(\tau - \tau_n^{(i)}(X_n))|X_n\r \,, \nn \\
\widehat \cM_{\bar n}^{(i)} |X_{\bar n}\r &= \delta(\tau - \tau_{\bar n}^{(i)}(X_{\bar n}))|X_{\bar n}\r \,, \nn \\
\widehat \cM_{us}^{(i)} |X_{us}\r &= \delta(\tau - \tau_{us}^{(i)}(X_{us}))|X_{us}\r \,.  
\end{align}
The subleading power measurement function has been derived for the thrust observable in \cite{Freedman:2013vya,Feige:2017zci} to $\cO(\lambda^2)$.

Note that subleading power corrections can also arise for measurement functions of observables, such as kinematic factors, that are not small in the $\lambda \ll 1$ limit. This is particularly important when multiple measurements are performed on the final states as occurs for \emph{Born measurements} in fully differential cross sections at hadron colliders. 

As an example, let us consider the beam thrust \cite{Stewart:2010tn} event shape $\Tau_0$ or the $q_T$ spectrum in color singlet production at the LHC. To obtain distributions that are fully differential in the momentum of the color singlet, one needs to include not only a measurement function for the observable $\Tau_0$ or $q_T$, but also a measurement $\hat\delta_Y \equiv  \delta(Y -\hat{Y}(X))$ for the rapidity and one for the invariant mass of the color singlet $\hat\delta_Q \equiv  \delta(Q^2 -\hat{Q^2}(X))$. To be precise, if we call $q^\mu$ the 4-momentum of the color singlet in the hadronic center of mass frame, the rapidity and the invariant mass measurements take the form
\be\label{eq:bornMeas}
	\hat\delta_Y = \delta\left(Y - \frac{1}{2}\log\frac{q^-}{q^+}\right)\,,\qquad\hat\delta_Q = \delta\Bigl(Q^2 - q^+q^- - {\vec{q}}_\perp^{\,\,2}\Bigr)\,.
\ee
At Born level the $\Tau_0$ or $q_T$ measurement gives $\delta(\Tau)$ or $\delta(q_T)$, however the observables defined in \eq{bornMeas} are in general non trivial already at the Born level. Hence, they are referred to as \emph{Born measurements}.
As shown in the fixed order calculations of \cite{Moult:2016fqy,Moult:2017jsg,Ebert:2018lzn,Ebert:2018gsn} and explained in detail in \cite{Ebert:2018lzn}, the power corrections to the Born measurements contribute significantly to the power correction of the differential distribution. In particular they introduce new non-perturbative functions, namely derivatives of the Parton Distribution Functions (PDFs), which do not appear at leading power.

Since we are interested in deriving factorization to $\cO(\lambda^2)$, and the first subleading power correction to the measurement function appears at $\cO(\lambda^2)$, contributions to the cross section whose power suppression arises from the measurement functions can be factorized just like at leading power by using the BPS field redefinition. Any insertion of subleading power Lagrangians or hard scattering operators would lead to further power suppression.

\subsection{Factorization with Lagrangian Insertions}\label{sec:subl_insert}

The most non-trivial aspect of subleading power factorization is the factorization of the subleading power Lagrangians. At leading power this is achieved in SCET through the BPS field redefinition, however, the BPS field redefinition does not decouple soft and collinear interactions beyond leading power. The Lagrangian governing the dynamics of the effective theory has the power expansion
\begin{align} 
\cL_\dyn=\sum_{i\geq0} \cL^{(i)} \,.
\end{align}
When working to any fixed power in $\lambda$ only a finite number of insertions of $\cL^{(i)}$, $i\geq1$ need to be considered. Explicitly, if we consider a time-ordered product ($T$-product) in the effective theory and we are interested in its expansion to $\cO(\lambda^2)$, we have
{\small
\begin{align}
&  \langle 0| T\{ \tilde O_j^{(k)}(0) \text{exp}[i \mbox{$\int$} d^4x \cL_\dyn]  \} |X \rangle        \\
&= \langle 0| T\{ \tilde O_j^{(k)}(0)  \text{exp}[i \mbox{$\int$} d^4x (\cL^{(0)} +\cL^{(1)} +\cL^{(2)}+\cdots )        ]  \} |X\rangle    \nn \\
&=\left \langle 0 \left| T\bigg\{ \tilde O_j^{(k)}(0)  \text{exp}[i \mbox{$\int$} d^4x \cL^{(0)}] \bigg (1 \! +\!  i \mbox{$\int$}d^4y\cL^{(1)}\!+\! \frac{1}{2}    \big( i\mbox{$\int$} d^4y\cL^{(1)} \big)  \big( i\mbox{$\int$} d^4z \cL^{(1)} \big) \!+\! i\mbox{$\int$} d^4z\cL^{(2)}   \!+\! \cdots  \bigg)          \bigg\} \right |X \right\rangle   \nn \\
&=  \left \langle 0 \left | T\bigg\{ \tilde O_j^{(k)}(0)   \bigg(1+i\mbox{$\int$} d^4y\cL^{(1)}+\frac{1}{2} \big( i\mbox{$\int$} d^4y\cL^{(1)} \big)  \big( i\mbox{$\int$} d^4z \cL^{(1)} \big) +i\mbox{$\int$} d^4z\cL^{(2)} \bigg)           \bigg\} \right |X \right\rangle_{\cL^{(0)}} +\cdots\,,  \nn
\end{align}}
where the dots represent higher power corrections.
In the final expression  all matrix elements are evaluated using the leading power SCET Lagrangian, and the subleading power Lagrangians appear only a finite number of times. From now on we will drop the ${\cal L}^{(0)}$ subscript. This expression highlights that to achieve factorization of the dynamics at any finite power in the power expansion, it is sufficient to show a decoupling of the leading power interactions. The insertions of the subleading power Lagrangians in the matrix elements will lead to the radiative functions in which we are interested.

The leading power interactions of ultrasoft and collinear degrees of freedom can be decoupled at the Lagrangian level using the BPS field redefinition of \Eq{eq:BPSfieldredefinition_v2}. After the BPS field redefinition, the leading power SCET Lagrangian decomposes as
\begin{align}
\cL^{(0)}= \sum_{n_i} \cL^{(0)}_{n_i} +\cL^{(0)}_{us}\,,
\end{align}
where the sum is over distinct collinear sectors. Since the leading power Lagrangian defines the time evolution,
states in the Hilbert space can then also be factorized as
\begin{align}
|X\r =|X_n \r |X_\bn \r |X_{us} \r\,.
\end{align}
Here we work in the interaction picture defined by the leading power Lagrangian, and considering perturbations in the power expansion. Note that these perturbations are in $\lambda$, unlike the interaction picture defining the perturbative expansion in $\alpha_s$. Here corrections in $\alpha_s$ are kept to all orders,. 

This allows hard-soft-collinear factorization to be achieved to any power in the effective field theory. Deriving the explicit structure of the factorization in the case of subleading power Lagrangian insertions will be the main focus of this paper, and will give rise to radiative functions.

\subsection{Factorized Cross Section to $\cO(\lambda^2)$}\label{sec:sum_fact}

Having understood the different sources of power corrections in the effective theory, we can now achieve a homogenous power expansion for the $e^+e^-\to$ dijet cross section, and give expressions at each order in the power expansion in terms of matrix elements of hard scattering operators, Lagrangian insertions, and measurement functions. Here we consider only the terms which arise up to $\mathcal{O}(\lambda^2)$.  At $\mathcal{O}(\lambda^0)$ we have the simple expression in terms of the different helicity configurations of the  leading power operator
\begin{align}\label{eq:xsec_lam0}
\frac{d\sigma}{d\tau}^{(0)}\!\! = N  \sum_X    \tilde \delta^{(4)}_q \sum_{\lambda_l}    
 \bra{0}   \sum_{\lambda_q} C^{(0)}_{(\lambda_l;\lambda_q)} \tO^{(0)\dg}_{(\lambda_q)}(0) \ket{X}
 \bra{X} \sum_{\lambda_q}  C^{(0)}_{(\lambda_l;\lambda_q)}  \tO^{(0)}_{(\lambda_q)}(0) \ket{0} 
 \, \delta\big( \tau - \tau^{(0)}(X) \big)\,.
\end{align}
Since all matrix elements are now evaluated with the leading power Lagrangian, there are no interactions between soft and collinear degrees of freedom, and the factorization into collinear and soft matrix elements is simply an algebraic exercise, leading to the well known factorization for the thrust observable \cite{Korchemsky:1999kt,Fleming:2007qr,Schwartz:2007ib}. We will review this factorization in more detail in \Sec{sec:fact_RadiativeFunction}.

At $\mathcal{O}(\lambda^1)$ we have potential contributions from $\mathcal{O}(\lambda^1)$ hard scattering operators, as well as subleading Lagrangian insertions,
{\begin{small}
\begin{align}\label{eq:xsec_lam}
\frac{d\sigma}{d\tau}^{(1)} &= N \sum_{X,i}  \tilde \delta^{(4)}_q  \bra{0} C_i^{(1)*} \tO_i^{(1) \dagger}(0) \ket{X}\bra{X} C^{(0)} \tO^{(0)}(0) \ket{0}   \delta\big( \tau - \tau^{(0)}(X) \big) +\text{h.c.} 
 \\
&\quad+ N \sum_X    \tilde \delta^{(4)}_q  \int d^4x \bra{0}  \ATO \big(-i\cL^{(1)}(x) \b
ig)  C^{(0)*} \tO^{(0)\dagger}(0) \ket{X}\bra{X}  C^{(0)} \tO^{(0)}0) \ket{0} \delta\big( \tau - \tau^{(0)}(X) \big)+\text{h.c.} \nn \\
&=0
\,. \nn
\end{align}
\end{small}}%
where here and in the following $\TO$($\ATO$) denotes (anti-)time ordering.
The vanishing of $\frac{d\sigma}{d\tau}^{(1)}$ from hard scattering operators was explained for thrust in \cite{Feige:2017zci}, and in \Sec{sec:RF_thrust} we will discuss the analogous explanation for the vanishing of Lagrangian insertion contributions at this order

At $\mathcal{O}(\lambda^2)$ all three sources of power corrections contribute
\begin{align}
\frac{d\sigma}{d\tau}^{(2)} =\frac{d\sigma}{d\tau}^{(2),\text{hard}} +\frac{d\sigma}{d\tau}^{(2),\text{measure}}+\left(\frac{d\sigma}{d\tau}^{(2),\text{radiative}}+\frac{d\sigma}{d\tau}^{(2),\text{mixed}}\right)\,,
\end{align}
or more explicitly
{\begin{footnotesize}
\begin{align}\label{eq:xsec_lam2}
&\frac{d\sigma}{d\tau}^{(2)} = N \sum_{X,i}  \tilde \delta^{(4)}_q  \bra{0} C_i^{(2)*} \tO_i^{(2)\dagger}(0) \ket{X}\bra{X} C^{(0)} \tO^{(0)}(0) \ket{0} \delta\big( \tau - \tau^{(0)}(X) \big) +\text{h.c.} \\
&+ N \sum_{X,i,j}  \tilde \delta^{(4)}_q  \bra{0} C_i^{(1)*} \tO_i^{(1)\dagger}(0) \ket{X}\bra{X} C_j^{(1)} \tO_j^{(1)}(0) \ket{0}   \delta\big( \tau - \tau^{(0)}(X) \big)   \nn\\
&- N \sum_X  \tilde \delta^{(4)}_q \bra{0} C^{(0)*} \tO^{(0)\dagger}(0) \ket{X}\bra{X}  C^{(0)} \tO^{(0)}(0) \ket{0}   \tau^{(2)}(X)\, \delta'\big( \tau - \tau^{(0)}(X) \big)\nn\\
&+ N \sum_{X,i}  \tilde \delta^{(4)}_q   \int d^4x  \bra{0} C_i^{(1)*} \tO_i^{(1)\dagger}(0) \ket{X}\bra{X}\TO ( i \cL^{(1)}(x)) C^{(0)} \tO^{(0)}(0) \ket{0}   \delta\big( \tau - \tau^{(0)}(X) \big)+\text{h.c.}    \nn\\
&+ N \sum_{X,i}  \tilde \delta^{(4)}_q   \int d^4x  \bra{0} \ATO( -i \cL^{(1)}(x)) C_i^{(1)*} \tO_i^{(1)\dagger}(0) \ket{X}\bra{X}  C^{(0)} \tO^{(0)}(0) \ket{0}   \delta\big( \tau - \tau^{(0)}(X) \big)+\text{h.c.}    \nn\\
&+ N \sum_X  \tilde \delta^{(4)}_q \int d^4x  \bra{0}  C^{(0)*} \tO^{(0)\dagger}(0) \ket{X}\bra{X} \TO( i \cL^{(2)}(x)) C^{(0)} \tO^{(0)}(0) \ket{0}   \delta\big( \tau - \tau^{(0)}(X) \big)  +\text{h.c.} \nn\\
&-\frac{N}{2} \sum_X    \tilde \delta^{(4)}_q  \int d^4x \int d^4y \bra{0}\ATO\cL^{(1)}(x) \cL^{(1)}(y)  C^{(0)*} \tO^{(0)\dagger}(0) \ket{X}\bra{X}  C^{(0)} \tO^{(0)}0) \ket{0}   \delta\big( \tau - \tau^{(0)}(X) \big)  +\text{h.c.} \nn\\
&-\frac{N}{2} \sum_X   \tilde \delta^{(4)}_q  \int d^4x  \int d^4y \bra{0}\ATO\cL^{(1)}(x)  C^{(0)*} \tO^{(0) \dagger}(0) \ket{X}\bra{X}\TO \cL^{(1)}(y) C^{(0)} \tO^{(0)}0) \ket{0}  \delta\big( \tau - \tau^{(0)}(X) \big) \,.\nn
\end{align}
\end{footnotesize}}%
Unlike the ${\cal O}(\lambda)$ power correction, the $\mathcal{O}(\lambda^2)$ correction to the cross section does not vanish. The $\mathcal{O}(\lambda^2)$ power correction for thrust was computed at fixed order to $\cO(\alpha_s)$ and to $\cO(\alpha_s^2)$ using SCET in \cite{Freedman:2013vya} and \cite{Moult:2016fqy,Moult:2017jsg}, respectively. Since the interactions between soft and collinear degrees of freedom have been decoupled, by algebraic manipulation of \Eq{eq:xsec_lam2}, the contributions to the cross section at each order in the power expansion can be expressed as a sum of vacuum matrix elements, involving a measurement function insertion, and each containing only collinear $n$, collinear $\bar n$, or ultrasoft fields.  To do this, we write the constraint on the final state as a sum of the measurement operators
\begin{align}
\delta(\tau-\hat \tau)= \int d\tau_n d\tau_{\bar n} d\tau_{us} \delta(\tau-\tau_n -\tau_{\bar n}-\tau_{us}) \delta(\tau_n -\hat \tau_n) \delta(\tau_{\bar n} -\hat \tau_{\bar n}) \delta(\tau_{us}-\hat \tau_{us})\,.
\end{align}
We can then perform the sum over the $|X_n\rangle \langle X_n|$,  $|X_{\bar n}\rangle \langle X_{\bar n}|$, and $|X_{us}\rangle \langle X_{us}|$ states to simplify all the matrix elements to vacuum matrix elements. The Lorentz, Dirac, and color structure can be simplified using Fierz relations, and the symmetry properties of the vacuum matrix elements, such that each matrix element is a scalar, and there are no index contractions between the soft and collinear functions, namely a completely factorized form.

In previous papers we have given complete bases of hard scattering operators \cite{Feige:2017zci,Moult:2017rpl,Chang:2017atu}, as well as the expansion of the measurement function \cite{Feige:2017zci}. Here we focus on the radiative type contributions, namely those term involving additional integrals over the position of Lagrangian insertions. We will formulate the factorization of the radiative contributions to the cross section as products of gauge invariant soft and collinear matrix elements involving either one or two convolutions, corresponding to the one or two Lagrangian insertions which can exist when working to $\cO(\lambda^2)$. 

Explicitly, we will be able to derive a representation of the form
{\small
\begin{align}\label{eq:sec3_convolution}
\frac{1}{\sigma_0}\frac{d\sigma^{(2),\text{radiative}} }{d\tau}&= Q^5\sum_j H_j(Q^2)  \int \frac{dr_1^+}{2\pi Q}  S_{j}(Q\tau_{us}, r_2^+) \otimes  J_{\bar n,j}(Q^2\tau_\bn) \otimes J_{n,j}(Q^2\tau_n, Q r_2^+) \\
&+Q^5\sum_j H_j(Q^2)  \int \frac{dr_1^+}{2\pi Q}   \int \frac{dr_2^+}{2\pi Q}  S_{j}(\tau_{us},r_1^+, r_2^+) \otimes  J_{\bar n,j}(Q^2\tau_\bn) \otimes J_{n,j}(Q^2\tau_n, Qr_1^+,Qr_2^+) \nn \\
&+Q^5\sum_j H_j(Q^2)  \int \frac{dr_1^+}{2\pi Q}   \int \frac{dr_2^+}{2\pi Q}  S_{j}(Q\tau_{us}, r_1^+,r_2^+) \otimes  J_{\bar n,j}(Q^2\tau_\bn,Qr_1^+) \otimes J_{n,j}(Q^2\tau_n, Qr_2^+)\,, \nn
\end{align}}%
where we choose to make the arguments of the soft functions dimension 1 and the arguments of the jet functions dimension 2 analogously to leading power.
Here $\otimes$ denotes the convolution in the thrust variable, $\tau$, 
\be
\int d\tau_n d\tau_{\bar n} d\tau_{us} \delta(\tau-\tau_n -\tau_{\bar n}-\tau_{us}) \,,
\ee
and we have used the symmetry under $n\leftrightarrow \bar n$ to combine several equivalent contributions. The derivation of this factorized form at the cross section level is the main goal of this paper. The factorization derived in this paper will be at the bare level, namely we do not consider the renormalization of the hard, jet and soft functions. To derive a renormalized factorization formula, one must show that the hard, jet and soft functions can be separately renormalized, and that the convolutions in the $r_1^+$ and $r_2^+$ variables are well defined. This is in general non-trivial, and even in simple cases the renormalization of the subleading power jet and soft functions involves mixing with additional operators that do not appear in the matching \cite{Paz:2009ut,Moult:2018jjd,Beneke:2018gvs}, with evanescent operators \cite{Buras:1989xd,Dugan:1990df,Herrlich:1994kh} and possibly with EOM operators \cite{Beneke:2019kgv} (though the particular EOM operators will differ from those found in \cite{Beneke:2019kgv} due to differences in the construction of the subleading power Lagrangians), and the convolutions do not naively converge  \cite{Beneke:2003pa}. However, the derivation of a bare factorization is the first step towards a complete, renormalized factorization.

After studying the structure of the subleading power Lagrangians in terms of gauge invariant quark and gluon fields in \Sec{sec:subleading_lagrangians}, in \Sec{sec:fact_RadiativeFunction} we will work out explicitly the structure of the factorization of the matrix elements for those contributions involving Lagrangian insertions, which give rise to the radiative jet functions. This will provide the ingredients needed to construct \Eq{eq:sec3_convolution} explicitly. This in turn yields all the pieces needed to explore the full factorization for subleading power thrust, which we plan to pursue in future work.

\section{Subleading Lagrangians for Gauge Invariant Fields}\label{sec:subleading_lagrangians}

Having identified the different sources of power corrections in SCET, we now focus on the structure of the radiative functions.  As has been emphasized, to achieve factorization into separately gauge invariant soft and collinear factors, it is essential that the radiative functions be formulated in terms of non-local gauge invariant fields, namely $\cB_{us}$ and $\psi_{us}$. We therefore will derive the subleading power Lagrangians describing the interactions of these non-local fields to all orders in $\alpha_s$.

The general form of the subleading power Lagrangians is quite complicated, since they describe the complete dynamics  of the soft and collinear sectors to all orders in $\alpha_s$. Nevertheless, due to the power counting and locality of the effective theory, there are a finite number of terms in each Lagrangian. Operationally, at a fixed order in perturbation theory, the number of terms in the Lagrangian which actually contribute is relatively small since most terms involve higher numbers of fields. Before proceeding to the full derivation of the subleading power Lagrangians, we give the structure of the Lagrangian in terms of field content, ignoring the detailed Dirac, Lorentz, and color structures. This is useful for understanding the general structure of the Lagrangians, and the order in perturbation theory at which different terms can contribute.

 At $\cO(\lambda)$ the field structure of the Lagrangian is given by
\begin{align}\label{eq:L1_fields}
\cL_n^{(1)\text{BPS}}&\sim\frac{1}{\bar \cP}\bar \chi_n \chi_n \cP_\perp \{  \partial_{us} \text{ or }\cB_{us(n)} \}   + \cB_{n\perp} \cB_{n\perp} \cP_\perp \{  \partial_{us} \text{ or }\cB_{us(n)} \}\nn \\
& + \frac{1}{\bar \cP} \bar \chi_n \chi_n \cB_{n\perp} \{  \partial_{us} \text{ or }\cB_{us(n)} \} +    \cB_{n\perp} \cB_{n\perp} \cB_{n\perp} \{  \partial_{us} \text{ or }\cB_{us(n)} \} \nn \\
&+ \frac{1}{\bar \cP}\bar \chi_n \cB_{n\perp} \psi_{us(n)} +\frac{1}{\bar \cP} \bar \psi_{us(n)} \cB_{n\perp}  \chi_n\,,
\end{align}
where we have organized the structure according to the collinear field content.
The number of fields appearing in the Lagrangian is fixed by power counting and locality, and at $\cO(\lambda)$ the Lagrangian involves up to three collinear fields. The operators that involve multiple collinear fields will not contribute at tree level to the emission of a soft parton from a single collinear parton, but are necessary to correctly reproduce the complete subleading power expression at loop level, or for multiple collinear emissions.

At $\cO(\lambda^2)$ the field structure of the Lagrangian is given by
\begin{align}\label{eq:L4_fields}
\cL_n^{(2)\text{BPS}}&\sim \cB_{n\perp} \cB_{n\perp} \{ \partial_{us} \partial_{us} \text{ or } \cB_{us(n)} \cB_{us(n)} \text{ or } \partial_{us} \cB_{us(n)} \} \\
&+\frac{1}{\bar \cP}\bar \chi_n \chi_n \{ \partial_{us} \partial_{us} \text{ or } \cB_{us(n)} \cB_{us(n)} \text{ or } \partial_{us} \cB_{us(n)} \}   \nn \\
&+\frac{1}{\bar \cP} \cB_{n\perp} \cB_{n\perp} \cB_{n\perp} \cB_{n\perp} \{  \partial_{us} \text{ or }\cB_{us(n)} \}  + \frac{1}{\bar \cP} \cB_{n\perp} \cB_{n\perp} \cP_\perp \cB_{n\perp}\{  \partial_{us} \text{ or }\cB_{us(n)} \}  \nn \\
&+  \frac{1}{\bar \cP} \cB_{n\perp} \cB_{n\perp}  \cP_\perp^2 \{  \partial_{us} \text{ or }\cB_{us(n)} \}\nn \\
&+ \frac{1}{\bar \cP^2}\bar \chi_n \chi_n \cB_{n\perp} \cB_{n\perp} \{  \partial_{us} \text{ or }\cB_{us(n)} \}    + \frac{1}{\bar \cP^2}\bar \chi_n \chi_n \cP_\perp \cB_{n\perp} \{  \partial_{us} \text{ or }\cB_{us(n)} \} \nn \\
&  + \frac{1}{\bar \cP^2}\bar \chi_n \chi_n \cP_\perp^2 \{  \partial_{us} \text{ or }\cB_{us(n)} \} + \frac{1}{\bar \cP^3}\bar \chi_n \chi_n \bar \chi_n \chi_n\{  \partial_{us} \text{ or }\cB_{us(n)} \}  \nn \\
&+ \frac{1}{\bar \cP}\bar \chi_n \chi_n \bar \chi_n \psi_{us(n)}   + \frac{1}{\bar \cP} \chi_n \cB_{n\perp}  \cB_{n\perp}    \psi_{us(n)}  + \bar \chi_n \partial_{us} \psi_{us(n)}  + \frac{1}{\bar \cP}\chi_n \cB_{n\perp}  \cP_\perp \psi_{us(n)} +\text{h.c.} \,, \nn
\end{align}
where we have again organized the terms based on their collinear field content, and we see that the $\cO(\lambda^2)$ Lagrangian involves up to four collinear fields.  

In this section we derive the exact form of the Lagrangians given in \Eqs{eq:L1_fields}{eq:L4_fields}.
We begin in \Sec{sec:sum_deriv} by summarizing the notation used in this section and the BPS transformations of different covariant derivative operators, which will allow us to write the subleading power Lagrangians in terms of gauge invariant quark and gluon fields. 
In \Sec{sec:eom} we discuss our reorganization of the Lagrangians using the equations of motion in the effective theory. 
Then, in Secs.~\ref{sec:lp_Lagrangian}-\ref{sec:lam2_Lagrangian} we present our simplified results for the BPS redefined Lagrangians in terms of gauge invariant quark and gluon fields, as well as relevant Feynman rules. 
These will be used to derive the structure of the radiative functions in \Secs{sec:RadiativeFunction_intro}{sec:fact_RadiativeFunction}.

\subsection{Field Redefinitions for Subleading Lagrangians}\label{sec:sum_deriv}

The subleading power Lagrangians in SCET are typically written in a local form, which still involve the interactions of soft and collinear partons \cite{Pirjol:2002km,Manohar:2002fd,Bauer:2003mga}. 
To derive subleading power factorization formulas involving radiative functions, we would like to rewrite them in terms of the non-local gauge invariant quark and gluon fields.
This can be achieved by performing the BPS field redefinition and manipulating the Wilson lines into gauge invariant combinations, which is the goal of this section.

Before BPS field redefinition the subleading power Lagrangians are written in terms of a variety of different covariant derivatives which we summarize here for convenience. The gauge covariant derivatives that we will use are defined by
\begin{align}
iD^\mu_n &= i\partial^\mu_n +g A^\mu_n\,, \qquad
 i \partial^\mu_n = \frac{\bn^\mu}{2} n \cdot \partial + \frac{n^\mu}{2} \overline{\cP} + \cP_\perp^\mu\,, \nn \\
 iD^\mu_{ns} &=i D^\mu_n +\frac{\bn^\mu}{2}gn \cdot A_{us}\,,\qquad
i\partial^\mu_{ns}=i \partial^\mu_n +\frac{\bn^\mu}{2} gn\cdot A_{us}\,, \nn \\
iD_{us}^\mu&=i\partial^\mu+gA^\mu_{us}\,,
\end{align}
and their gauge invariant versions are given by
\begin{align}
i\cD^\mu_{n}&=W_n^\dagger iD^\mu_{n} W_n\,, \qquad i\cD^\mu_{n\perp}=W_n^\dagger iD^\mu_{n\perp} W_n= \cP^\mu_{n\perp}+gB^\mu_{n\perp}\,, \qquad i \cD^\mu_{ns}=W_n^\dagger iD^\mu_{ns} W_n\,.
\end{align}
It is also useful to summarize the transformation of the different derivative operators under the BPS field redefinition. These are all derived using the defining relations of the Wilson line, 
\begin{align}
Y_n^\dagger Y_n=1, \qquad in\cdot D_{us} Y_n=0\,,
\end{align}
which imply the relations
\begin{align}
Y_n^\dagger in\cdot D_{us} Y_n = i n \cdot \partial_{us}\,, \qquad &&Y_n^\dagger gn \cdot A_{us} Y_n = i n \cdot \partial_{us} - Y_n^\dagger in\cdot \partial_{us} Y_n \,.
\end{align}
In addition, the ultrasoft Wilson lines commute with the label momentum operators
\begin{align}
[Y_n,\cP_\perp^\mu] = 0\,, \qquad [Y_n,\bar \cP] = 0\,.
\end{align}
Denoting the BPS transformation of an operator $\hat{O}$ as $\BPS[\hat{O}]$, we then have the following transformations for the derivative operators
\begin{align}
	\BPS[iD^\mu_{n \perp}] &=  Y_n iD^{\mu}_{n \perp} Y_n^\dagger\,, \qquad \BPS[i\cD^\mu_{n\perp}]  = Y_n i\cD^{\mu}_{n \perp} Y_n^\dagger\,, \qquad
    \BPS[i\cD^\mu_{ns}] =Y_n i\cD^{ \mu}_{n } Y_n^\dagger\,.
\end{align}
Additional useful relations are given in \App{app:BPS_identities}.

Given these identities, it is now a straightforward algebraic exercise to compute the BPS field redefinitions of the Lagrangians. By applying the unitarity condition on the ultrasoft Wilson lines, all ultrasoft Wilson lines can either be cancelled, or absorbed into gauge invariant soft quark or gluon fields, as defined in \Eqs{eq:usgaugeinvdef}{eq:softgluondef}. To illustrate explicitly how this works, we consider two simple examples. First, consider a term from the leading power collinear gluon Lagrangian,
\begin{align}
	&\BPS\left[\tr \big\{ ([i \cD^\mu_{ns}, i \cD^\nu_{ns}])^2\big\} \right]    
	= \tr \Big\{ \left([ Y_n i \cD^{(0)\mu}_{n} Y_n^\dagger, Y_n i \cD_n^{(0)\nu} Y_n^\dagger]\right)^2\Big\}  \nn \\
	&\hspace{2cm}= \tr \Big\{ \left(Y_n [ i \cD^{(0)\mu}_{n} ,  i \cD_n^{(0)\nu} ] Y_n^\dagger\right)^2\Big\}
	= \tr \Big\{ \left([i \cD^{(0)\mu}_{n} , i \cD_n^{(0)\nu} ]\right)^2\Big\}\,.
\end{align}
In this case,  all the soft Wilson lines explicitly cancel, decoupling the interactions of the ultrasoft and collinear gluons. As a second example we consider a term from $\cL^{(1)}$ which contains an explicit $\cD_{us}$. Here we find that the ultrasoft gluons do not decouple
\begin{align}
	&\BPS\left[\tr\Big\{ \big[ i {\cal D}_{ns}^\mu,i {\cal D}_{n\perp}^\nu \big]\big[i {\cal D}_{ns\mu},iD^\perp_{us\,\nu} \big]\Big\} \right] \,=\, \tr\Big\{ \big[ Y_n i \cD_{n}^{(0)\mu} Y^\dagger_n,\, Y_n i \cD^{(0)\, \nu}_{n\perp}Y^\dagger_n \big]\big[Y_n i \cD^{(0)}_{n\,\mu} Y^\dagger_n,iD^\perp_{us\,\nu} \big]\Big\}  \nn\\[0.2cm]
	&= \tr\Big\{ \big[  i \cD_{n}^{(0)\mu} ,\, i \cD^{(0)\, \nu}_{n\perp} \big] \big[ i \cD^{(0)}_{n\,\mu},Y^\dagger_n iD^\perp_{us\,\nu} Y_n \big]\Big\} \, = \, \tr\Big\{ \big[  i \cD_{n}^{(0)\mu} ,\, i \cD^{(0)\, \nu}_{n\perp} \big] \big[ i \cD^{(0)}_{n\,\mu},i\partial_{us\, \nu}^\perp + g\cB^\perp_{us\,\nu} \big]\Big\}\,.
\end{align}
In the last step we used the definition of the gauge invariant ultrasoft gluon field. The derivation of the BPS field redefinition for other terms in the Lagrangian proceeds similarly, so in the following sections we will simply state the final results for the BPS redefined Lagrangians.

\subsection{Simplifications Using the Equations of Motion}\label{sec:eom}

In addition to writing the subleading power Lagrangians in terms of the non-local gauge invariant quark and gluon fields, we can also simplify their structure using the equations of motion. Recall that when building bases of hard scattering operators, only the gauge invariant building blocks in \Tab{tab:PC} are required. In particular, for the collinear gluon field, only the two degrees of freedom in $\cB_{n\perp}$ appear explicitly, and not the other components of $\cB_{n}$. In particular, the large components of the gauge field $\bar n \cdot A_n$ appear entirely in Wilson lines, and the small components have been eliminated using the equations of motion. We begin by reviewing how this is achieved, following the results of \cite{Marcantonini:2008qn}, and then apply the same simplifications to the subleading power Lagrangians.

In SCET the collinear gauge invariant covariant derivative is given by
\begin{align}
\cD^\mu_n=W_n^\dagger D^\mu_n W_n\,.
\end{align} 
which can be broken into components as
\begin{align}
i\cD^{\perp \mu}_n &=\cP^\mu_{n\perp}+g\cB^\mu_{n\perp}\,, \qquad i\overleftarrow{\cD}^{\perp \mu}_n =-\cP^{\dagger\mu}_{n\perp}-g\cB^\mu_{n\perp}\,, \nn \\
in\cdot \cD_n&=in\cdot \partial +gn\cdot \cB_n\,,  \qquad in\cdot \overleftarrow{\cD_n}=in\cdot \overleftarrow{\partial} -gn\cdot \cB_n\,, \qquad
i\bar n \cdot \cD_n=\bar \cP\,,
\end{align}
where we have defined the gauge invariant fields for the different components as
\begin{align}
g\cB^\mu_{n\perp}=\left[  \frac{1}{\bar \cP_n}   [i\bar n \cdot \cD_n, i\cD_n^{\perp \mu}]  \right]\,, \qquad gn \cdot \cB_n = \left[  \frac{1}{\bar \cP_n}   [i\bar n \cdot \cD_n, in\cdot \cD_n]  \right]\,.
\end{align}
Here the $\bar \cP_n$ operators act only within the external square brackets.
We can now eliminate the $n\cdot \cB_n$ component of the gluon field using the equation of motion
\begin{align}\label{eq:eomndotb}
\bar n \cdot \cP g n\cdot \cB_n=-2\cP_{\perp} \cdot \cB^{n\perp}_\nu +\frac{4}{\bar n \cdot \cP} g^2 T^A \sum\limits_f  \bar \chi_n^f T^A \frac{\Sl{\bn}}{2} \chi_n^f +\frac{2}{\bar n \cdot \cP}[\cB^\perp_{n\nu},[\bar n \cdot \cP g\cB^\perp_{n\nu}] ]\,.
\end{align}
This allows the Lagrangian to be written entirely in terms of $\cB_{n\perp}$ fields. From the form of \Eq{eq:eomndotb} we can see why this will lead to significant simplifications when studying soft emissions from a single collinear gluon, since all terms on the right hand side involve either a higher number of fields, or the $\cP_\perp$ operator. When studying soft emission at lowest order and lowest multiplicity, any term of the form $\cB_{us} \cB_{n} n\cdot \cB_{n}$ can therefore be dropped, which will simplify our discussion of the radiative functions.

Additionally, it is also possible to eliminate from the Lagrangian all instances of the ultrasoft derivative operator $n\cdot \partial_n$ acting on $n$-collinear fields. This is achieved for the collinear quark field using the equation of motion
\begin{align}
in\cdot \partial_n \chi_n=-g n\cdot \cB_n \chi_n -i \Sl{\cD}_n^\perp \frac{1}{\bar \cP}  i \Sl{\cD}_n^\perp \chi_n\,,
\end{align}
and for the collinear gluon field using
\begin{align} \label{eq:eomndotpartialB}
&\bar \cP [in\cdot \partial_n g\cB^\mu_{n\perp}]=-\left[ \cP^\perp_\nu [g\cB^\mu_{n\perp},g\cB^\nu_{n\perp}]   \right]     -  \left[ \cB^{n\perp}_\nu [g\cP^{[\mu}_\perp,g\cB^{\nu]}_{n\perp}]   \right]    -\left[ \cB^\perp_{n\nu} [g\cB^\mu_{n\perp},g\cB^\nu_{n\perp}]   \right] \\
&+\frac{\bar \cP}{2}[\cP^\mu_\perp g n\cdot \cB_n]   -\left[ \cP^\perp_\nu \cP^{[\mu}_\perp g\cB_n^{\nu ]}    \right]   +\frac{\bar \cP}{2}[\cB^\mu_{n\perp} g n\cdot \cB_n]    -\frac{1}{2}   \left[  gn\cdot \cB_n,[\bar \cP g\cB_{n\perp}^\mu]  \right]  \nn \\
&-g^2 T^A \sum\limits_f \left[ \bar \chi_n^f T^A \gamma^\mu_\perp \frac{1}{\bar \cP^\dagger}  (\Sl{\cP}_\perp^\dagger +g \Sl{\cB}_{n\perp})   \frac{\Sl{\bn}}{2}  \chi_n^f   \right] -g^2 T^A \sum\limits_f   \left[  \bar \chi_n^f \frac{\Sl{\bn}}{2}(\Sl{\cP}_\perp^\dagger +g \Sl{\cB}_{n\perp}) \frac{1}{\bar \cP^\dagger} T^A \gamma_\perp^\mu \chi_n^f  \right]\,. \nn
\end{align}
These equations of motion, particularly for the gluon case are considerably more cumbersome. When writing the full Lagrangian, as well as for performing fixed order calculations, we therefore find it simpler to work with ultrasoft derivatives. However, we note that if we are interested in tree level soft emissions off of a single collinear line, an identical discussion as for $n\cdot B_n$ applies, and we can ignore all appearances of $n \cdot \partial_{us}$ acting on $n$ collinear fields in the Lagrangian. By using these equations of motion, we are therefore able to greatly simplify the structure of the radiative functions we consider.

\subsection{Lagrangian at $\cO(\lambda^0)$}\label{sec:lp_Lagrangian}

For completeness, we begin by considering the leading power SCET Lagrangian. Those familiar with the leading power BPS field redefinition and SCET Lagrangian can skip to the next section. Before BPS field redefinition, the leading power Lagrangian involves interactions between collinear and ultrasoft particles. It can be written as \cite{Bauer:2000ew, Bauer:2000yr, Bauer:2001ct, Bauer:2001yt}
\begin{align} \label{eq:leadingLag_2}
\cL_{\dyn}^{(0)} &= \cL^{(0)}_{n \xi} + \cL^{(0)}_{n g} +  \cL^{(0)}_{us}\,, 
\end{align}
where
\begin{align}
\cL^{(0)}_{n \xi} &= \bar{\xi}_n\big(i n \cdot D_{ns} + i \slashed{D}_{n \perp} W_n \frac{1}{\overline{\cP}_n} W_n^\dagger i \slashed{D}_{n \perp} \big)  \frac{\slashed{\bar{n}}}{2} \xi_n\,, \\
\cL^{(0)}_{n g} &= \frac{1}{2 g^2} \tr \big\{ ([i D^\mu_{ns}, i D^\nu_{ns}])^2\big\} + \zeta \tr \big\{ ([i \partial^\mu_{ns},A_{n \mu}])^2\big\}+2 \tr \big\{\bar{c}_n [i \partial_\mu^{ns}, [i D^\mu_{ns},c_n]]\big\} \,,\nn 
\end{align}
and the ultrasoft Lagrangian, $\cL^{(0)}_{us}$, is simply the QCD Lagrangian. Throughout this paper, we use a general covariant gauge with gauge fixing parameter $\zeta$ for the collinear gluons, and $c_n$ are the corresponding ghosts.

After performing the BPS field redefinition we have
\begin{align}
\cL^{(0)\text{BPS}} &= \cL^{(0)\text{BPS}}_{n \xi} + \cL^{(0)\text{BPS}}_{n g} +  \cL^{(0)}_{us}\,, 
\end{align}
where the ultrasoft Lagrangian is unchanged. The collinear quark Lagrangian is given by
\begin{align}
\cL^{(0)\text{BPS}}_{n \xi   }&=\bar \chi_n  \left(  i n \cdot \cD_{n}  + i \slashed{\cD}_{n \perp}  \frac{1}{\overline{\cP}_n}  i \slashed{\cD}_{n \perp}  \right)  \frac{\Sl \bn}{2} \chi_n\,,
\end{align}
and the collinear gluon Lagrangian is given by\footnote{Note that $\tr \big\{ ([i \cD^\mu_{n}, i \cD^\nu_{n}])^2\big\} = \tr \big\{ W_n^\dagger([i D^\mu_{n}, i D^\nu_{n}])^2W_n\big\} = \tr \big\{ ([i D^\mu_{n}, i D^\nu_{n}])^2\big\} $ which is the form sometimes used in the literature to write down this term of the collinear leading power Lagrangian.}
\begin{align}
\cL^{(0)\text{BPS}}_{n g} &=  \frac{1}{2 g^2} \tr \big\{ ([i \cD^\mu_{n}, i \cD^\nu_{n}])^2\big\} + \zeta \tr \big\{ ([i \partial^\mu_{n},A_{n \mu}])^2\big\}+2 \tr \big\{\bar{c}_n [i \partial_\mu^{n}, [i D^\mu_{n},c_n]]\big\} \,,
\end{align}
explicitly showing that ultrasoft and collinear interactions have been decoupled to leading power. 

\subsection{Lagrangian at $\cO(\lambda)$}\label{sec:lam_Lagrangian}

Before BPS field redefinition, the $\mathcal{O}(\lambda)$ Lagrangian can be written 
\begin{align}
\cL^{(1)}={\cal L}_{\chi_n}^{(1)}+{\cal L}_{A_n}^{(1)}+{\cal L}_{\chi_n q_{us}}^{(1)} \,,
\end{align}
where \cite{Chay:2002vy,Pirjol:2002km,Manohar:2002fd,Bauer:2003mga}
\begin{align}\label{eq:sublagcollq_2}
{\cal L}_{\chi_n}^{(1)} &= \bar \chi_n \Big(
i \slashed{D}_{us\perp}\frac{1}{ \bar \cP} 
i \slashed{\cal D}_{n\perp}
+i \slashed{\cal D}_{n\perp}\, \frac{1}{\bar \cP} 
i \slashed{D}_{us\perp} 
\Big)\frac{\slashed{\bar{n}}}{2} \chi_n \ ,
 \end{align}
describes the interactions between collinear quarks and gluons, and
\begin{align}
{\cal L}_{A_n}^{(1)}&= \frac{2}{g^2}\text{Tr}\Big(
\big[ i {\cal D}_{ns}^\mu,i {\cal D}_{n\perp}^\nu \big]\big[
i {\cal D}_{ns\mu},iD^\perp_{us\,\nu} 
\big]
\Big)  
+ 2 \zeta \text{Tr} \left(   [i D_{us\perp}^\mu, A_{n\perp \mu}] [i\partial^\nu_{ns},A_{n\nu}] \right) \nn \\
&+2 \text{Tr}   \left( \bar c_n [iD_{us\perp}^\mu, [iD^\perp_{n\mu}, c_n  ]]  \right)          +2 \text{Tr}   \left( \bar c_n [\cP_\perp^\mu,[ W_n iD^\perp_{us\,\mu} W_n^\dagger, c_n   ]]  \right)    \,,  
\end{align}
describes the dynamics of the pure gluon sector, including gauge fixing terms\footnote{Note that the presence of power suppressed gauge fixing Lagrangians is necessary due to the fact that RPI symmetry connects Lagrangians at different orders in the power counting, and would be broken if they were not included.  For example, these subleading power gauge fixing Lagrangians have been shown to give important contributions to the derivation of the LBK theorem for gluons in SCET, see Appendix D of \cite{Larkoski:2014bxa}.}, and
\begin{align}
{\cal L}_{\chi_n q_{us}}^{(1)} 
&= \bar{\chi}_n  g \slashed{\cB}_{n\perp} q_{us}+\text{h.c.,}
\end{align}  
describes the coupling of soft and collinear quarks.

We now wish to express the subleading power Lagrangians in a simplified form in terms of the gauge invariant building blocks, which will be one of the main results of this paper. This organization of the Lagrangians after BPS field redefinition was also considered in \cite{Larkoski:2014bxa}, although there it was performed schematically. Here we will provide explicit expressions for all components, as well as use the equations of motion to simplify the result so that it can easily be used for subleading power factorization.

After performing the BPS field redefinition, we can perform the same division of the Lagrangian as above,  
\begin{align}
\cL^{(1)\text{BPS}}={\cal L}_{\chi_n}^{(1)\text{BPS}}+{\cal L}_{A_n}^{(1)\text{BPS}}+{\cal L}_{\chi_n q_{us}}^{(1)\text{BPS}} \,,
\end{align}
where the collinear quark Lagrangian is given by
\begin{align} 
{\cal L}_{\chi_n}^{(1) \text{BPS}} &=g \bar \chi_n  \cB^\perp_{us(n)} \cdot \cP_\perp \frac{\Sl \bn}{\bar \cP} \chi_n   +   g \bar \chi_n  \partial^\perp_{us(n)} \cdot \cP_\perp \frac{\Sl \bn}{\bar \cP} \chi_n 
 +\bar \chi_n \left( i\Sl{\partial}_{\perp us}  \frac{1}{\bar \cP} g \Sl{\cB}_{n\perp} +g\Sl{\cB}_{n\perp}\frac{1}{\bar \cP} i\Sl{\partial}_{\perp us}   \right) \frac{\Sl{\bar n}}{2} \chi_n  \nn \\  
&+\bar \chi_n \left( i\Sl{\cB}_{\perp us(n)}  \frac{1}{\bar \cP} g \Sl{\cB}_{n\perp} +g\Sl{\cB}_{n\perp}\frac{1}{\bar \cP} i\Sl{\cB}_{\perp us(n)}   \right) \frac{\Sl{\bar n}}{2} \chi_n \,, 
\end{align}
the collinear gluon Lagrangian is divided into three pieces
\begin{align}
{\cal L}_{A_n}^{(1)\text{BPS}}={\cal L}_{g_n}^{(1) \text{BPS}}  +  {\cal L}_{\text{ghost}}^{(1) \text{BPS}}  +  {\cal L}_{\text{gf}}^{(1) \text{BPS}}\,,
\end{align}
which are given by
\begin{align}
{\cal L}_{g_n}^{(1) \text{BPS}}&=[\cP^{\nu}_{\perp} \cB^{\mu a}_\perp][i\partial^\perp_{us \nu}  \cB^{\mu a}_{n\perp}]-[\cP^\mu_\perp \cB^{\nu a}_\perp][i\partial^\perp_{us \nu} \cB^{\mu a}_{n\perp}] \nn \\
&~~~\,-igf_{abc} \left\{ \cB^{\nu a}_{us \perp} \cB^{\mu b}_{n\perp}[\cP^\mu_\perp \cB^{\nu c}_{n\perp}] -\cB^{\nu a}_{us \perp} \cB^{\mu b}_{n\perp}[\cP^\nu_\perp \cB^{\mu c}_{n\perp}] + \cB^{\mu a}_{n\perp} \cB^{\nu b}_{n\perp}  [i\partial_{us \nu}^\perp g \cB_{n\perp}^{\mu c}] \right\}\nn \\
&~~~\,+g^2 f_{abe} f_{cde} \cB^{\mu a}_{n\perp}   \cB^{\nu b}_{n\perp} \cB^{\nu c}_{us \perp} \cB^{\mu d}_{n\perp}  \nn \\
&~~~\,+ [\bar \cP \cB^{\nu a}_{n\perp}] [i\partial^\perp_{us \nu} n\cdot \cB^a_n] + ig f^{abc}[\bar \cP \cB^{\nu a}_{n\perp}] n\cdot \cB^b_n \cB_{us \perp}^{c\nu}\,,\nn \\
{\cal L}_{\text{ghost}}^{(1) \text{BPS}} &= 2 \text{Tr}  \left(    \bar c_n [ i \partial^\mu_{us\perp},[i D^\perp_{n\mu}, c_n]] \right)    +2 \text{Tr}  \left(    \bar c_n [ T^a,[i D^\perp_{n\mu}, c_n]]        \right)   g \cB_{us(n)}^{a\mu} \,   \nn \\
&+2 \text{Tr}  \left(    \bar c_n  [ \cP^\mu_\perp  , [W_n   i \partial_{us\perp\mu}   W^\dagger_n, c_n] ]   \right)         +2 \text{Tr}  \left(    \bar c_n  [ \cP_{\perp \mu}  , [W_n   T^a   W^\dagger_n, c_n] ]   \right)   g \cB_{us(n)}^{a\mu} \,, \nn \\
{\cal L}_{\text{gf}}^{(1) \text{BPS}} &= 2\zeta \text{Tr}\left([i\partial^\mu_{us\perp},A_{n\perp \mu}][i \partial_n^\nu,A_{n\nu}]    \right)    + 2\zeta \text{Tr}\left( [T^a,A_{n\perp \mu}][i \partial_n^\nu,A_{n\nu}]     \right) g \cB_{us(n)}^{a\mu} \,.
\end{align}
Finally, the interaction of soft quarks is described by the Lagrangian
\begin{align}
{\cal L}_{\chi_n \psi_{us}}^{(1)\text{BPS}} 
&= \bar{\chi}_n g \slashed{\cB}_{n\perp} \psi_{us (n)}+\text{h.c.} \,.
\end{align}

The structure of the $\cO(\lambda)$ Lagrangian is quite complicated, since it describes the complete dynamics of the subleading power corrections to the soft and collinear dynamics, including ghost and gauge fixing terms. In its current form, it also involves multiple polarizations of the collinear gluon field.  To simplify its structure, we use the equations of motion\footnote{Note that the EOM are homogeneus in the power counting $\lambda$, but not in the coupling constant. Therefore the use of the EOM can reshuffle terms among different orders in $g$, but it won't move terms between Lagrangians at different orders.}, as discussed in \Sec{sec:eom}. Simplifying the result to focus only on ultrasoft emissions out of two collinear fields we find the structure
\begin{align}\label{eq:L1_complete}
{\cal L}_{n}^{(1)\text{BPS}} =&  -2g[\cP_\perp^\mu \, \cB_{n\perp }^\nu] [\cB_{n  \nu}^\perp , \cB^{\perp}_{us \,\mu}] +g \bar \chi_n  \cB^\perp_{us(n)} \cdot \cP_\perp \frac{\Sl \bn}{\bar \cP} \chi_n + \bar{\chi}_n g \slashed{\cB}_{n\perp} \psi_{us (n)}+\text{h.c.} \nn \\
&+ \cB_{us} \cdot K^{(1)}_{\cB_{us}} + K^{(1)}_{\partial_{us}}\,,
\end{align}
where $K^{(1)}_{\cB_{us}}$ and $K^{(1)}_{\partial_{us}}$ contain $\geq 3$ collinear fields, and are therefore not relevant for our current analysis. 
In this form, the Lagrangian is written entirely in terms gauge invariant fields, and due to the organization in terms of fields, it is clear at which order in perturbation theory each term can contribute. After performing the BPS field redefinition, and writing the result in terms of collinear and soft gauge invariant fields, the soft and collinear fields are only coupled through Lorentz and color indices, as well as through potential derivative operators. Since each of the building blocks appearing in the Lagrangian is separately gauge invariant, this will allow for a simple factorization into collinear and soft components, tied together through Lorentz and color indices, which will give rise to the radiative functions.

\begin{figure}[t!]
\begin{center}
\begin{align}
	\fd{3cm}{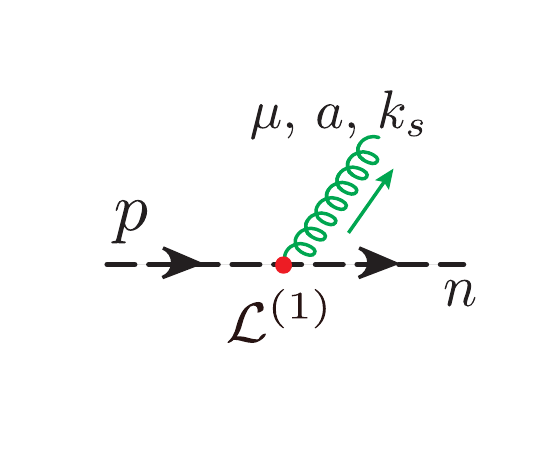}~&=~\frac{2igT^a}{\bn \cdot p}\left(p_\perp^\mu - \frac{k_s^\perp \cdot p_\perp}{n\cdot k_s} n^\mu\right)\frac{\bnslash}{2}\nn \\
	\fd{3cm}{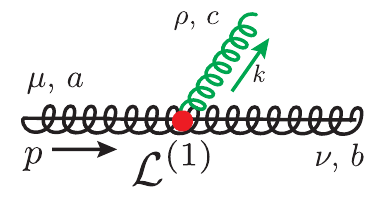}&=2gf^{abc}\left[  g_\perp^{\mu \nu} p_\perp^\rho -g_\perp^{\mu \nu} p_\perp\cdot k_\perp  \frac{n^\rho}{n\cdot k} \right.  \nn\\
&\hspace{2cm}\left.- \left(p_\perp^\rho - p_\perp\cdot k_\perp  \frac{n^\rho}{n\cdot k} \right)\left( p_\perp^\mu \frac{\bn^\nu}{\bn \cdot p} + p_\perp^\nu \frac{\bn^\mu}{\bn \cdot p} -p_\perp^2 \frac{\bn^\mu\bn^\nu}{(\bn \cdot p)^2}  \right) \right]\nn \\
\fd{3cm}{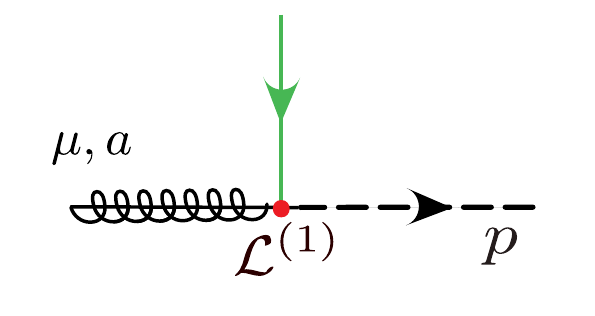}&=igT^a\left( \gamma_\perp^\mu - \frac{\Sl{p}_{\perp}\, \bar{n}^\mu}{\bar n \cdot p} \right)	\nn
\end{align}
\end{center}
\vspace{-0.4cm}
\caption{ Feynman rules for the $\cO(\lambda)$ Lagrangian describing the emission of a soft gluon or quark.
 } 
\label{fig:lam_feynrules}
\end{figure}

The first three terms of \Eq{eq:L1_complete} describe the $\cO(\lambda)$ emission of a soft gluon from a collinear gluon, a soft gluon from a collinear quark, and a soft quark from a collinear quark or gluon, respectively. Using the Lagrangian, we can derive the tree level Feynman rules, which are given in \Fig{fig:lam_feynrules}. Note that in accord with the LBK theorem, the single ultrasoft gluon Feynman rule of $\cL^{(1)}$ vanishes when the label $\perp$ momentum of the collinear leg is set to zero. Unlike for the emission of a soft gluon, the Feynman rule for a soft quark emission does not vanish when the $\cP_\perp$ of the collinear line vanishes.

Since the Lagrangian is defined in terms of gauge invariant soft quark and gluon fields, which involve ultrasoft Wilson lines, they also give the Feynman rules for an arbitrary number of additional leading power soft gluon emissions.  Similarly, the gauge invariant collinear fields also involve collinear Wilson lines, which describe collinear radiative corrections to the above Feynman rules.   
The $K^{(1)}_{\cB_{us}}$ and $K^{(1)}_{\partial_{us}}$ in \Eq{eq:L1_complete} involve additional collinear fields. For a single soft emission from a collinear line, these can first appear at loop level. We will not work out the explicit form of these loop contributions in this initial paper, however,  we will discuss their contributions in later sections.

\subsection{Lagrangian at $\cO(\lambda^2)$}\label{sec:lam2_Lagrangian}

At $\mathcal{O}(\lambda^2)$ the SCET Lagrangian before BPS field redefinition can be written as \cite{Pirjol:2002km,Manohar:2002fd,Bauer:2003mga}
\begin{align}
	{\cal L}^{(2)} &= {\cal L}_{\chi_n}^{(2)}  + {\cal L}_{A_n}^{(2)} 
	+ {\cal L}_{\chi_n q_{us}}^{(2)} \,,
\end{align}
where for convenience, we further decompose the gluon Lagrangian as
\begin{align}
{\cal L}_{A_n}^{(2)}={\cal L}_{g_n}^{(2) }  +  {\cal L}_{\text{ghost}}^{(2) }  +  {\cal L}_{\text{gf}}^{(2) }\,.
\end{align}
The different components of the Lagrangian are given by
\begin{align}
	\cL_{\chi_n q_{us}}^{(2)}&=  \bar \chi_n \frac{\Sl \bn}{2} [ W_n^\dagger in\cdot D W_n]  q_{us}    + \bar \chi_n \frac{\Sl \bn}{2}  i\Sl \cD_{n\perp}    \frac{1}{\overline{\cP}}   ig \Sl \cB_{n\perp}  q_{us}+ \text{h.c.} \,, \nn \\
	\cL_{\chi_n}^{(2)}&=  \bar \chi_n   \left( i\Sl D_{us\perp} \frac{1}{\overline{\cP}}  i\Sl D_{us\perp}-  i\Sl \cD_{n\perp}   \frac{i\bn \cdot D_{us}}{(\overline{\cP})^2}   i\Sl \cD_{n\perp}   \right)    \frac{\Sl \bn}{2} \chi_n\,, \nn \\
	\cL_{ng}^{(2)}&=\frac{1}{g^2} \text{Tr} \left(  [  i\cD^\mu_{ns} , iD_{us}^{\perp \nu}   ]    [ i \cD_{ns\mu}   ,i D^\perp_{us\nu}   ] \right)   +\frac{1}{g^2} \text{Tr} \left(  [ iD^\mu_{us\perp}  ,i D^\nu_{us\perp}   ]    [ i\cD^\perp_{n\mu}  , i\cD^\perp_{n\nu}   ] \right) \nn \\
	&\hspace{-0.4cm}+ \frac{1}{g^2} \text{Tr} \left(  [ i\cD_{ns}^{\mu}  , i n\cdot \cD_{ns}  ]    [  i\cD_{ns\mu} , i\bn \cdot D_{us}   ] \right)+\frac{1}{g^2} \text{Tr} \left(  [ iD^\mu_{us\perp}  ,  i\cD^\perp_{n\nu}  ]    [  i\cD^\perp_{n\mu}  ,  iD_{us \perp}^\nu  ] \right)\,, \nn \\
	\cL_{\text{gf}}^{(2)}&= \zeta \text{Tr} \left(   [ i D^\mu_{us\perp}, A_{n\perp \mu}]   [i D^\nu_{us\perp}, A_{n\perp \nu}] \right)   + \zeta \text{Tr} \left(   [i \bn \cdot D_{us},n \cdot A_n]   [i \partial^\mu_{ns}, A_{n\mu}] \right) \,, \nn \\
	\cL_{\text{ghost}}^{(2)}&= 2 \text{Tr} \left(  \bar c_n [iD^\mu_{us\perp},[ W_n iD^\perp_{us\mu}W_n^\dagger ,c_n]] \right)+ \text{Tr} \left( \bar c_n [i \bn \cdot D_{us} ,[ i n\cdot D_{ns},c_n]] \right) \nn \\
	&+ \text{Tr} \left( \bar c_n [\overline{\cP},[W_n i \bn \cdot D_{us} W_n^\dagger,c_n]] \right)\,.
\end{align}
After performing the BPS field redefinition, and writing the result in terms of ultrasoft gauge invariant fields, we find that the Lagrangians involving quark fields can be written
\begin{align}
{\cal L}_{\chi_n}^{(2) \text{BPS}} &= \bar \chi_n  \left( i \Sl \partial_{us\perp}  \frac{1}{ \bar \cP} i \Sl \partial_{us\perp}    \right)   \frac{\Sl \bn}{2} \chi_n \nn \\
&-\bar \chi_n  \frac{i\bar n \cdot \partial_{us} }{\bar \cP^2} \cP_\perp^2 \frac{\Sl{\bar n}}{2} \chi_n              -\bar \chi_n g \Sl{\cB}_\perp  \frac{i\bar n \cdot \partial_{us} }{\bar \cP^2} \Sl{\cP}_\perp \frac{\Sl{\bar n}}{2} \chi_n    \nn \\
&-\bar \chi_n  \frac{i\bar n \cdot \partial_{us} }{\bar \cP^2} \Sl{\cP}_\perp \Sl{\cB}_\perp \frac{\Sl{\bar n}}{2} \chi_n              -\bar \chi_n g \Sl{\cB}_\perp  \frac{i\bar n \cdot \partial_{us} }{\bar \cP^2} \Sl{\cB}_\perp \frac{\Sl{\bar n}}{2} \chi_n \nn \\
&+   \bar \chi_n  \left(  T^a \gamma^\mu_\perp \frac{1}{\bar \cP}  i \Sl \partial_{us\perp} - i  {\overleftarrow{\Sl \partial}}_{us\perp} \frac{1}{\bar \cP} T^a \gamma^\mu_\perp    \right)   \frac{\Sl \bn}{2} \chi_n  \,   g \cB_{us(n)}^{a\mu} \nn \\
&-\bar \chi_n T^a \frac{\cP_\perp^2 }{\bar \cP^2}  \frac{\Sl{\bar n}}{2} \chi_n    g \bar n \cdot\cB_{us(n)}^{a}        -\bar \chi_n g \Sl{\cB}_\perp  \frac{T^a }{\bar \cP^2} \Sl{\cP}_\perp \frac{\Sl{\bar n}}{2} \chi_n   g \bn \cdot \cB_{us(n)}^{a} \nn \\
&-\bar \chi_n \Sl{\cP}_\perp  \frac{T^a  }{\bar \cP^2}  \Sl{\cB}_\perp \frac{\Sl{\bar n}}{2} \chi_n      g\bn \cdot \cB_{us(n)}^{a}        -\bar \chi_n g \Sl{\cB}_\perp  \frac{T^a  }{\bar \cP^2} \Sl{\cB}_\perp \frac{\Sl{\bar n}}{2} \chi_n g \bn \cdot \cB_{us(n)}^{a} \nn \\
&+    \bar \chi_n  \left( T^a \gamma^\mu_\perp \frac{1}{\bar \cP} T^b \gamma^\nu_\perp   \right)   \frac{\Sl \bn}{2} \chi_n  \,   g \cB_{us(n)}^{a\mu}   g \cB_{us(n)}^{b\nu}\nn \\
&+\bar \chi_n \frac{\Sl \bn}{2}  in\cdot \cB_n  \psi^{(n)}_{us}    + \bar \chi_n \frac{\Sl \bn}{2}  i\Sl \cD_{n\perp}    \frac{1}{\overline{\cP}}   ig \Sl \cB_{n\perp}  \psi^{(n)}_{us}+ \text{h.c.}\,.
\end{align}

The Lagrangians describing the pure glue sector are more complicated, involving both ghost and gauge fixing terms. We find that they can be written
\begin{align}\label{eq:subsubgluonlagr}
	\cL_{ng}^{(2)\text{BPS}}&=\text{Tr} \left(i n\cdot \partial\cB_{n \mu}^\perp    i \bn\cdot \partial\cB^{\perp\mu}_{n} - [ \cPbar n\cdot \cB_{n}  ]    i\bn \cdot \partial n\cdot\cB_{n} -[\cP_\perp^\mu n \cdot \cB_n ] i\bn \cdot \partial  \cB_{n\mu}^\perp \right) \nn \\
	&+ g \text{Tr} \left( \partial_\perp^{[\mu} \cB_{us \perp}^{\nu]}[ \cB^\perp_{n\mu}  , \cB^\perp_{n\nu}   ] - in\cdot \partial\cB_{n\mu}^\perp [ \cB_{n\mu}^\perp , \bn \cdot \cB^{(n)}_{us}  ]  \right. \nn \\
	&\left.\qquad\quad +\cP_\perp^\mu n \cdot \cB_n [ \cB_{n\mu}^\perp , \bn \cdot \cB^{(n)}_{us}  ] + \cPbar n\cdot \cB_{n}     [  n\cdot\cB_{n} , \bn \cdot \cB^{(n)}_{us}   ]   \right)\nn \\
	&+  g^2\text{Tr} \left([ \cB_{us}^{\perp (n) \mu}  ,\cB_{us}^{\perp (n) \nu}   ]    [ \cB^\perp_{n\mu}  , \cB^\perp_{n\nu}   ]   \right) \,, \nn\\
	\cL_{{\text{gf}}}^{(2)\text{BPS}}&= \zeta \text{Tr} \left(   [ i \partial^\mu_{us\perp}+T^a g \cB^{a\mu}_{us(n)} , A_{n\perp \mu}]   [i \partial^\nu_{us\perp}+T^a g \cB^{a\nu}_{us(n)} , A_{n\perp \nu}] \right)   \nn \\
	&+ \zeta \text{Tr} \left(   [i \bn \cdot \partial_{us} +T^a g \bn\cdot \cB^{a}_{us(n)}  ,n \cdot A_n]   [i \partial^\mu_{n}, A_{n\mu}] \right) \,, \nn \\
	\cL_{\text{ghost}}^{(2)\text{BPS}}&= 
	 2 \text{Tr} \left(  \bar c_n [i\partial^\mu_{us\perp}+T^a g  \cB^{a\mu}_{us(n) \perp},[ W_n (i\partial^\perp_{us\,\mu} +T^b g  \cB^{b\perp}_{us(n)\mu})W_n^\dagger ,c_n]] \right) \nn \\
	&+ \text{Tr} \left( \bar c_n [i \bn \cdot \partial_{us}+ T^a g  \bn \cdot\cB^{a}_{us(n)},[ i n\cdot D_{n},c_n]] \right) \nn \\
	&+ \text{Tr} \left( \bar c_n [\overline{\cP},[W_n( i \bn \cdot \partial_{us}+T^a g \bn\cdot \cB^{a}_{us(n)} ) W_n^\dagger,c_n]] \right)\,.
\end{align}

To make the $\cO(\lambda^2)$ Lagrangian more tractable, we can use the equations of motion to write it entirely in terms of our basis of gauge invariant building blocks. This is a straightforward, but tedious algebraic exercise, and therefore we simply present the final result. Using the equations of motion to rewrite the Lagrangian in terms of our operator basis, we find 
\begin{align}\label{eq:lam2_BPS}
	\cL^{(2)\text{BPS}}&= \text{Tr} \left(i n\cdot \partial\cB_{n \mu}^\perp    i \bn\cdot \partial\cB^{\perp\mu}_{n} - 4[ \cP_{n\perp}\cdot \cB_{n\perp} ]    i\bn \cdot \partial \frac{\cP_{n\perp}\cdot \cB_{n\perp}}{\bar n \cdot \cP}  +2\left[\cP_\perp^\mu\frac{\cP_{n\perp}\cdot \cB_{n\perp}}{\bar n \cdot \cP} \right] i\bn \cdot \partial  \cB_{n\mu}^\perp \right)  \nn \\
	&- \bar \chi_n   \frac{\partial^2_{\perp}}{\overline{\cP}}\frac{\Sl \bn}{2} \chi_n+ \nn \\
	&+g\bar \chi_n \frac{\Sl \bn}{2}  i\Sl \cP_{\perp}    \frac{1}{\overline{\cP}}   \Sl \cB_{n\perp}  \psi_{us(n)} - g\bar \chi_n \frac{\Sl \bn}{2} \frac{2}{\overline{\cP}}\cP_\perp \cdot \cB_n  \psi_{us(n)} + \text{h.c.} \nn \\
	&+ \text{Tr} \left(2g\left[i\partial^{\mu}_\perp  \cB^{\nu}_{us}\right]\,[ \cB^\perp_{n\mu}  , \cB^\perp_{n\nu}   ] +g^2 [ \cB_{us}^{\perp \mu}  ,\cB_{us}^{\perp \nu}   ]    [ \cB^\perp_{n\mu}  , \cB^\perp_{n\nu}   ]  \right)  \nn \\ 
	&+g\bar \chi_n   \left( [i\Sl \partial_{us\perp}  \Sl \cB_{us\perp}] + 2 \cB_{us\perp} \cdot   i{\partial}_{us\perp}  \right) \frac{1}{\overline{\cP}}  \frac{\Sl \bn}{2} \chi_n + g^2 \bar \chi_n  \Sl \cB_{us\perp} \Sl \cB_{us\perp}  \frac{1}{\overline{\cP}}\frac{\Sl \bn}{2} \chi_n \nn \\
	&+ \cB_{us} \cdot K^{(2)}_{\cB_{us}} + K^{(2)}_{\partial_{us}}\,, 
\end{align}
where, as in \Eq{eq:L1_complete}, the $K$ contain $\geq3$ collinear fields, which will not be relevant for the discussion in this paper. This gives the Lagrangian at $\cO(\lambda^2)$ in terms of gauge invariant soft and collinear quark and gluon fields in such a way that it is clear at which order in perturbation theory each term can contribute. We have used the EOM to write it entirely in terms of the $\cB_{n\perp}$ field, eliminating the other polarizations. For practical applications, we can also apply the EOM of \Eq{eq:eomndotpartialB}, however, this significantly complicates the structure of the Lagrangian, and therefore we have not written it out explicitly. This simplified form of the $\cO(\lambda^2)$ Lagrangian is one of the key results of this paper. We again emphasize that its highly non-trivial non-local structure, involving a multitude of soft and collinear Wilson lines, is fully determined by the structure of the BPS field redefinition, and the local $\SCETi$ Lagrangians, allowing it to be constructed systematically. This form, in terms of gauge invariant building blocks linked only by Lorentz and color indices, will allow for a straightforward factorization into radiative functions.

\begin{figure}[t!]
\begin{center}
\small
\begin{align}
\fd{5cm}{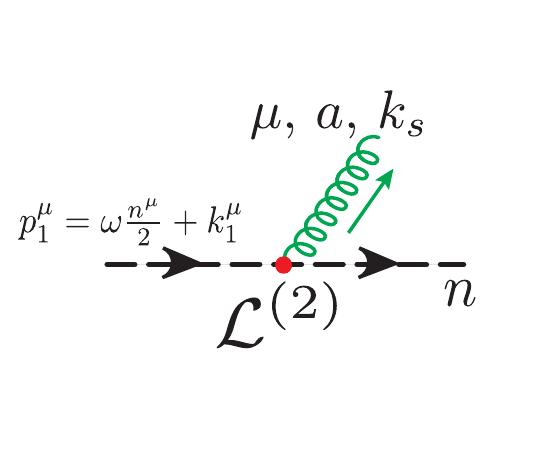}\hspace{-.8cm}&=igT^a\left[ 2 k^\perp_{1 \mu} -\Sl k^\perp_s \gamma^\perp_\mu -\left( 2 k_s^\perp \cdot k_1^\perp -(k_s^\perp)^2 \right) \frac{n_\mu}{n\cdot k_s} \right]\frac{\bnslash}{2\bar n \cdot p}\nn \,,\\
\fd{4cm}{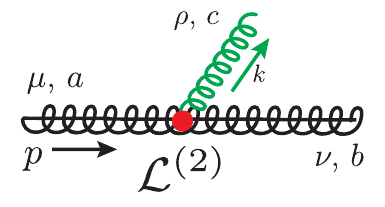}&=2 g f^{abc}\left(k_\perp^\mu g_\perp^{\nu \rho } - k_\perp^\nu g_\perp^{\mu \rho } \right)\nn\,, \\[3mm]
\fd{4cm}{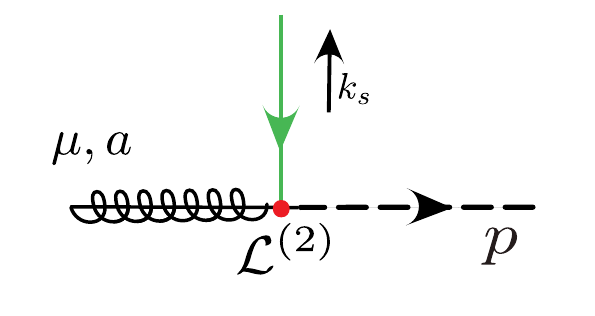}&=-\frac{gT^a}{n \cdot k_s}\left[n^\mu -\frac{n \cdot p }{\bar n \cdot p}\,\bar{n}^\mu  \right]\nn \,,\\
	\fd{5cm}{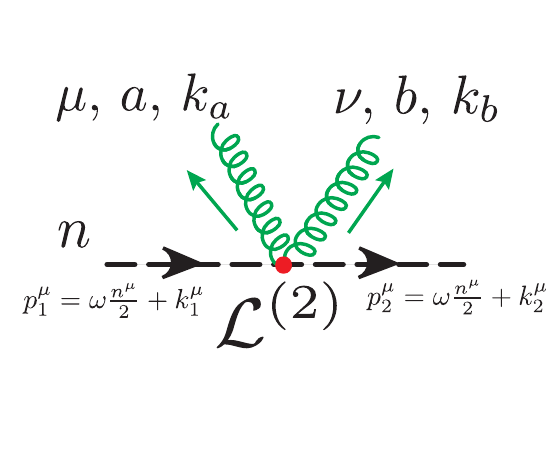}\hspace{-5mm}&=-\frac{g^2 \bnslash}{2\bn\cdot p}\left(\frac{1}{4}[T^a,T^b][\gamma_\perp^\alpha, \gamma_\perp^\beta] + \frac{1}{2}\{T^a,T^b\}g_\perp^{\alpha \beta}\right)\left(g_\perp^{\mu \alpha} g_\perp^{\nu\beta} - \frac{k^\alpha_{a\perp} g_\perp^{\nu\beta}}{n\cdot k_a} n^\mu \right.    \nn \\[-1cm]
	& \left.\quad- \frac{g_\perp^{\mu \alpha}k^\beta_{b\perp}}{n\cdot k_b} n^\nu  +n^\mu n^\nu \frac{k^\alpha_{a\perp}  k^\beta_{b\perp}}{n\cdot k_a\, n\cdot k_b} \right) + \big(a\leftrightarrow b,\mu \leftrightarrow \nu, k_a \leftrightarrow k_b \big)\nn \\
	&=-\frac{g^2 \bnslash}{2\bn\cdot p}T^a T^b\left(\gamma_\perp^\mu \gamma_\perp^\nu - \frac{\Sl{k}_{a\perp} \gamma_\perp^\nu}{n\cdot k_a} n^\mu - \frac{\gamma_\perp^\mu\Sl{k}_{b\perp}}{n\cdot k_b} n^\nu  +n^\mu n^\nu \frac{\Sl{k}_{a\perp}  \Sl{k}_{b\perp}}{n\cdot k_a\, n\cdot k_b} \right) \nn \\
	&\quad+ \big(a\leftrightarrow b,\mu \leftrightarrow \nu, k_a \leftrightarrow k_b \big)\nn\,, \\[2mm]
	\fd{4cm}{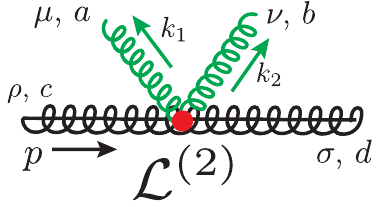}&=-2g^4 f^{abc}f^{cde} \left[ g^\perp_{\mu \rho} g^\perp_{\nu \sigma} -g^\perp_{\mu \sigma} g^\perp_{\nu \rho} + \frac{n^\mu}{n\cdot k_1}\left(g^\perp_{\nu \rho} k_{1 \sigma}^\perp -g^\perp_{\nu \sigma} k_{1 \rho}^\perp \right)  \nn \right. \\
	&\left.\quad - \frac{n^\nu}{n\cdot k_2}\left(g^\perp_{\mu \rho} k_{2 \sigma}^\perp -g^\perp_{\mu \sigma} k_{2 \rho}^\perp \right) + \frac{n^\mu n^\nu}{n\cdot k_1 n\cdot k_2}\left( k_{1 \rho}^\perp k_{2 \sigma}^\perp - k_{1 \sigma}^\perp k_{2 \rho}^\perp \right) \right]\nn\,.
\end{align}
\end{center}
\vspace{-0.4cm}
\caption{ Feynman rules for the $\cO(\lambda^2)$ Lagrangian when $p_\perp^n=0$ describing the emission of a soft gluon or quark, or the double non-eikonal emission of soft gluons.
 } 
\label{fig:lam2_feynrules}
\end{figure}

In \Eq{eq:lam2_BPS}, terms appear involving $0,1$ and $2$ $\cB_{us}$ fields, as well as the gauge invariant ultrasoft quark field. Since the $\cO(\lambda^2)$ Lagrangians contain various terms we are going to give the Feynman rules under the common assumption of vanishing  label perpendicular momentum of all collinear fields to zero $\cP_\perp =0$. Under this assumption the Feynman rules are given in \Fig{fig:lam2_feynrules}.

It is important to emphasize that since $\cB_{us(n)}^{b\nu}$ has Feynman rules with an infinite number of soft emissions, the terms involving one and two $\cB_{us}$ fields will both contribute to the complete two gluon Feynman rule. In the Feynman rules in \Fig{fig:lam2_feynrules} we have given only the contribution from the Lagrangian insertion involving two $\cB_{us}$ fields. For simplicity we have not given the two soft gluon Feynman rule from $\cL_{\cB}^{(2)}$ which can be straightforwardly derived using the two gluon Feynman rule of the ultrasoft gauge invariant gluon field in \eq{twogluonBus}. These contributions are separately gauge invariant.

The other terms in \Eq{eq:lam2_BPS} involve additional collinear fields. For soft emissions from a collinear line, they can first contribute when there are collinear loops. We will not explicitly compute their loop level contributions, but will further discuss their structure in later sections. Note that at $\cO(\lambda^2)$, we also have contributions from two insertions of the $\cO(\lambda)$ operator. These will be discussed when we consider the complete classification of radiative function for $e^+e^-\to$ dijets in \Sec{sec:RF_thrust}.

\section{Amplitude Level Factorization with Radiative Functions}\label{sec:RadiativeFunction_intro}

In this section we derive factorization formulas in terms of radiative functions for soft emissions at amplitude level. While our goal is to study the factorization of event shapes, and the structure of radiative functions at cross section level, initiating our studies at amplitude level is useful for several reasons. First, it is useful for connecting to the study of the subleading power soft behavior of amplitudes, which in itself is an interesting subject to which the factorization theorems that we derive can be applied. Second, it allows us to connect to other work in the literature on radiative functions, which have typically been formulated at amplitude level. Finally, it also provides a slightly simpler situation to illustrate the general features of factorization involving radiative functions, which will persist at the cross section level. In particular, we will illustrate how radiative functions are defined as integrals along the lightcone of Lagrangian insertions, which dress the leading power Wilson lines, giving rise to a breakdown of eikonalization. 

In this section we will only consider those radiative functions that are relevant for describing tree level soft emissions.  In particular, this eliminates all contributions involving more than two collinear fields. Furthermore, we will use RPI to take each collinear sector in the amplitude to have a total $\cP_\perp=0$, which eliminates $\cO(\lambda)$ contributions to radiative soft gluon emission, as is guaranteed by the LBK theorem \cite{Low:1958sn,Burnett:1967km}. See \cite{Larkoski:2014bxa} for a detailed discussion in the context of SCET.
This leaves us with the following cases of interest
\begin{itemize}
\item Single $\psi_{us}$ emission at $\cO(\lambda)$,
\item Single $\cB_{us}$ emission at $\cO(\lambda^2)$,
\item Double $\cB_{us}$ emission at  $\cO(\lambda^2)$,
\end{itemize}
each of which will be studied in this section. Single $\psi_{us}$ emission could also be studied at $\cO(\lambda^2)$, however, due to fermion number conservation in the soft sector, this can first contribute at cross section level at $\cO(\lambda^4)$, and is therefore not of interest to us here. Terms with additional collinear fields, that do not contribute at tree level, can be treated in an identical manner, and we will briefly comment on loop level contributions at the end of this section.

For convenience a summary of the radiative functions is given in \Tab{tab:rad_func_amp}, which shows the schematic factorization of the amplitude, the tree level Feynman rule for the radiative function, as well as the equation number where its definition can be found. The derivation of the factorizations are given in the text.

After extending the formalism of this section to cross section level factorization in \Sec{sec:fact_RadiativeFunction}, a complete classification of the field content of all radiative jet functions contributing to a physical observable, namely thrust in $e^+e^-\to $ dijets, will be given in \Sec{sec:RF_thrust}. This includes those which first contribute at loop level. By fixing a particular physical process, we will be able to exploit the symmetries of the problem to slightly simplify the structure of the radiative contributions.

{
\renewcommand{\arraystretch}{1.4}
\begin{table}[t!]
\scalebox{0.842}{
\hspace{-1.5cm}\begin{tabular}{| c | c | c |c |c|c| r| }
  \hline                       
  $\begin{matrix} \text{Radiative} \\[-3mm] \text{Function} \end{matrix}$& Factorization & Tree Level Feynman Rule& Equation \\
  \hline
  $\left[\cJ_{\psi {n_j}}^{\balpha i} (k^+)\right]^{\beta_j,s_j}$ & $\fd{3cm}{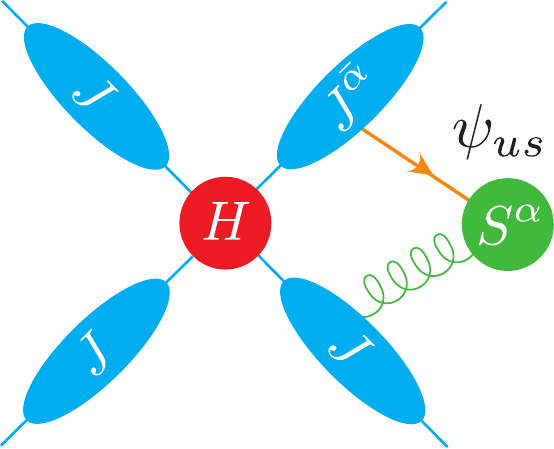}$ &  $\fd{3cm}{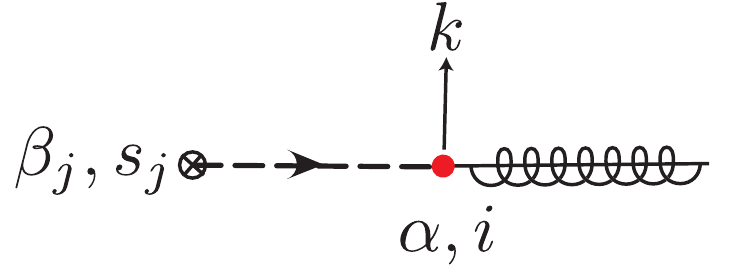} = -\left[ \left( \Sl{\epsilon}^A_\perp - \frac{\Sl{p}_{\perp}\, n\cdot \epsilon^A(p)}{\bar n \cdot p} \right)\frac{\Sl{n}}{2 k^+}\right]_{\balpha, s} T^A_{i, \beta_j}$& \Eq{eq:rad_jet_soft_quark_amplitude} \\
  \hline
   $\left[\cJ^{\mu \nu}_{q,A}(k^+)\right]^{i,s}$ & $\fd{3cm}{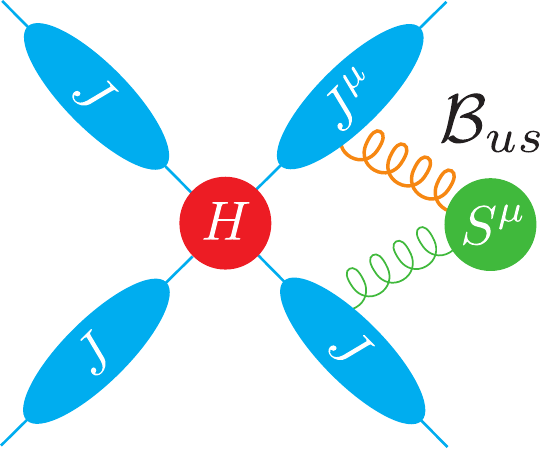}$ & $\fd{3cm}{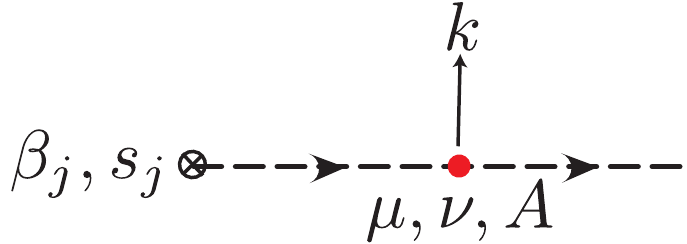}= -\left[\bar{u}_n(p)\frac{gT^A  \gamma_\perp^\mu \gamma_\perp^\nu}{\bar n \cdot p\, n\cdot k}\right]^{i,s}$ &  \Eq{eq:rad_jet_soft_gluon_amplitude}\\
  \hline  
   $\left[\cJ_{g, A}^{\mu \nu}(k)\right]^{M \rho} $ & $\fd{3cm}{figures_amp/amp_sub_gluon_fac_low.pdf}$ & $\fd{3cm}{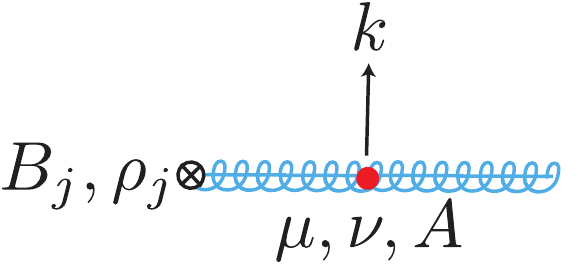}= \frac{-igf^{ABM}}{\bar n \cdot p\, n\cdot k} \left( g_\perp^{\mu \rho}\epsilon_\perp^{\nu B} - g_\perp^{\nu \rho}\epsilon_\perp^{\mu B}\right)\,$ &  \Eq{eq:rad_amp_gluon}\\
   \hline
      $\left[\cJ^{\mu \nu }_{q, A B,n}(k)\right]^{i,s} $ & $\fd{3cm}{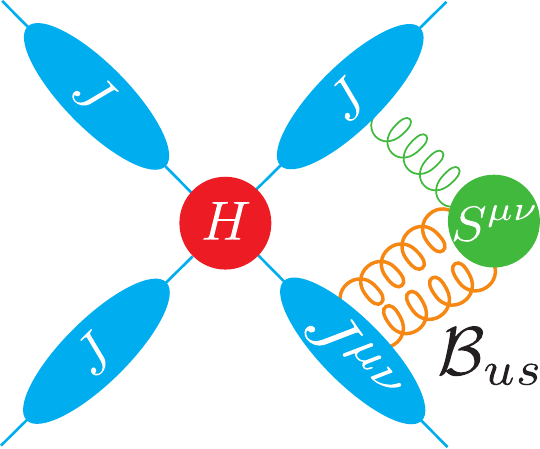}$ & $\fd{3cm}{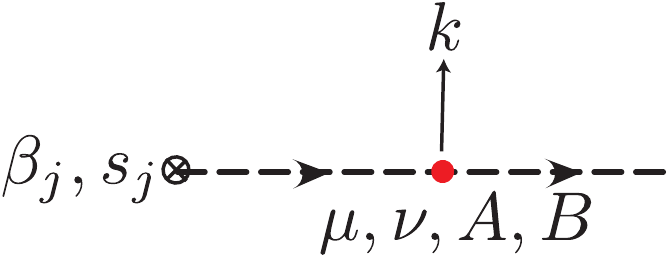}= \frac{i g^2}{\nbar \cdot p n\cdot k}  \left [\bar{u}_n(p)\left([\gamma^\mu_\perp, \gamma^\nu_\perp] [T^A,T^B]+ g^{\mu \nu}_\perp \{T^A,T^B\} \right) \right]^{i,s}$ &  \Eq{eq:amp_quark_doublegluon}\\
      \hline
         $\left[\cJ_{g, A}^{\mu \nu}(k)\right]^{} $ & $\fd{3cm}{figures_amp/amp_doubleB_low.pdf}$ & $\fd{3cm}{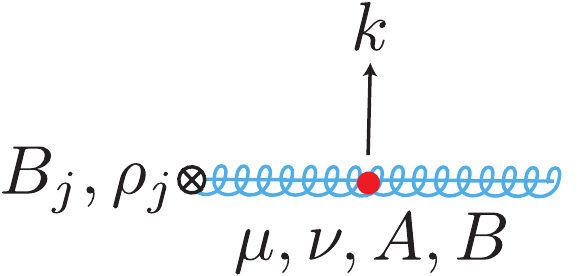}=\frac{g^2f_{ABC}f_{CDM }}{\nbar \cdot p n\cdot k} \left( g_\perp^{\mu \rho} \epsilon_\perp^{\nu D} - g_\perp^{\nu \rho} \epsilon_\perp^{\mu D} \right) = gf_{ABC}\cJ_{g,C}^{\mu \nu}\,$ &  \Eq{eq:amp_gluon_doublegluon}\\
         \hline
\end{tabular}}
\caption{A summary of the radiative functions with tree level Feynman rules, showing also the schematic factorization of the amplitude, and the equation where the definition of the radiative function can be found. Derivations of the factorizations are given in the text.
}
\label{tab:rad_func_amp}
\end{table}
}

\subsection{Leading Power Amplitude Factorization}\label{sec:LP_amp}

Before considering radiative functions at the amplitude level, we begin by briefly reviewing the well known leading power amplitude level factorization. This will help to establish our notation, as well as to emphasize distinctions when we consider the subleading power case.

To study factorization at the amplitude level, we can proceed as in \Sec{sec:sub-fact}, however, we study only the matrix element 
\begin{align}
\cA_N=\langle X | \cO(0) | 0\rangle\,,
\end{align}
instead of the squared matrix element.
Here $\cO$ is a full theory QCD operator, and $X$ is an $N$-jet state in the full theory. In the soft and collinear limits, we can proceed identically to the factorization at the cross section level, namely we match to the leading power $N$-jet operator in the EFT, which we assume has a single collinear field, $X^{k_i}_{n_i}$, in each collinear sector
\begin{align}
\cO_N^{(0)}=C_N^{(0)}\left( \{ Q_i \} \right) \otimes \prod\limits_{i=1}^N \left[ \delta (\bar n_i Q_i -\bar n\cdot i \partial_n ) X_{n_i}^{\kappa_i}(0)  \right]\,.
\end{align}
Here the $\kappa_i$ labels the parton identity of the $n_i$ collinear field that can either be a quark jet field $\chi_{n_i}$ or a gluon jet field $\cB_{n_i}^\perp$.
Throughout this section, we will assume for simplicity that there is a single such operator, since the structure of the leading power operator will not play a significant role in our discussion. Furthermore, we will suppress explicit contractions of color indices, since they are standard.
The BPS field redefinition factorizes the Hilbert space, and hence the state
\begin{align}
\langle X |=\prod\limits_i  \langle X_{n_i} |  \langle X_{us} | \,,
\end{align}
into collinear states $\langle X_{n_i} | $ and an ultrasoft state $\langle X_{us} |$. With the interactions in the Lagrangian decoupled, the leading power factorization of the matrix element is then simply an algebraic exercise, and we obtain the factorized expression
\begin{align}\label{eq:LP_amp}
\cA^{(0)}_N=\langle X | \cO_N^{(0)} |0\rangle = C^{(0)}_N\left( \{ Q_i \} \right) \prod\limits_i \langle X_{n_i} | \delta (\bar n_i Q_i -\bar n\cdot i \partial_n ) X_{n_i}^{\kappa_i}(0) |0\rangle   \left \langle X_{us} \left | \TO  \prod _{i}Y^{\kappa_i}_{n_i} (0) \right |0 \right\rangle\,.
\end{align}
Here $\kappa_i$ labels both the parton identity of the $n_i$ collinear field, as well as the representation of the Wilson line, as determined by the BPS field redefinition, and $\TO$ denotes time ordering.
This gives rise to the familiar factorization into a hard matching coefficient, coefficient functions describing the collinear radiation along each lightlike direction, and a soft amplitude, 
\begin{align}
\cA^{(0)}_N=C^{(0)}_N\left( \{ Q_i \} \right) \cdot \left( \prod\limits_i \cJ^{\kappa_i}_{n_i} \right ) \cdot \cS_N=\fd{3cm}{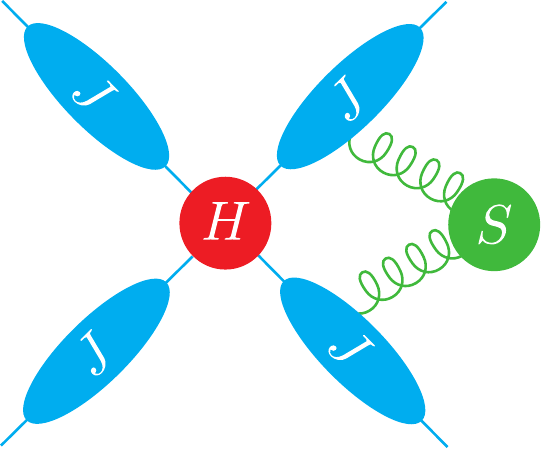}\,.
\end{align}
The leading power collinear function and soft amplitude are defined as
\begin{align}
\cJ^{\kappa_i}_{n_i}=\langle X_{n_i} | \delta (\bar n_i Q_i -\bar n\cdot i \partial_n ) X_{n_i}^{\kappa_i}(0) |0\rangle\,, \qquad \cS_N= \left \langle X_{us} \left | \TO  \prod\limits_{i} Y^{\kappa_i}_{n_i} (0)\right |0 \right \rangle\,.
\end{align}
At leading power the soft function is defined as a matrix element of Wilson lines, which are generated in SCET through the BPS field redefinition. The soft emissions therefore only resolve the color and direction of the collinear legs. To simplify our notation, in \Eq{eq:LP_amp} we have left implicit the contraction of all color indices, and denote it simply by the ``dot" symbol.   No Lorentz or Dirac indices are passed between the jet and soft functions, and therefore the soft degrees of freedom have no sensitivity to the spin of the collinear particles. Furthermore, the factorization is multiplicative, with no convolution in the soft momentum. We will see that when we consider the factorization involving Lagrangian insertions at subleading power, these features no longer hold.

\subsection{Definition of Radiative Functions}\label{sec:RF_amp}
We now  consider  amplitude level factorization at subleading power. Here we will focus solely on contributions from Lagrangian insertions, which will give rise to radiative functions. We have studied the structure of subleading power hard scattering operators extensively in previous papers \cite{Feige:2017zci,Moult:2017rpl}. 
After performing the BPS field redefinition,  the contributions from subleading power Lagrangian insertions to the amplitude take the form
\begin{align}
\cA_N^{(j),\text{rad.}}=C^{(0)}_N \int d^4 x ~ \prod\limits_{n_i} \left \langle X_{n_i} \left |  \left \langle X_{us} \left | \TO\left\{ \cL_{n_i}^{(j)\text{BPS}}(x) \cO_N^{(0)\text{BPS}} \right\} \right |0 \right\rangle \right. \right. \,.
\end{align}
Here $ \cO_N^{(0)\text{BPS}}$ is the leading power BPS redefined $N$ jet operator, and $\cL_{n_i}^{(j)}(x)$ is the $\cO(\lambda^j)$, $j\geq 1$, Lagrangian for the $n_i$ sector, after BPS field redefinition. The ``$\text{rad.}$" superscript on the matrix element emphasizes that this is only the radiative contribution to the amplitude. More generally, one must consider multiple Lagrangian insertions, or Lagrangian insertions with subleading power hard scattering operators, as detailed in  \Eq{eq:xsec_lam2}. These will factorize in a similar manner, and will be discussed in \Sec{sec:fact_RadiativeFunction}.

Unlike the leading power case of \Eq{eq:LP_amp}, where the amplitude factorized into a product of functions, at subleading power this factorization will include  integration variables linking the jet and soft functions. These integration variables will parametrize the position along the light cone direction, and describes the position of the soft emission from the collinear sector. Furthermore, at subleading power  the soft function no longer couples just to the color charge and direction of the jet functions, but can instead couple via Lorentz or Dirac indices in a manner which depends on the spin of the collinear particle.

\subsubsection{Soft Quark Emission}\label{sec:RF_amp_squark}

The simplest case for which to consider the factorization is the $\cL^{(1)}$ emission of a soft quark. Unlike the $\cL^{(1)}$ emission of a soft gluon, this does not vanish when $\cP_\perp$ vanishes, and it has a simpler structure than the $\cL^{(2)}$ insertions. It therefore provides a simple demonstration of the convolution structure which will appear at subleading powers. At the amplitude level, fermionic soft theorems in supersymmetric field theories and supergravity, and their relation to asymptotic symmetries have been considered \cite{Chen:2014xoa,Dumitrescu:2015fej,Avery:2015iix,Lysov:2015jrs}. At the cross section level soft quarks were found to give a leading logarithmic contribution to $B$-physics process \cite{Lee:2004ja} and event shape observables  \cite{Moult:2016fqy,Boughezal:2016zws,Moult:2017jsg}. The cross section level factorization will be discussed in \Sec{sec:fact_RadiativeFunction}.

For radiative soft quark emission at $\cO(\lambda)$, the relevant Lagrangian is 
\begin{align}
{\cal L}_{\chi_n \psi_{us}}^{(1)\text{BPS}} 
&= \bar{\chi}_n g \slashed{\cB}_{n\perp} \psi_{us (n)}+\text{h.c.} \,.
\end{align}
We are therefore interested in the factorization of the matrix element
\begin{align}
\cA_{N,\psi}^{(1),\text{rad.}}= C^{(0)}_N \int d^4 x ~ \prod\limits_{n_i} \left \langle X_{n_i}  \left |  \left \langle X_{us} \left | \TO \left \{ \bar{\chi}_{n_j} g \slashed{\cB}_{n_j\perp} \psi_{us (n_j)}(x) \cO_N^{(0)\text{BPS}} \right \} \right |0 \right \rangle \right. \right. \,.
\end{align}
The subscript $\psi$ labeling the amplitude indicates that a soft fermion is radiated.
Since the Lagrangian insertion appears only in the collinear sector $n_j$, the factorization of the other collinear sectors proceeds exactly as at leading power, giving rise to the leading power jet functions discussed in \Sec{sec:LP_amp}. For concreteness, we assume that the field in the $n_j$ collinear sector is a collinear quark. To simplify the notation in intermediate steps, we will drop the explicit time ordering, and reinstate it only in the final factorized formula. We then have
\begin{align}
\cA_{N,\psi}^{(1),\text{rad.}}=C^{(0)}_N \int d^4 x ~ \langle X_{n_j}|  \Big( \bar{\chi}_{n_j}  g \slashed{\cB}_{n_j\perp} \Big)^\balpha(x) \chi_{n_j}(0) |0\rangle      \langle X_{us} |  \psi^{\alpha}_{us(n_j)} (x)  \prod_i Y^{\kappa_i}_{n_i} (0) |0\rangle \prod\limits_{i\neq j} J^{\kappa_i}_{n_i}\,,
\end{align}
and it remains only to factorize the $n_j$ collinear sector from the soft sector.  To facilitate a comparison with definitions of radiative functions given in the literature, it will be convenient to formulate the convolution between the jet and soft functions in momentum space. Inserting $1=\int d^4y \delta^{(4)}(x-y)=\int d^4y \int \frac{d^4k}{(2\pi)^4} e^{ik\cdot(x-y)}$, we obtain
{
\begin{align}\label{eq:fact_sq_partial}
\cA_{N,\psi}^{(1),\text{rad.}}=&C^{(0)}_N \int \frac{d^4k}{(2\pi)^4}   \left[\int d^4 x ~ e^{ik \cdot x} \left \langle X_{n_j} \left| \ \Big( \bar{\chi}_{n_j}  g \slashed{\cB}_{{n_j}\perp} \Big)^\balpha(x)  \chi_{n_j}(0) \right |0 \right\rangle \right] \nn \\
&\hspace{2cm} \cdot \left[  \int d^4 y ~ e^{-ik \cdot y}  \left  \langle X_{us}  \left |  \psi_{us}^\alpha (y)  \prod_i Y^{\kappa_i}_{n_i} (0)   \right |0 \right \rangle \right] 
\cdot \prod\limits_{i\neq j} J^{\kappa_i}_{n_i}\,.
\end{align}}
In its current form, \Eq{eq:fact_sq_partial} is written as a four dimensional convolution. To regulate this expression in dimensional regularization, one must extend this to a $d=4-2\epsilon$ dimensional convolution. This implies that one cannot separately consider the soft and collinear functions after expansion in dimensional regularization, and therefore that one cannot achieve a renormalized factorization. Furthermore, as written, the factorized expression has not yet made manifest the physical picture of decorating a Wilson line via an insertion of an operator along the light cone. To simplify the convolution structure, we make use of the multipole expansion that has been implemented in the effective theory. Due to the multipole expansion the collinear matrix elements in the effective theory are local in the $n\cdot x$ and $x_\perp$ components, 
\begin{align}\label{eq:local_collinear}
\left \langle X_{n_j} \left | \Big( \bar{\chi}_{n_j}  g \slashed{\cB}_{{n_j}\perp} \Big)^\balpha(x) \chi_{n_j} (0)(0) \right |0 \right\rangle\sim \delta^2(x_\perp) \delta(x^+)\,,
\end{align}
where the $\perp$ and $+$ are in the light cone coordinates with respect to $n_j$.
This can also be seen from the multipole expanded propagator \Eq{eq:prop}. Using this property, we can simplify the \Eq{eq:fact_sq_partial} to a single variable convolution. Focusing just on the $n_j$ and soft sectors, we have
{\scriptsize
\begin{align}
&\int \frac{d^4k}{(2\pi)^4}    \left[  \int d^4 x ~ e^{ik \cdot x} \langle X_{n_j}| \Big( \bar{\chi}_{n_j}  g \slashed{\cB}_{{n_j}\perp} \Big)^\balpha(x) \chi_{n_j}(0) |0\rangle \right]   \left[  \int d^4 y ~ e^{-ik \cdot y}   \langle X_{us} |  \psi_{us({n_j})}^\alpha (y)   \prod_i Y^{\kappa_i}_{n_i} (0)  |0\rangle \right]\nn \\
&=\int \frac{d^4k}{(2\pi)^4}    \left[  \int d x^- ~ e^{ik^+x^-/2} \langle X_{n_j}| \Big( \bar{\chi}_{n_j}  g \slashed{\cB}_{{n_j}\perp} \Big)^\balpha(x) \chi_{n_j}(0) |0\rangle \right]  \left[  \int d^4 y ~ e^{-ik\cdot y}   \langle X_{us} |  \psi_{us({n_j})}^\alpha (y)   \prod_i Y^{\kappa_i}_{n_i} (0)  |0\rangle \right] \nn \\
&=\int \frac{dk^+}{(2\pi)}    \left[  \int d x^- ~ e^{ik^+x^-/2} \langle X_{n_j}| \Big( \bar{\chi}_{n_j}  g \slashed{\cB}_{{n_j}\perp} \Big)^\balpha(x^-) \chi_{n_j}(0) |0\rangle \right]  \left[\frac{dk^-}{2\pi}\frac{d^2 k_\perp}{(2\pi)^2}  \int d^4 y ~ e^{-ik\cdot y}   \langle X_{us} |  \psi_{us({n_j})}^\alpha (y)   \prod_i Y^{\kappa_i}_{n_i} (0)  |0\rangle \right] \nn \\
&=\int \frac{dk^+}{(2\pi)}    \left[  \int d x^- ~ e^{ik^+x^-/2} \langle X_{n_j}| \Big( \bar{\chi}_{n_j}  g \slashed{\cB}_{{n_j}\perp} \Big)^\balpha(x^-) \chi_{n_j}(0) |0\rangle \right]  \left[ \int dy^- ~ e^{-ik^+y^-/2}   \langle X_{us} | \psi_{us({n_j})}^\alpha (y^-)   \prod_i Y^{\kappa_i}_{n_i} (0)  |0\rangle \right] \nn \\
& \normalsize= \int \frac{dk^+}{(2\pi)} \cJ_{\psi {n_j}}^\balpha (k^+)S_{N\psi {n_j}}^\alpha (k^+)\,.
\end{align}
}
Note that here we use the lightcone definition for the $x$ variable as
\begin{align}
x^\mu=x^-\frac{n^\mu}{2}+x^+\frac{\bar n^\mu}{2}+x_\perp\,.
\end{align}
This then gives the momentum space definition of the radiative jet function $\left[\cJ_{\psi {n_j}}^\balpha (k^+)\right]^{i,s}$, and the corresponding soft function $S_{N\psi {n_j}}^\alpha (k^+)$
\be \label{eq:rad_jet_soft_quark_amplitude}
	\left[\cJ_{\psi {n_j}}^\balpha (k^+)\right]^{i,s} = \int dx^- ~ e^{ik^+ x^-/2}\langle X_{n_j}| \Big( \bar{\chi}_{n_j}  g \slashed{\cB}_{{n_j}\perp} \Big)^\balpha(x^-) \chi^{i,s}_{n_j}(0) |0\rangle\,.
\ee
In position space, we have
\begin{align}
&\int \frac{dk^+}{(2\pi)} \cJ_\psi^\balpha (k^+)S_\psi^\alpha (k^+)\nn \\
&=\int d x^-   \left[    \left \langle X_{n_j} \left| \TO\Big( \bar{\chi}_{n_j}  g \slashed{\cB}_{{n_j}\perp} \Big)^\balpha(x^-) \chi_{n_j}(0)  \right |0\right\rangle \right]   \left[   \left \langle X_{us} \left | \TO  \psi_{us({n_j})}^\alpha (x^-)   \prod_i Y^{\kappa_i}_{n_i} (0)  \right |0 \right\rangle \right] \nn \\
&= \int dx^- \cJ_{\psi {n_j}}^\balpha (x^-)S_{N\psi {n_j}}^\balpha (x^-)\,,
\end{align}
where in this final form, we have explicitly reinstated the time ordering.
The factorization for this contribution to the full amplitude is then given by
\begin{align}
\cA_{N,\psi}^{(1),\text{rad.}}= C^{(0)}_N\left( \{ Q_i \} \right)  \prod\limits_{i\neq j} J^{\kappa_i}_{n_i} \int dx^- \cJ_{\psi {n_j}}^\balpha (x^-)S_{N\psi {n_j}}^\alpha (x^-) =\fd{3cm}{figures_amp/amp_sub_quark_fac_low.pdf}\,.
\end{align}
This factorization gives the physical picture of dressing the Wilson line with an operator at a position $x^-$ along the light cone. Due to the multipole expansion, the soft sector still sees the collinear sector as lying exactly on the light cone. Unlike the leading power case, the soft function and jet function both carry a fermionic index. The soft degrees of freedom are therefore aware of the identity of the collinear partons. Other radiative functions will have an analogous structure, but can involve more complicated contractions between the soft and collinear sectors and additional integrals.

The tree level result for the radiative function is given by
\begin{align}\label{eq:1quarklam_feyn_rad}
&\fd{3cm}{figures_b/Kfactor_L1_quark_noWilson_rad_low}= \left[\cJ_{\psi {n_j}}^{\balpha i} (k^+)\right]^{\beta_j,s_j}_{|\text{LO}} = -\left[ \left( \Sl{\epsilon}^A_\perp - \frac{\Sl{p}_{\perp}\, n\cdot \epsilon^A(p)}{\bar n \cdot p} \right)\frac{\Sl{n}}{2 k^+}\right]_{\balpha, s} T^A_{i, \beta_j}\,.
\end{align}
In this initial investigation, we will not consider loop corrections. Although here we have treated an outgoing quark which emits a soft quark and becomes a collinear gluon, the opposite case of a gluon field converting to a quark field can be treated in an identical fashion. 

\subsubsection{Soft Gluon Emission from a Collinear Quark}\label{sec:amp_softgluon}

We now consider the $\cO(\lambda^2)$ emission of a soft gluon insertion. We will consider first the case of the emission from a collinear quark, where we will work through the convolution structure in detail. We will then state the result for the emission from a collinear gluon, which has a similar structure.   For the emission of a soft gluon from a collinear quark, the relevant Lagrangian is
\begin{align}
{\cal L}_{n \xi}^{(2) \text{BPS}} &=  \bar \chi_n  \left(  T^a \gamma^\mu_\perp \frac{1}{\bar \cP}  i \Sl \partial_{us\perp} - i  {\overleftarrow{\Sl \partial}}_{us\perp} \frac{1}{\bar \cP} T^a \gamma^\mu_\perp   \right)   \frac{\Sl \bn}{2} \chi_n  \,   g \cB_{us(n)}^{a\mu} \,.
\end{align}
We therefore must consider the factorization of the matrix element
\begin{align}
\cA_{N_q,\cB_{us}}^{(2),\text{rad.}}=  C^{(0)}_N \int d^4 x ~ \prod\limits_{n_i} \left \langle X_{n_i} \left|  \langle X_{us} |\TO {\cal L}_{n_j \xi}^{(2) \text{BPS}}(x) \cO_N^{(0)\text{BPS}}  \right|0 \right\rangle\,.
\end{align}
As in the previous section, we will drop the explicit time ordering until the final formula.
Due to the presence of the ultrasoft derivative in the Lagrangian, the factorization is slightly more complicated than for the case of a soft quark emission. 

As motivation for the structure that the factorization should take, we can look at the LBK theorem for soft gluon emission at $\cO(\lambda^2)$. For the emission of a soft gluon off of a collinear quark, the LBK theorem can be written as \cite{Larkoski:2014bxa}
\begin{align}
S^{(2)}_{i\psi} \cA_N &=g \frac{2\epsilon_{s\mu} p_{s\nu} }{(\bar n_i \cdot p_i)(n_i\cdot p_s)}  \bar u(p_i) T_i \left \{ n_i^{[\mu} \bar n_i^{\nu]}\frac{\bar n_i \cdot p_i}{2} \frac{\partial}{\partial(\bar n_i \cdot p_i)} \right. \nn \\
&\hspace{5cm}\left. +\gamma_\perp^{[\mu} n_i^{\nu]} \frac{\Sl{\bar n}_i}{4} + p_{s\perp}^{[\mu} \frac{n_i^{\nu]}}{2n_i\cdot p_s} +\frac{1}{4} [\gamma^\mu_\perp,\gamma^\nu_\perp]   \right\} \tilde \cA_N\,,
\end{align}
or in terms of the angular momentum generator $J^{\mu \nu}$, as
\begin{align}
S^{(2)}_{i\psi} \cA_N &=T^i \frac{\epsilon_{s\mu} p_{s\nu} J_i^{\mu \nu}}{p_i \cdot p_s} \cA_N\,.
\end{align}
This expression holds only for as on-shell emission, which cannot be assumed of  the group momentum flowing into the $\cB_{us(n)}$ field, and furthermore it is the complete result for the amplitude, not just the contribution from the radiative functions. Nevertheless, we would like our radiative functions to have a structure which mimics this as closely as possible. In particular, we would like that the ultrasoft derivatives appearing in the subleading power Lagrangian act only on the soft sector.  Furthermore, it suggests that the radiative jet function should carry two Lorentz indices, which are contracted with a $\cB_{us(n)}$ field, and  an ultrasoft derivative. 

To perform the factorization, we assume for concreteness that there is a collinear quark field in the $n_j$ collinear sector. Since the Lagrangian insertion is in the $n_j$ sector we can immediately factorize the other collinear sectors, giving the leading power jet functions, and we obtain
\begin{align}
\cA_{N_q,\cB_{us}}^{(2),\text{rad.}}&= C^{(0)}_N \prod\limits_{i\neq j} J^{\kappa_i}_{n_i} \int d^4 x ~ e^{ik \cdot x}  \\
&\hspace{-1.5cm}\cdot \left \langle X_{n_j} \left | \left \langle X_{us} \left| \bar \chi_{n_j} \left(  T^a \Sl{\cB}_{us({n_j}) \perp} \frac{1}{\bar \cP} i\Sl{\partial}_{us\perp}-  i{\overleftarrow{\Sl{\partial}}}_{us\perp}  \frac{1}{\bar \cP}  T^a  \Sl{\cB}_{us({n_j}) \perp}    \right) \frac{\Sl{\bar n}}{2}\chi_{n_j}(x) \chi_{n_j}^{(0)}(0) \prod_i Y^{\kappa_i}_{n_i} (0)  \right |0 \right \rangle \right. \right. \,. \nn
\end{align}
To achieve a factorization of the ultrasoft derivatives, we can choose the external states to have no $\perp$ residual momentum. After BPS field redefinition,  there are no soft collinear interactions, other than through the single Lagrangian insertions. Therefore, momentum is only passed between the soft and collinear sectors at this vertex. The ultrasoft derivative operator can either act on the group momentum that flows out through the $\cB_{us(n)}$ field, or on a residual momentum component coming from a collinear loop. In dimensional regularization, we have \cite{iain_notes}
\begin{align}
	\sum\limits_{q_l} \int d^d q_r (q_r)^j F(q_l^-, q_l^\perp, q_r^+)=0\,,
\end{align}
where $(q_r)^j$ with $j>0$ denotes positive powers of the $q_r^-$ and $q_r^\perp$ momenta, which are the only residual momenta which appear in the subleading power Lagrangians. Any residual momentum from a collinear loop momentum picked up by the derivative is therefore set to zero. Therefore the derivative picks up just the group momentum of the $\cB_{us(n)}$ field.  Ultimately, this is due to the fact that the ultrasoft momentum of the $\cB_{us(n)}$ field is the only physical ultrasoft momentum flowing in the graphs. At loop level, this statement is slightly non-trivial since the subleading power Lagrangian can couple to closed fermion loops. However, given a $\bar \chi_n$ which produces a fermion in the hard scatter in the $n$ collinear sector, it is possible to choose the momentum routing so that the soft momentum is routed only along the direction of fermion number flow of the collinear operator.  Therefore, the ultrasoft derivative acts just on the soft gluon field, and only one of the tensor structures appears.

For the particular matrix elements of interest, we can therefore perform the following simplifications
{\begin{small}
\begin{align}
& \int d^4 x ~ e^{ik \cdot x} \left \langle X_{n_j}\left| \left\langle X_{us}\left| \bar \chi_n \left(  T^a \Sl{\cB}^a_{us({n_j}) \perp} \frac{1}{\bar \cP} i\Sl{\partial}_{us\perp}-  i{\overleftarrow{\Sl{\partial}}}_{us\perp}  \frac{1}{\bar \cP}  T^a  \Sl{\cB}^a_{us({n_j}) \perp}    \right) \frac{\Sl{\bar n}}{2} \chi_{n_j}(x) \chi_{n_j}^{(0)}(0) \prod_i Y^{\kappa_i}_{n_i} (0) \right |0 \right \rangle \right. \right. \nn \\
&= \int d^4 x ~ e^{ik \cdot x} \left \langle X_{n_j} \left | \left \langle X_{us} \left| \bar \chi_n \left( -  i   \frac{1}{\bar \cP}  T^a  \left[ \Sl{\partial}_{us\perp}    \Sl{\cB}^a_{us({n_j}) \perp} \right]    \right) \frac{\Sl{\bar n}}{2}  \chi_{n_j}(x) \chi_{n_j}^{(0)}(0) \prod_i Y^{\kappa_i}_{n_i} (0)  \right|0 \right \rangle \right. \right. \,.
\end{align}
\end{small}}
Here the square brackets indicate that the derivatives act only within those brackets.
It is now straightforward to factorize this contribution to the amplitude, following the steps in \Sec{sec:RF_amp_squark}, so we will not repeat them explicitly. After simplifying the convolution structure, we find
\begin{align}
\cA_{N,\cB_{us}}^{(2),\text{rad.}}= C^{(0)}_N \int \frac{dk^+}{(2\pi)}    &\left[  \int dx^- ~ e^{ik^+ x^-/2} \left \langle X_{n_j} \left | \bar \chi_n \left(  -  \gamma^\perp_\nu \frac{1}{\bar \cP}  T^A  \gamma^\perp_\mu    \right)\frac{\Sl{\bar n}}{2} \chi_{n_j}(x)\chi_{n_j}(0) \right |0 \right\rangle \right] \nn \\
 &\hspace{-0.2cm}\cdot \left[  \int dy^- ~ e^{-ik^+ y^-/2}  \left \langle X_{us} \left |  \left[i\partial_{\perp}^\nu\cB_{us({n_j})}^{\mu A} (y) \right]   \prod_i Y^{\kappa_i}_{n_i} (0)  \right |0 \right \rangle \right]  \prod\limits_{i\neq j} J^{\kappa_i}_{n_i}  \nn \\
  &= C^{(0)}_N \int \frac{dk^+}{(2\pi)} \cJ_{q,{n_j}}^{\mu \nu A} (k)S_{q,{n_j}}^{\mu \nu A}(k) \prod\limits_{i\neq j} J^{\kappa_i}_{n_i}\,.
\end{align}
Here, we have defined the radiative function $\left[\cJ^{\mu \nu A}_{q, n_j}(k^+)\right]^{i,s}$ as
\be \label{eq:rad_jet_soft_gluon_amplitude}
	\left[\cJ^{\mu \nu A }_{q,n_j}(k^+)\right]^{i,s} = \int dx^- ~ e^{ik^+ x^-/2}  \left \langle X_{n_j} \left | \TO \bar \chi_{n_j} \left(  -  \gamma^\perp_\nu \frac{1}{\bar \cP}  T^A  \gamma^\perp_\mu    \right)\frac{\Sl{\bar n}}{2}\chi_{n_j}(x) \chi^{i,s}_{n_j}(0)  \right |0 \right \rangle\,.
\ee
We note that the subscript $q$ denotes that this describes the radiative emission from a quark, with the fact that it is the radiative emission of a single gluon being specified by the single free adjoint color index.

This radiative function exhibits an explicit coupling of $i\partial_{\perp}^\nu\cB_{us}^\mu$ to something that is reminiscent of the spin orbital angular momentum, inserted at a position along the light cone.  In position space, we have
\begin{align}
\cA_{N,\cB_{us}}^{(2),\text{rad.}}= & C^{(0)}_N\int dx^-    \left[  \left \langle X_{n_j} \left | \TO \bar \chi_{n_j} \left(  -  \gamma^\perp_\nu \frac{1}{\bar \cP}  T^A  \gamma^\perp_\mu    \right)\frac{\Sl{\bar n}}{2}\chi_{n_j}(x) \chi^{i,s}_{n_j}(0)  \right |0 \right \rangle \right] \nn \\
&\hspace{1.65cm}\cdot \left[     \left \langle X_{us} \left | \TO  \left[i\partial_{\perp}^\nu\cB_{us({n_j})}^{\mu A} (x) \right]   \prod_i Y^{\kappa_i}_{n_i} (0)  \right |0 \right\rangle \right] \prod\limits_{i\neq j} J^{\kappa_i}_{n_i}   \nn \\
  &=  C^{(0)}_N \int dx^- \cJ_{q A}^{\mu \nu } (x^-)S_{A}^{\mu \nu}(x^-) \prod\limits_{i\neq j} J^{\kappa_i}_{n_i} = \fd{3cm}{figures_amp/amp_sub_gluon_fac_low.pdf}\,,
\end{align}
where we have reinstated the explicit time ordering.

The tree level result for the radiative function is
\begin{align}\label{eq:1Blam2_fromquark_feyn_rad}
	\fd{3cm}{figures_b/Kfactor_L2_gluon_noWilson_rad_low}~= \left[\cJ^{\mu \nu}_{q,A}(k^+)\right]^{i,s}_{|LO} &= -\left[\bar{u}_n(p)\frac{gT^A  \gamma_\perp^\mu \gamma_\perp^\nu}{\bar n \cdot p\, n\cdot k}\right]^{i,s}\nn \\
	&= -\left[\bar{u}_n(p)\frac{gT^A }{\bar n \cdot p\, n \cdot k}\left( g_\perp^{\mu \nu} + i \sigma_\perp^{\mu \nu} \right)\right]^{i,s}\,.
\end{align}
In \Sec{sec:comp_laenen} we will compare this with results in the literature.

We have focused on the factorization for the particular insertion of the subleading power quark Lagrangian onto a collinear line, which contributes at tree level, and have shown how this gives rise to a radiative function.  At loop level, one can also consider contributions from the other terms in the Lagrangian. For example, one can insert the gluon component of the $\cL^{(2)}$ Lagrangian into an outgoing quark leg. This will contribute, for example through the diagram
\begin{align}
 \fd{3cm}{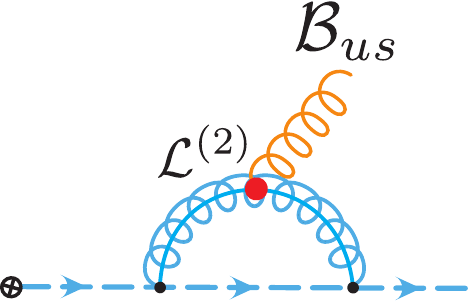}\,.
\end{align}
The calculation of contributions of this form were performed in the threshold limit in \cite{Bonocore:2016awd}.
The factorization for this is identical to the case of the gluon, which will be discussed in the next section, \sec{amp_softgluon_from_gluon}. As a matter of fact, since the coupling of the Lagrangian is the same, one must just exchange the field at the origin $\cB_{n\perp}\to \chi_n$. Since the goal of this section is to show in detail for several examples how radiative functions at amplitude level arise in SCET, we will not perform a complete classification of the radiative functions at loop level. In \Sec{sec:RF_thrust}, we will perform this classification for the case of the thrust observable in $e^+e^-\to$ dijets, using symmetries specific to the problem to simplify the number of distinct contributions.

\subsubsection{Soft Gluon Emission from a Collinear Gluon}\label{sec:amp_softgluon_from_gluon}

We can also consider the emission of a gauge invariant soft gluon field from a collinear gluon. Since the derivation of the factorization is similar, here we skip most of the steps to quickly get to the final result. However, since the gluon fields carry both color and Lorentz indices and the resulting radiative functions definition will depend on them, in this paragraph we treat these indices explicitly.\\
Even though the $\cO(\lambda^2)$ Lagrangian in the gluon sector is somewhat complicated, it strongly simplifies under the assumption of no label perpendicular momentum flowing through the collinear fields, as explained in \Sec{sec:lam2_Lagrangian}. Therefore, for the emission of a soft gluon from a collinear gluon with $\cP_\perp =0$, the relevant Lagrangian is simply
\begin{align}
	{\cal L}_{n g}^{(2) \text{BPS}} &=  g \text{Tr} \left( \partial_\perp^{[\mu} \cB_{us \perp}^{\nu]}[ \cB^\perp_{n\mu}  , \cB^\perp_{n\nu}   ] \right)\,.
\end{align}
Since we want to study the factorization involving a gluon lagrangian insertion, our hard scattering operator $\cO_{N}^{(0)}$ must contain at least one gluon field. For concreteness let's take $\cO_{N}^{(0)}$ to have a collinear gluon field in the $n_j$ direction. We therefore must consider the factorization of the matrix element
\begin{align}
\cA_{\alpha_1,\dots,\alpha_N}^{(2),\text{rad.}}=  C^{(0)}_{N} \int d^4 x ~ \prod\limits_{n_i} \left \langle X_{n_i} \left|    \langle X_{us} | \TO {\cal L}_{n_j g}^{(2) }(x) \cO_{N \, \alpha_1,\dots,\alpha_N}^{(0)}(0)  \right|0 \right\rangle\,.
\end{align}
where $\alpha_i$ are color indices of the external legs of the amplitude and they belong to the representation determined by the parton identity.\footnote{We use the labels $k_i = q,\bar{q},g$ for the fundamental, anti-fundamental and adjoint representation respectively.}
As usual, after BPS field redefinition we have 
\be
	\cO_{N \alpha_1,\dots,\alpha_N}^{(0)} \to \cO_{N \beta_1, \dots,\beta_N}^{(0) \text{BPS}} \prod_{i=1}^N \left(Y_{n_i }^{k_i}\right)_{\beta_i \alpha_i}\,,\qquad {\cal L}_{n_j g}^{(2) } \to {\cal L}_{n_j g}^{(2) \text{BPS}}\,.
\ee
It is convenient to isolate the gluon leg on which we want to insert the subleading lagrangian by singling out the gluon field $\cB_{n_j \perp}^{\rho \beta_j}$, where $\rho$ is a Lorentz index and $\beta_j$ a color one, from the hard scattering operator, as follow
\be
	\cO_{N \beta_1, \dots,\beta_N}^{(0) \text{BPS}} \prod_{i=1}^N \left(Y_{n_i }^{k_i}\right)_{\beta_i \alpha_i} = \cO_{N \, \beta_{i\neq j}}^{(0)\text{BPS}}\cB_{n_j \perp}^{\rho \beta_j} (Y^g_{n_j})_{\beta_j \alpha_j} \prod_{i\neq j} \left(Y_{n_i }^{k_i}\right)_{\beta_i \alpha_i} \, ,
\ee
where $(Y^g_{n_j})_{\beta_j \alpha_j}$ is the Wilson line\footnote{Note that this is an adjoint Wilson line, so that $(Y^g_{n_j})_{\beta_j \alpha_j} \equiv \left(\cY_{n_j}\right)_{\beta_j \alpha_j}$. In our notation this information is carried by the parton label $g$.} resulting from the BPS field redefinition of the gluon field $\cB_{n_j}^{\mu \beta_j}$.
Since all the gauge invariant gluon fields appearing in this section are perpendicular, we will drop the $\perp$ label on $\cB_{n_j}^\perp$ and $\cB_{us}^\perp$ to lighten the notation. Therefore, after BPS field redefinition, the amplitude reads
{\small
\begin{align}
	\cA_{\alpha_1,\dots,\alpha_N}^{(2),\text{rad.}}=  C^{(0)}_{N} \int d^4 x ~ \left(\prod\limits_{n_i} \Big \langle X_{n_i} \Big|\right) \Big\langle X_{us} \Big|  \TO {\cal L}_{n_j g}^{(2) \text{BPS}}(x) \cO_{N \, \beta_{i\neq j}}^{(0) \text{BPS} }(0) \cB_{n_j}^{\rho \beta_j} (Y^g_{n_j})_{\beta_j \alpha_j} \prod_{i\neq j} \left(Y_{n_i }^{k_i}\right)_{\beta_i \alpha_i}  \Big|0 \Big\rangle\,.
\end{align}}%
Having made the color indices explicit, we can start by factorizing all the $n_i$ collinear jets with $i\neq j$, where $n_j$ is the direction of the collinear jet on which we insert the subleading Lagrangian
{\footnotesize
\begin{align}
\cA_{N_g,\cB_{us}}^{(2),\text{rad.}}&= C^{(0)}_N \prod\limits_{i\neq j} J^{\kappa_i \, \beta_i}_{n_i} \int d^4 x \, e^{ik \cdot x} \left \langle X_{n_j} \left | \left \langle X_{us} \left| g f^{ABC} \left[\partial_\perp^{[\mu} \cB_{us}^{\nu]A}\right] \cB^{B}_{n_j\mu}  \cB^{C}_{n_j\nu}  (x) \cB_{n_j  \rho}^{a}(0) \prod_{i} \left(Y_{n_i }^{k_i} (0)\right)_{\beta_i \alpha_i}   \right |0 \right \rangle \right. \right. \,. \nn
\end{align}}%
After simplifying the convolution structure, we find
\begin{align}
\cA_{N_g,\cB_{us}}^{(2),\text{rad.}}&= C^{(0)}_N  \int \frac{dk^+}{(2\pi)}    \left[  \int dx^- ~ e^{ik^+ x^-/2} \left \langle X_{n_j} \left | g f^{ABC}  \cB^{B}_{n_j\mu} (x) \cB^{C}_{n_j\nu}  (x) \cB_{n_j  \rho}^{M}(0) \right |0 \right\rangle \right] \nn \\
 &\hspace{1.85cm}\cdot \left[  \int dy^- ~ e^{-ik^+ y^-/2}  \left \langle X_{us} \left |  \partial_\perp^{[\mu} \cB_{us}^{\nu]A}(y)  \prod_i \cY^{\kappa_i}_{n_i} (0)  \right |0 \right \rangle \right]  \prod\limits_{i\neq j} J^{\kappa_i \beta_i}_{n_i}  \nn \\
  &= C^{(0)}_N \int \frac{dk^+}{(2\pi)} \cJ_{g, A,n_j}^{\mu \nu } (k^+) S_{g, A,n_j}^{\mu \nu }(k^+) \prod\limits_{i\neq j} J^{\kappa_i}_{n_i}\,.
\end{align}
where we have defined 
\be\label{eq:rad_amp_gluon}
	\left[\cJ_{g, A ,n_j}^{\mu \nu}(k^+)\right]^{M \rho} \equiv  \int dx^- ~ e^{ik^+ x^-/2} \langle X_{n_j}| g f_{ABC} \cB_{n_j}^{\mu B}(x^-)\cB_{n_j}^{\nu C}(x^-)\cB_{n_j}^{M \rho}(0)  |0\rangle \,.
\ee
As before, the subscript $g$ indicates that this describes emission from a gluon, while the adjoint color index indicates that it is describing the emission of a single gluon field.

The tree level Feynman rule for this radiative function is
\begin{align}\label{eq:1Blam_fromgluon_feyn_rad}
&\fd{3cm}{figures_b/Kfactor_L2_gluon_gluon_noWilson_rad_low}= \left[\cJ_{g, A}^{\mu \nu}(k^+)\right]^{M \rho}_{|{LO}} = \frac{-igf^{ABM}}{\bar n \cdot p\, n\cdot k} \left( g_\perp^{\mu \rho}\epsilon_\perp^{\nu B} - g_\perp^{\nu \rho}\epsilon_\perp^{\mu B}\right)\,.
\end{align}

\subsubsection{Double Non-Eikonal Soft Gluon Emission}

In addition to the radiative functions which couple a single $\cB_{us(n)}$ field to the collinear line, the $\cL^{(2)}$ Lagrangian also contains a term that couples to a product $\cB_{us(n)}\cB_{us(n)}$ describing a double emission. It is important to emphasize that the radiative functions of \Sec{sec:amp_softgluon} with a single ultrasoft insertion also have two gluon Feynman rules arising from the Wilson lines present in the gauge invariant definition of the ultrasoft gluon field. However, such additional emissions are effectively eikonal. The double soft radiative function to be studied in this sections describes a genuinely double non-eikonal emission. 

Double soft theorems have attracted some attention in the literature \cite{Low:2015ogb,DiVecchia:2015bfa,Georgiou:2015jfa,McLoughlin:2016uwa}. We note that as compared to some treatments in the literature, the SCET power counting is such that we always take a limit such that the soft gluons are becoming simultaneously soft. This is in contrast to consecutive limits, where the soft limit for two gluons is taken consecutively. See \cite{Klose:2015xoa} for a discussion of the difference between these limits. As emphasized in \cite{ArkaniHamed:2008gz}, double soft limits are interesting as they allow one to probe the group structure of the theory, and are proportional to the structure constants of the theory. We will see that this is indeed true for the double radiative function.

	\paragraph{Emissions from collinear quarks.}

We start with the emission from a fermionic leg. Therefore, we are interested in the factorization of the term in the $\cL_{\chi_n}^{(2)}$ Lagrangian involving two $\cB_{us(n)}$ fields
\begin{align}
	\cL_{\chi_n ,\cB_{us} \cB_{us}}^{(2) } &= \bar \chi_n  \left[ T^a \gamma^\mu_\perp \frac{1}{\bar \cP} T^b \gamma^\nu_\perp   \right]   \frac{\Sl \bn}{2} \chi_n  g \cB_{us(n)}^{a\mu} g \cB_{us(n)}^{b\nu} \nn \\
	& =\bar \chi_n  \left[  \frac{1}{\bar \cP}  [\gamma^\mu_\perp, \gamma^\nu_\perp] [T^a,T^b]+ g^{\mu \nu}_\perp \{T^a,T^b\} \right]   \frac{\Sl \bn}{2} \chi_n  g \cB_{us(n)}^{a\mu} g \cB_{us(n)}^{b\nu} \,,
\end{align}
and in particular, the factorization of the matrix element
\begin{align}
\cA_{N,\cB_{us}\cB_{us}}^{(2),\text{rad.}}= C^{(0)}_N\int d^4 x ~ \prod\limits_{n_i} \left \langle X_{n_i} \left |  \left \langle X_{us} \left| \TO  \cL_{\chi_n ,\cB_{us} \cB_{us}}^{(2) } (x) \cO_N^{(0)\text{BPS}} \right |0 \right\rangle \right. \right.\,.
\end{align}

Since this matrix element contains no derivatives, the derivation of the convolution structure is identical to that for the soft quark emission in \Sec{sec:RF_amp_squark}. We therefore simply give the final result, skipping intermediate steps. We find 
{\normalsize
\begin{align}
\cA_{N,\cB_{us}\cB_{us}}^{(2),\text{rad.}} &=C^{(0)}_N \int \frac{dk^+}{(2\pi)} \nn \\
&\medmath{\cdot    \left[  \int dx^- ~ e^{ik^+ x^-/2} \left \langle X_{n_j} \left| \bar \chi_{n_j}  \left[  \frac{1}{\bar \cP}  [\gamma^\mu_\perp, \gamma^\nu_\perp] [T^A,T^B]+ g^{\mu \nu}_\perp \{T^A,T^B\} \right]   \frac{\Sl \bn}{2} \chi_{n_j}(x^-) \chi^{i, s}_{n_j}(0) \right |0 \right \rangle \right]} \nn \\
 & \medmath{\cdot \left[  \int dy^- ~ e^{-ik^+ y^-/2}   \left \langle X_{us} \left |  \left[\cB_{us({n_j})\perp}^{\mu A}  \cB_{us({n_j})\perp}^{\nu B} (y^-) \right]   \prod_i Y^{\kappa_i}_{n_i} (0)  \right |0 \right\rangle \right] \prod\limits_{i\neq j} J^{\kappa_i}_{n_i} } \nn \\
  &=C^{(0)}_N \int \frac{dk^+}{(2\pi)} \cJ^{\mu \nu}_{q, AB,n_j}(k^+)\, S_{q, AB,n_j}^{\mu \nu} (k^+)\prod\limits_{i\neq j} J^{\kappa_i}_{n_i}\,.
\end{align}}%
We have defined the radiative function as
{\small
\be \label{eq:amp_quark_doublegluon}
	\left[\cJ^{\mu \nu }_{q ,A B,n_j}(k^+)\right]^{i,s} = \int dx^- ~ e^{ik^+ x^-/2} \left \langle X_{n_j} \left| \bar \chi_{n_j}  \left[  \frac{1}{\bar \cP}  [\gamma^\mu_\perp, \gamma^\nu_\perp] [T_A,T_B]+ g^{\mu \nu}_\perp \{T_A,T_B\} \right]   \frac{\Sl \bn}{2} \chi_{n_j}(x) \chi^{i, s}_{n_j}(0) \right |0 \right \rangle\,.
\ee}%
The subscript $q$ and the two free adjoint indices are meant to indicate that this radiative function describes the double emission of soft gluons from a quark field.

In position space, we have
\begin{align}
\cA_{N,\cB_{us}\cB_{us}}^{(2),\text{rad.}}=C^{(0)}_N\int dx^-    &\left[   \langle X_{n_j}| \TO  \bar \chi_{n_j}  \medmath{\left[  \frac{1}{\bar \cP}  [\gamma^\mu_\perp, \gamma^\nu_\perp] [T^a,T^b]+ g^{\mu \nu}_\perp \{T^a,T^b\} \right] }  \frac{\Sl \bn}{2} \chi_{n_j}(x^-) \chi_{n_j}(0)  |0\rangle \right] \nn \\
 & \left[   \langle X_{us} | \TO  \cB_{us({n_j})\perp}^{\mu a} \cB_{us({n_j})\perp}^{\nu b} (x^-)    \prod_i Y^{\kappa_i}_{n_i} (0)  |0\rangle \right]  \prod\limits_{i\neq j} J^{\kappa_i}_{n_i}  \\
  &=C^{(0)}_N \int dx^-  \cJ^{\mu \nu}_{q, AB,n_j}(x^-)\, S_{q, AB,n_j}^{\mu \nu} (x^-)\prod\limits_{i\neq j} J^{\kappa_i}_{n_i} =\fd{3cm}{figures_amp/amp_doubleB_low.pdf}\,. \nn
\end{align}
The form of the $\cJ_{\cB\cB}$ current is quite interesting. In particular, it involves both the symmetric and anti-symmetric structure constants, each coupling to different Lorentz structures. The antisymmetric color structure constants are associated with the perp components of the orbital momentum generator. 

Evaluating the radiative function at tree level, we find
\begin{align}\label{eq:1Blam_fromquark_feyn_doublerad}
&\fd{3cm}{figures_b/Kfactor_L2_gluon_noWilson_doublerad_low}=\left[\cJ^{\mu \nu }_{q, A B,n}(k^+)\right]^{i,s}_{|LO} = \frac{i g^2}{\nbar \cdot p n\cdot k}  \left [\bar{u}_n(p)\left([\gamma^\mu_\perp, \gamma^\nu_\perp] [T^A,T^B]+ g^{\mu \nu}_\perp \{T^A,T^B\} \right) \right]^{i,s}\,.
\end{align}
It is important to emphasize that this radiative function does not completely describe double soft gluon emission at $\cO(\lambda^2)$. One must also consider the two gluon Feynman rule of the radiative function of \Sec{sec:amp_softgluon} involving a single $\cB_{us(n)}$. Interestingly, since the $\cL^{(1)}$ are directly proportional to $\cP_\perp$, at tree level, there is no contribution to the double soft limit from a product $\cL^{(1)} \cdot \cL^{(1)}$, or from the combination of a $\cO(\lambda)$ hard scattering operator and a $\cL^{(1)}$ insertion.

	\paragraph{Emissions from collinear gluons.}

The final tree level contribution is the emission of two gauge invariant gluon fields from a collinear gluon field. Since the derivation is similar, here we present only the final result. The relevant term in the $\cO(\lambda^2)$ Lagrangian is
\begin{align}
	\cL^{(2)\text{BPS}}&\supset  \text{Tr} \left(g^2 [ \cB_{us}^{\perp \mu}  ,\cB_{us}^{\perp \nu}   ]    [ \cB^\perp_{n\mu}  , \cB^\perp_{n\nu}   ]  \right)  \,.
\end{align}
From this we can derive the factorization of the amplitude using the same procedure as in \Sec{sec:amp_softgluon_from_gluon} to get
\begin{align}
\cA_{N,\cB_{us}\cB_{us}}^{(2),\text{rad.}}&=C^{(0)}_N\int dx^-  \left[   \langle X_{n_j}| f_{ABC}f_{CDE} \cB_{n_j}^{\mu D}\cB_{n_j}^{\nu E}(x)\cB_{n_j}^{M \rho}(0)  |0\rangle \right] \nn \\
 &\hspace{2.1cm}\cdot \left[   \langle X_{us} | \TO  \cB_{us({n_j})\perp}^{\mu A} \cB_{us({n_j})\perp}^{\nu B} (x)    \prod_i Y^{\kappa_i}_{n_i} (0)  |0\rangle \right]  \prod\limits_{i\neq j} J^{\kappa_i}_{n_i}  \\
  &=C^{(0)}_N \int dx^- \cJ_{\cB\cB{n_j}}^{\mu \nu AB}(x^-) S_{\cB\cB{n_j}}^{\mu \nu AB}(x^-)\prod\limits_{i\neq j} J^{\kappa_i}_{n_i}\,. \nn
\end{align}
where we have defined 
\be\label{eq:amp_gluon_doublegluon}
	\left[\cJ_{g, AB,n_j}^{\mu \nu}\right]^{M \rho} \equiv \langle X_{n_j}| g^2 f_{ABC}f_{CDE} \cB_{n_j}^{\mu D}\cB_{n_j}^{\nu E}(x)\cB_{n_j}^{M \rho}(0)  |0\rangle \,.
\ee
Note that 
\be 
	\left[\cJ_{g, AB,n_j}^{\mu \nu}\right]^{M \rho} = gf^{ABC}\left[\cJ_{g, C,n_j }^{\mu \nu}\right]^{M \rho}\,.
\ee
Therefore we have just shown  that the double emission radiative function is completely determined by the single emission one. Since we have proven this relation at the operator level, it is true at all orders in perturbation theory.

Evaluating the radiative function at tree level, we have
\begin{align}\label{eq:1Blam_fromgluon_feyn_doublerad}
&\fd{3cm}{figures_b/Kfactor_L2_gluon_gluon_noWilson_doublerad_low}=\frac{g^2f_{ABC}f_{CDM }}{\nbar \cdot p n\cdot k} \left( g_\perp^{\mu \rho} \epsilon_\perp^{\nu D} - g_\perp^{\nu \rho} \epsilon_\perp^{\mu D} \right) = gf_{ABC}\cJ_{g,C}^{\mu \nu}\,.
\end{align}

\subsection{Comparison to the Literature}\label{sec:comp_laenen}

In this section we make contact with other definitions of radiative functions given in the literature. Radiative functions have been studied in the context of threshold resummation \cite{Laenen:2008gt,Laenen:2008ux,Laenen:2010uz,Bonocore:2014wua,White:2014qia,Bonocore:2015esa,Bonocore:2016awd}, where they have been defined for the case of the emission of a single soft gluon. If partonic initial states are used, then the radiative jet functions defined in the threshold limit are most similar to what we have referred to as amplitude level radiative functions in this section, but with incoming instead of outgoing conventions. This is due to the fact that the kinematics of the threshold limit imply that collinear emissions cannot cross the cut, and therefore the factorization is distinct from the cross section level factorizations we study in \Sec{sec:fact_RadiativeFunction}.

In \Refs{DelDuca:1990gz,Bonocore:2015esa,Bonocore:2016awd} a radiative function\footnote{Note that in~\cite{Bonocore:2015esa} this is called \emph{radiative jet function}. As explained in \Sec{sec:intro} we have preferred to keep the term \emph{radiative jet function} only for objects entering the factorization at the cross section level, in analogy with the leading power \emph{jet functions}. The amplitude level functions are identified as \emph{radiative functions}. Since the comparison in this section is done at the amplitude level we will use the term \emph{radiative function} throughout the section.} was defined as  
\begin{align}\label{eq:Bonocore_function}
J_{\mu,a}(p,n,k) u(p)= \int d^d y e^{-i(p-k)\cdot y} \langle 0 | \Phi_n(\infty, y) \psi(y) j_{\mu,a}(0) |p \rangle\,,
\end{align}
with the non-abelian current, $j_{\mu,a}$, given by \cite{Bonocore:2016awd}
\begin{align}\label{eq:Bonocore_current}
j^\mu_a (x) =g \left( -\bar \psi(x) \gamma^\mu T_a \psi(x) + f_a^{bc} \left[  F^{\mu \nu}_c (x) A_{\nu v}(x) +\partial_\nu (A^\mu_b(x) A^\nu_c(x)) \right]    \right)\,.
\end{align}
Here $\Phi_n(\infty, y)$ is a Wilson line, so that the combination $\Phi_n(\infty, y) \psi(y)$ is equivalent to the gauge invariant field $\chi_n$ in SCET.
This structure is clearly recognized to be of the same general form as the amplitude level radiative functions defined in the previous sections. Following~\cite{Bonocore:2015esa}, the radiative function defined in \eq{Bonocore_function} can be expanded in powers of $\alpha_s$ and at tree level its Feynman rule reads 
\be\label{eq:Bonocore_tree_level}
	J^{\nu(0)} (p,k) = \frac{\Sl{k}\gamma^\nu}{2 p\cdot k} - \frac{p^\nu}{p\cdot k}\,.
\ee
Using this radiative function, the next-to-leading power corrections in the threshold limit were computed to NNLO. We would now like to show that the Feynman rule of the radiative function we have derived in \eq{1Blam2_fromquark_feyn} matches \eq{Bonocore_tree_level} after expanding homogeneously in the SCET power counting and taking the ultrasoft emission to be on-shell and the external quark to be purely collinear.

These two assumptions translate into the following kinematics
\be\label{eq:kinematic_comparison}
	k^\mu = n\cdot k \frac{\bn^\mu}{2} + \bn\cdot k \frac{n^\mu}{2} + k_{\perp}^\mu\,,\qquad k^2 = 0 \,,\qquad p^\mu = \omega \frac{n^\mu}{2} + k^\mu\,.
\ee
We can then kinematically expand the current of \eq{Bonocore_tree_level} as
\begin{align}\label{eq:Bonocore_comparison}
	J^{\nu(0)} (p,k) &= \frac{\left(\Sl{k}_\perp  + \frac{\bnslash}{2} n \cdot k + \frac{\nslash}{2} \bn \cdot k \right) \left(\gamma^\nu_\perp  + \frac{\bnslash}{2} n^\nu + \frac{\nslash}{2} \bn^\nu \right)}{ \omega  n\cdot k +\dots} - 2\frac{n\cdot k \frac{\bn^\mu}{2} + (\cancel{\omega} + \bn\cdot k) \frac{n^\mu}{2} + k_{\perp}^\mu}{ \omega\, n\cdot k +\dots} \nn \\
	&= \frac{1}{\omega\, n\cdot k}\left[\left(\Sl{k}_\perp \gamma_\perp^\nu + P_n \bn \cdot k n^\nu + P_\bn n\cdot k \bn^\nu\right) - (n\cdot k \bn^\mu + \bn\cdot k n^\mu + 2k_{\perp}^\mu) \right] \nn \\
	&= \frac{1}{\omega\, n\cdot k}\left(\Sl{k}_\perp \gamma_\perp^\nu - P_\bn \bn \cdot k n^\nu - \cancel{P_n n\cdot k \bn^\nu} - 2k_{\perp}^\mu \right) \nn \\
	&= \frac{1}{\omega\, n\cdot k}\left(\Sl{k}_\perp \gamma_\perp^\nu + \frac{k_\perp^2}{n\cdot k} n^\nu - 2k_{\perp}^\mu \right)\,.
\end{align}
In the first line we neglected $\omega \frac{n^\mu}{2}$ since we are focusing on the $\cO(\lambda^2)$ expansion of this object. In the third and fourth line we used the properties of the Dirac projectors $P_n, \, P_\bn$ acting on outgoing collinear spinors
\be 
	P_n = \frac{\nslash\bnslash}{4} \,,\qquad P_\bn = \frac{\bnslash\nslash}{4}\,,\qquad \id =P_n + P_\bn \,,\qquad \bar{u}_n P_\bn = \bar{u}_n \,,\qquad \bar{u}_n P_n = 0 \,.
\ee

We can now show how this result is reproduced in our framework. We first note, that in our definition of the radiative function for the emission of a soft gluon from a quark, as discussed in \Sec{sec:amp_softgluon}, we have factorized the ultrasoft derivative and ultrasoft gluon field into a soft function. Therefore, we do not have a correspondence at the level of the radiative function itself with \Eq{eq:Bonocore_comparison}. Instead, we must contract the radiative jet function with the corresponding soft function. Evaluating this at tree level, we have
\be\label{eq:1Blam2_fromquark_feyn}
	\fd{3cm}{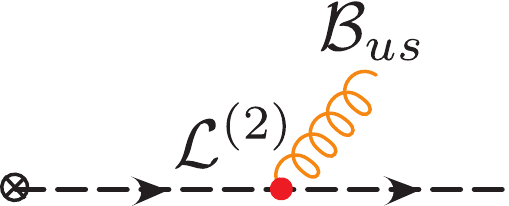}~=~\frac{1}{\omega\, n \cdot k}\left[\Sl k_\perp \gamma_\perp^\nu+ \frac{k_\perp^2}{n\cdot k} n^\nu -2 k_\perp^{\nu}  \right] \,.
\ee
We see that this indeed matches \eq{1Blam2_fromquark_feyn}. This shows, that despite the slightly different organization, the radiative functions in the two approaches are describing the same physics.

While the general form of the radiative functions in~\cite{Bonocore:2016awd} and those defined in this paper are similar, and indeed we have shown they give the same tree level result, the radiative functions defined in this paper differ in many aspects from the ones of~\cite{Bonocore:2016awd}. In addition to several trivial differences that are conventional, namely  \cite{Bonocore:2016awd} uses an incoming, instead of outgoing collinear state, and that the Lagrangian insertion occurs at $y=0$, there are some more major differences, which we now elaborate on.

The definitions of \Refs{DelDuca:1990gz,Bonocore:2015esa,Bonocore:2016awd} involve  a four dimensional convolution (which in dimensional regularization must be extended to a $d=4-2\epsilon$ dimensional convolution) instead of the one dimensional convolution along the light cone direction  derived in the effective theory. The ability to simplify the definition to involve a single variable convolution along the light cone relied on the multipole expansion in SCET, which renders the collinear matrix elements local in certain directions. The multipole expansion also implies that no expansions need to be performed after performing any integrals, and that all results are automatically homogeneous in the power counting. We believe that this is essential for achieving a true factorization. In particular, to claim a factorization, it must be that no divergences are generated by the integral in the final convolution variable, and that the convolution variable must be gauge invariant. This seems difficult to achieve if the convolution variable is a $d=4-2\epsilon$ dimensional momenta in dimensional regularization. While we are not yet able to prove that the convolutions in the single scalar lightcone variables in our formulation converge, we believe that the reduction to single variable convolutions that are not dimensionally regularized is an essential first step.

A second major difference in the definitions relates to gauge invariance. The current of \Eq{eq:Bonocore_current} is not by itself a gauge invariant object, and hence neither is the jet function of \Eq{eq:Bonocore_function}. However, as shown in \Refs{DelDuca:1990gz,Bonocore:2015esa,Bonocore:2016awd} it will give a gauge invariant result for the cross section with a single soft emission. This lack of gauge invariance in a radiative function defined in the full theory is perhaps not surprising, as one is attempting to factorize a gluon emission from a quark-antiquark current, but a full theory gluon is not gauge invariant. In the context of QED, where the radiative function was originally defined \cite{DelDuca:1990gz}, this issue was not present, since the photon is not charged, and therefore the radiative function is itself gauge invariant. It is ultimately this difference which makes the extension to the QCD case more difficult. To achieve a true factorization into objects which can be separately renormalized, it seems desirable that each of the objects be separately gauge invariant.

In the effective theory approach to defining the radiative functions the radiative jet and soft functions are each separately gauge invariant, since they are constructed from the gauge invariant gluon, $\cB_{us(n)}$ and quark, $\psi_{us(n)}$ fields which couple to the radiative functions. This leads to a fairly intricate structure of Wilson lines in the definition of the currents, which ensures their gauge invariance. Without the presence of these soft Wilson lines, it is also not clear how the current of \Refs{DelDuca:1990gz,Bonocore:2015esa,Bonocore:2016awd} can describe multiple soft gluon emissions.

Finally, another difference between the two approaches is that due to the manifest power counting in the effective theory, there are a large number of distinct field structures present in the SCET $\cO(\lambda)$ and $\cO(\lambda^2)$ Lagrangians, as compared with only the two terms present in the current of \Eq{eq:Bonocore_current}. These additional terms in the SCET Lagrangian have more than two collinear fields, or have an additional $\cP_\perp$ insertion, so that they first contribute when their is an additional collinear loop. For example, with a collinear loop, we have the distinct diagrams
\begin{align}
\fd{3.7cm}{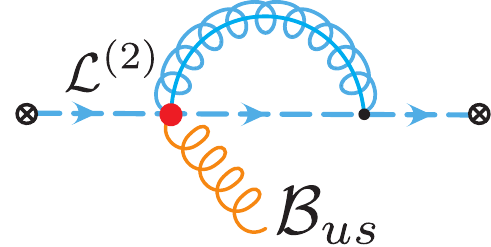}\,,\qquad
\fd{3.7cm}{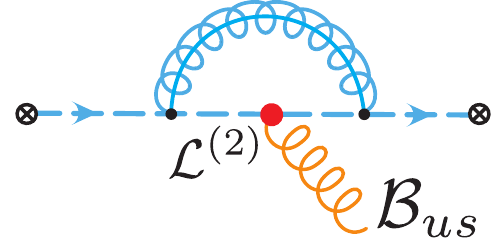}\,,\qquad
\fd{3.7cm}{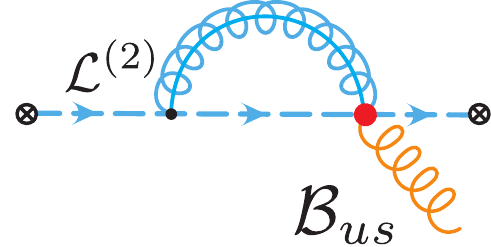}\,.
\end{align}
This is forced upon us in the effective theory by manifest power counting, which does not hold for the current of \Eq{eq:Bonocore_current}, which must be expanded after performing loop integrals. While this admittedly requires more distinct terms to be considered, we believe that they will have a simpler structure, and will facilitate an all orders understanding. Furthermore, it is hoped that only a subset will contribute at a given logarithmic accuracy. The additional terms involving multiple collinear fields will be discussed in more detail in \Sec{sec:RF_thrust} where we will consider the field structure of the complete set of radiative functions which contribute to the thrust event shape.

\section{Cross Section Level Factorization with Radiative Jet Functions}\label{sec:fact_RadiativeFunction}

In this section we show in detail how to perform subleading power factorization at the cross section level for radiative contributions in a gauge invariant manner to all orders in $\alpha_s$ in a non-abelian gauge theory. We derive the explicit structure of the factorization for the insertion of: 
\begin{itemize}
\item A single insertion of a $\cB_{us}$ field from the $\cL^{(2)}$ Lagrangian (\Sec{sec:sub_cross_quark}),
\item A double insertion of the $\psi_{us}$ field from two $\cL^{(1)}$ insertions (\Sec{sec:sub_cross_gluon}).
\end{itemize}
Here we have given the sections where the factorization is given to aid the reader.
Furthermore, we consider when this insertion is on both a collinear quark, or collinear gluon leg.  These are the two contributions to a factorized cross section that are non-vanishing at lowest order in $\alpha_s$, and allow a factorized description for the contributions that have been computed to fixed order in \cite{Moult:2016fqy,Moult:2017jsg}. Since they describe the lowest order contributions to the cross section, their renormalization should capture the LL series. The key result of this section is a factorized expression for these contributions to the cross section as a convolution between separately gauge invariant soft and collinear factors. The convolution is in terms of either one or two one-dimensional variables, which represent the position of operator insertions along the lightcone, similar to the convolution over the observable variable in the leading power factorization review in \Sec{sec:fact_LP}. 

After having worked out these two examples in detail, and illustrating how a gauge invariant factorization can be achieved, in \Sec{sec:RF_thrust}, we then provide an enumeration of the different radiative jet functions which contribute to the event shape thrust in $e^+e^-\to$ dijets to $\cO(\lambda^2)$, including those which first contribute at loop level. This will illustrate the complexity of subleading power factorization in an example of interest.

\subsection{Convolution Structure at Leading Power}\label{sec:fact_LP}

We begin by reviewing the well known leading power factorization for an SCET$_\text{I}$ event shape, $\tau$, in $e^+e^-\to$ dijets. Here the unique leading power operator is
\begin{align}\label{eq:LP_op_fact}
\cO^{(0)\mu}=\bar \chi_n  \gamma^\mu_\perp \chi _{\bar n}\,.
\end{align}
Our focus is on the convolution structure in the observable and momentum, as it is this aspect which will change at subleading power for the radiative jet functions. Therefore we will often suppress explicit Dirac or color indices.

Following the discussion of \Sec{sec:sub-fact}, after performing the BPS field redefinition, factorizing the observable, and Fierzing the Lorentz and Dirac structure (which we will suppress, as it is not important to our current discussion), the leading power cross section can be written as
\begin{align}
	\frac{1}{\sigma_0}\frac{d\sigma^{(0)}}{d\tau}&=H(Q^2) \int d^4x  \int d \tau_n d\tau_\bn d\tau_{us} \delta(\tau -\tau_n -\tau_\bn -\tau_{us}) \cdot\frac{1}{Q N_c} \tr \langle 0 | \bar \chi_n(x)_\alpha \delta(\tau_n-\hat \tau^{(0)}_n)   \chi_n(0)_\delta |0 \rangle \nn \\
&\hspace{-0.4cm}\cdot\frac{1}{Q N_c} \tr \langle 0 |   \chi_\bn(x)_\beta \delta(\tau_{\bar n}-\hat \tau^{(0)}_{\bar n})  \bar \chi_\bn(0)_\gamma   |0 \rangle \cdot \tr \langle 0 |  Y_\bn (x) Y_n^\dagger(x) \delta(\tau_{us}-\hat \tau^{(0)}_{us})  Y_n(0)  Y^\dagger_\bn(0) |0 \rangle\,.
\end{align}
This is derived by considering the factorization of the squared matrix element of \Eq{eq:LP_op_fact}, as was described in \Sec{sec:sum_fact}.
We would now like to simplify this expression, and remove the convolution over the variable $x$, which couples the different functions. To achieve this, we first define Fourier transforms of each of the functions,
\begin{align}
	\frac{1}{QN_c}\tr \langle 0 |  Y_\bn (x) Y_n^\dagger(x) \delta(\tau_{us}-\hat \tau^{(0)}_{us}) Y_n(0)  Y^\dagger_\bn(0) |0 \rangle &= \int \frac{d^4 r}{(2\pi)^4} e^{-ir \cdot x}  \cS(Q\tau_{us}, r)\,, \\
\frac{1}{QN_c} \tr \langle 0 | \bar \chi_n(x)_\alpha \delta(\tau_n-\hat \tau^{(0)}_n)  \chi_n(0)_\delta |0 \rangle&=\int \frac{d^4 l}{(2\pi)^4}   e^{-il\cdot x}  \cJ_n(\tau_n,l,Q) \left(\frac{\Sl{n}}{2} \right)_{\delta \alpha}\,, \nn\\
\frac{1}{QN_c} \tr \langle 0 |   \chi_\bn(x)_\beta \delta(\tau_{\bar n}-\hat \tau^{(0)}_{\bar n})  \bar \chi_\bn(0)_\gamma   |0 \rangle &=\int \frac{d^4 k}{(2\pi)^4} e^{-ik\cdot x} \cJ_\bn(\tau_\bn,k,Q)   \left(\frac{\Sl{\bar n}}{2} \right)_{\beta \gamma}  \,, \nn
\end{align}
and writing the result in terms of light cone coordinates, we can express the cross section as
\begin{align}
	\frac{1}{\sigma_0}\frac{d\sigma^{(0)}}{d\tau}&= H(Q^2) \int d^4x \int d \tau_n d\tau_\bn d\tau_{us} \delta(\tau -\tau_n -\tau_\bn -\tau_{us}) \cdot   Q\int \frac{d^4 r}{(2\pi)^4}  e^{-ir_1 \cdot x}  \nn \\
&\cdot \int \frac{dk^+ dk^- d^2 k_\perp}{2 (2\pi)^4} e^{-i(k^+x^-/2+ k^-x^+/2 -k_\perp \cdot x_\perp)} \cdot \int \frac{dl^+ dl^- d^2 l_{\perp}}{2 (2\pi)^4} e^{-i(l^+x^-/2+ l^-x^+/2 -l_{\perp} \cdot x_\perp)} \nn \\ 
&\cdot   \cS(\tau_{us}, r)     \cJ_n(\tau_n, l,Q)      \cJ_\bn(\tau_\bn, k,Q)\,.
\end{align}
We now use that  $\cJ_\bn$ depends only on $k^-$, and $\cJ_n$ depends only on $k^+$, which follows from the multipole expansion in the effective theory.
Performing the integrals in $k^+, k^\perp$, $l^-, l^\perp$, we then find
\begin{align}
	\frac{1}{\sigma_0}\frac{d\sigma^{(0)}}{d\tau}&=H(Q^2) \int d \tau_n d\tau_\bn d\tau_{us} \delta(\tau -\tau_n -\tau_\bn -\tau_{us})    \nn \\
&\cdot \left[  \int \frac{d^4 r}{(2\pi)^4}    \cS(Q\tau_{us}, r)  \right] \cdot \left[  \int \frac{dk^-}{2\pi Q} \cJ_\bn(\tau_\bn, k^-,Q)  \right] \cdot   \left[   \int \frac{dl^+}{2\pi Q} \cJ_n(\tau_n, l^+,Q)  \right]\,,
\end{align}
which defines  the standard leading power factorization
\begin{align}\label{eq:fact_tau_LP}
	\frac{1}{\sigma_0}\frac{d\sigma^{(0)}}{d\tau}&= Q^5 H(Q^2) \int d \tau_n d\tau_\bn d\tau_{us} \delta(\tau -\tau_n -\tau_\bn -\tau_{us})    S(Q\tau_{us}) J_{\bar n}(Q^2\tau_\bn) J_n(Q^2\tau_n)\,.
\end{align}
Each of $S$, $J_n$ and $J_{\bar n}$ are infrared finite\footnote{We assume that all functions are defined with their appropriate zero-bin subtractions \cite{Manohar:2006nz}.} and gauge invariant. The only coupling between the soft and collinear degrees of freedom is the convolution in the physical observable $\tau$. Diagramatically, we have
\begin{align}
\fd{3cm}{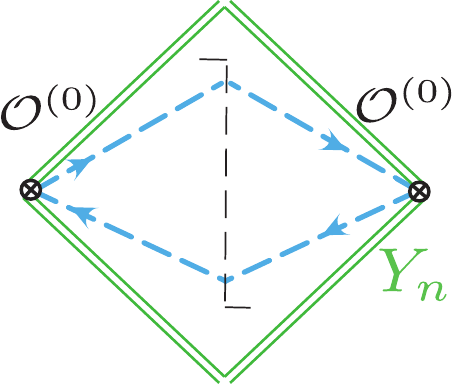}= \fd{3cm}{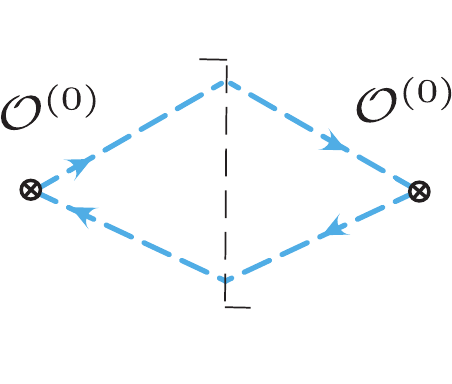} \otimes  \fd{3cm}{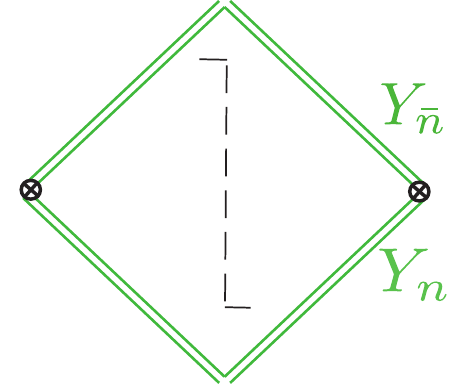}\,,
\end{align}
where the blue lines indicated collinear fields, the double green lines indicate Wilson lines, and the $\otimes$ indicates the convolution in $\tau$.
We have not drawn explicitly the loop corrections, as it is understood that this holds to all orders in $\alpha_s$.
Here, for simplicity, we have only illustrated the factorization of the soft and collinear pieces of the diagrams, since it is this aspect which is relevant for the radiative jet functions. 
This picture will be contrasted with that for the radiative functions derived in the next section.

\subsection{Convolution Structure at Subleading Power}\label{sec:fact_NLP}

We now consider the subleading factorization at cross section level for the contributions coming from Lagrangian insertions and show how they lead to radiative jet functions. We discuss in detail the convolution structure for how these radiative jet functions enter the factorization.

\subsubsection{Soft Quark Emission}\label{sec:sub_cross_quark}

 At the cross section level, two insertions of the $\cL^{(1)}$ soft quark Lagrangian are required to give a non-vanishing contribution. At tree level, such contributions will describe single soft quark emission. Subleading power shape functions arising from two insertions of the $\cL^{(1)}$ soft quark Lagrangian were considered in \cite{Beneke:2002ph,Bosch:2004cb,Lee:2004ja}. They have a similar structure to the matrix elements considered in this section, although the shape functions are defined as matrix elements of $B$ meson states, instead of vacuum matrix elements.

The Lagrangian of interest for soft quark emission is 
\begin{align}
\cL^{(1)\text{BPS}}_{\chi \psi}=\bar \chi_n g \Sl{\cB}_{n\perp} \psi_{us(n)} +\bar \psi_{us(n)} g\Sl{\cB}_{n\perp} \chi_n\,. 
\end{align}
We first consider the insertion of this operator onto a collinear quark line, which at lowest perturbative order corresponds to the emission of a soft quark from a collinear quark. To illustrate this, we use as an example $e^+e^-\to$ dijets with the hard scattering operator of \Eq{eq:LP_op_fact}. The derivation of the factorization requires both a factorization of the spin and color structures, as well as of the momentum convolutions. The factorization of the spin and color structures is in principle a straightforward excerise, which requires, in the present case, the repeated application of Fierz identities to the matrix element
\begin{align}
\int d^4x \int d^4y \int d^4z\, \bigl\langle 0 \bigr|\cO^{(0)\mu \dagger}(x)  \cL^{(1)\text{BPS}}_{\chi \psi}(y) \cL^{(1)\text{BPS}}_{\chi \psi}(z) \cO^{(0)\mu}(0) \bigl|0\bigr\rangle\,.
\end{align}
The details of this are presented in \App{app:Fierzing}, and here we focus on the convolution structure. We note that in general one must consider the presence of evanescent operators, however, since in this paper we do not consider the renormalization of the matrix elements appearing in the factorization formula, we leave this to future work. After factorizing the spin and color structure, we arrive at the following expression for the radiative contribution to the cross section
\begin{align}
&\frac{1}{\sigma_0}\frac{d\sigma^{(2),\text{rad}}_{\psi, n,q} }{d\tau}= H(Q^2) \int d^4x \int d^4y \int d^4z\int d \tau_n d\tau_\bn d\tau_{us} \delta(\tau -\tau_n -\tau_\bn -\tau_{us}) \nn \\
&\cdot \frac{1}{N_c}\tr \langle 0 |  Y_\bn^\dagger (x) Y_n^\dagger(x) \psi_{us(n)}(y)_\beta \delta(\tau_{us}-\hat \tau^{(0)}_{us}) \bar \psi_{us(n)}(z)_\gamma Y_n(0)  Y_\bn(0) |0 \rangle  \left(\frac{\Sl{n}}{2} \right)_{\beta \gamma} \nn \\
&\cdot \frac{1}{Q N_c} \tr \langle 0 | \chi_n(x)_\beta \bar \chi_n(y)_\gamma \cB^\mu_{n\perp} (y) \delta(\tau_n-\hat \tau^{(0)}_n)\cB^\nu_{n\perp}(z) \chi_n(z)_\alpha \bar \chi_n(0)_\delta |0 \rangle  g^{\mu\nu}_\perp  \left(\frac{\Sl{ n}}{2} \right)_{\beta \gamma}  \left(\frac{\Sl{ n}}{2} \right)_{\alpha\delta}  \nn \\
&\cdot \frac{1}{Q N_c} \tr \langle 0 |   \chi_\bn(x)_\beta \delta(\tau_{\bar n}-\hat \tau^{(0)}_{\bar n})  \bar \chi_\bn(0)_\gamma   |0 \rangle \left(\frac{\Sl{\bar n}}{2} \right)_{\beta \gamma}\,,
\end{align}
which we would like to factorize into hard, jet and soft functions. The superscripts and subscripts labeling the cross section indicate that this is the $\cO(\lambda^2)$ contribution arising from the radiative emission of a soft quark from the $n$ collinear sector. Fourier transforming each of the matrix elements, we have
\begin{align}
&\frac{1}{QN_c}\tr \langle 0 |  Y_\bn^\dagger (x) Y_n^\dagger(x) \psi_{us(n)}(y)_\beta \delta(\tau_{us}-\hat \tau^{(0)}_{us}) \bar \psi_{us(n)}(z)_\gamma Y_n(0)  Y_\bn(0) |0 \rangle \nn \\
&\hspace{1cm}= \int \frac{d^4 r_1}{(2\pi)^4}  \frac{d^4 r_2}{(2\pi)^4}  \frac{d^4 r_3}{(2\pi)^4} e^{-ir_1 \cdot x} e^{-ir_2 \cdot y} e^{-ir_3 \cdot z} S_{\chi n\psi}^{(2)}(\tau_{us}, r_1, r_2, r_3,Q) \left(\frac{\Sl{\bar n}}{2} \right)_{\beta \gamma}\,,\nn \\
&\frac{1}{Q N_c} \tr \langle 0 | \chi_n(x)_\beta \bar \chi_n(y)_\gamma \cB^\mu_{n\perp} (y) \delta(\tau_n-\hat \tau^{(0)}_n)\cB^\nu_{n\perp}(z) \chi_n(z)_\alpha \bar \chi_n(0)_\delta |0 \rangle \nn \\
&\hspace{1cm}=\int \frac{d^4 l_1}{(2\pi)^4}  \frac{d^4 l_2}{(2\pi)^4}  \frac{d^4 l_3}{(2\pi)^4}  e^{-il_1\cdot x} e^{-il_2\cdot y} e^{-il_3\cdot y} \cJ_{\chi n\psi}^{(2)}(\tau_n, l_1, l_2,l_3,Q)g^{\mu\nu}_\perp  \left(\frac{\Sl{\bar n}}{2} \right)_{\beta \gamma}  \left(\frac{\Sl{\bar n}}{2} \right)_{\alpha\delta}\,,\nn\\
&\frac{1}{QN_c} \tr \langle 0 |   \chi_\bn(x)_\beta \delta(\tau_{\bar n}-\hat \tau^{(0)}_{\bar n})  \bar \chi_\bn(0)_\gamma   |0 \rangle=\int \frac{d^4 k}{(2\pi)^4} e^{-ik\cdot x} \cJ_\bn(\tau_\bn, k,Q) \left(\frac{\Sl{n}}{2} \right)_{\beta \gamma}  \,.
\end{align}
Using the locality of the collinear matrix elements, as discussed in our derivation of the amplitude level radiative functions (see \Eq{eq:local_collinear} and the surrounding discussion), we find
{\small
\begin{align}
	&\hspace{-0.25cm}\frac{1}{\sigma_0}\frac{d\sigma^{(2),\text{rad.}}_{\psi, n,q} }{d\tau}=Q^5 H(Q^2) \int d \tau_n d\tau_\bn d\tau_{us} \delta(\tau -\tau_n -\tau_\bn -\tau_{us})  \left[  \int \frac{dk^-}{2\pi Q} \cJ_\bn(\tau_\bn, k^-,Q)  \right] \! \int \frac{dr_2^+}{2\pi Q} \frac{dr_3^+}{2\pi Q} \\
	&\quad\cdot \left[  Q\int \frac{d^4 r_1}{(2\pi)^4}   \int \frac{dr_2^-}{2\pi}  \frac{d^2r_2^\perp}{(2\pi)^2}   \int \frac{dr_3^-}{2\pi}  \frac{d^2r_3^\perp}{(2\pi)^2}       \frac{ S_{\chi n\psi}^{(2)}(\tau_{us}, r_1, r_2,r_3,Q)}  {r_2^+ r_3^+}  \right]   \cdot   \left[  Q^2 \int \frac{dl_1^+}{2\pi} \cJ_{\chi n \psi}(\tau_n, l_1^+, r_2^+,r_3^+) \cdot r_2^+ r_3^+ \right]\nn\\
	&\equiv Q^5 H(Q^2)  \int d \tau_n d\tau_\bn d\tau_{us} \delta(\tau -\tau_n -\tau_\bn -\tau_{us}) J_\bn(Q^2\tau_\bn) \int \frac{dr_2^+}{2\pi Q} \frac{dr_3^+}{2\pi Q}  S_{\chi n\psi}^{(2)}(Q\tau_{us}, r_2^+,r_3^+) J_{\chi n\psi}^{(2)}(\tau_n, r_2^+,r_3^+,Q)\,.\nn
\end{align}}%
Here $J_{n\psi}^{(2)}(\tau_n, r_2^+,r_3^+)$ defines the radiative jet function at the cross section level. Note that it is defined as a collinear matrix element, so all loop corrections are collinear in nature. Diagramatically, we have
\begin{align}
\fd{3cm}{figures_b/soft_quark_diagram_BPS_low.pdf} =  \int dr_2^+ dr_3^+ \fd{3cm}{figures_b/soft_quark_collinear_piece_low.pdf} \otimes  \fd{3cm}{figures_b/soft_quark_diagram_wilsonframe_low.pdf}\,.
\end{align}
At lowest order in $\alpha_s$, the soft function is proportional to $\delta(r_2^+-r_3^+)$, which simplifies the structure of the convolution to a single variable. In the presence of radiative corrections, the full convolution structure is required.
As expected at subleading power, this contribution is first non-vanishing crossing the cut, namely the soft quark. Purely virtual contributions are proportional to $\delta(\tau)$, and hence leading power.  In the case that the soft quark is radiated from the $\bar n$ collinear sector, an identical factorization applies.

The ability to formulate this factorization in a gauge invariant way relies on the use of non-local gauge invariant collinear quark and gluon and soft quark and gluon fields. These operators have a highly intricate Wilson line structure, involving both soft and collinear Wilson lines situated at a variety of positions along different light cones. However, this structure is completely dictated by the symmetries of the effective theory. 

As compared with the leading power factorization of \Eq{eq:fact_tau_LP}, there is a convolution structure in the $+$ component of the soft momentum, in addition to the convolution in the observable $\tau$. This couples the collinear and soft sectors in a more non-trivial way, describing the ``radiation" of a soft parton from the collinear jet at a position along the light cone, as was the case at the amplitude level. Note that while we can perform the factorization of the jet and soft functions into this convolution structure, it is not a priori guaranteed that such convolutions converge. This would indicate a naive breakdown of the factorized expression, or at least that a reorganization is required. This has been explicitly observed in subleading power factorization formulae for $B$-meson decays \cite{Beneke:2003pa}. In defining the jet and soft functions, we have inserted factors of the convolution variables $r_2^+$ and $r_3^+$. This is done so that the lowest order divergence appears entirely in the soft function, allowing it to be extracted using standard plus distributions. We can therefore make sense of the convolutions appearing in the factorization at lowest order in $\alpha_s$.  An understanding of the convolutions appearing in the factorization formulas in this paper is left to future work. In this paper we will not consider radiative corrections to the soft and jet functions $S_{\chi n \psi}^{(2)}(\tau_{us}, r_2^+,r_3^+)$, $J_{\chi n\psi}^{(2)}(\tau_n, r_2^+,r_3^+)$, leaving this to future work.  The renormalization of these operators, and the corresponding renormalization group evolution will resum subleading power logarithms in the thrust variable.

We can also perform an identical factorization when the soft quark Lagrangian is inserted on a collinear gluon leg. Here we take as a concrete example the case of thrust in $H\to gg$ with the leading power operator 
\begin{align}
\cO_\cB^{(0)}=-2\omega_1 \omega_2 \delta^{ab} \cB_{\perp \bar n, \omega_2}^a \cdot \cB_{\perp \bar n, \omega_1}^b H\,.
\end{align}
Going through an identical procedure as for the previous case, after performing the Fierzing we arrive at the following expression
\begin{align}
&\frac{1}{\sigma_0}\frac{d\sigma^{(2),\text{rad}}_{\psi, n,g} }{d\tau}= H(Q^2) \int d^4x \int d^4y \int d^4z\int d \tau_n d\tau_\bn d\tau_{us} \delta(\tau -\tau_n -\tau_\bn -\tau_{us}) \nn \\
&\cdot  \langle 0 |  \cY_\bn^{ac} (x) \cY_n^{ad}(x) \bar \psi_{us(n)}(y)_\beta \delta(\tau_{us}-\hat \tau^{(0)}_{us})  \psi_{us(n)}(z)_\gamma \cY_\bn^{\dagger bc}(0)  \cY_n^{\dagger bd}(0) |0 \rangle  \left(\frac{\Sl{n}}{2} \right)_{\beta \gamma} \nn \\
&\cdot \frac{1}{N_c} \tr \langle 0 |  \cB^{d\mu}_{n\perp} (x)   (\bar \chi_n(z)  \Sl{\cB}_{n\perp} (z)  )_\alpha  \delta(\tau_n-\hat \tau^{(0)}_n)   (g\Sl{\cB}_{n\perp} (y) \chi_n(y)   )_\delta \cB^{d\nu}_{n\perp}(0) |0 \rangle  g^{\mu\nu}_\perp  \left(\frac{\Sl{ n}}{2} \right)_{\alpha\delta}  \nn \\
&\cdot \frac{1}{N_c} \tr \langle 0 |   \cB^{c\mu}_{\bn \perp}(x) \delta(\tau_{\bar n}-\hat \tau^{(0)}_{\bar n})   \cB^{c\nu}_{\bn \perp}(0)   |0 \rangle g^{\mu \nu}_\perp\,.
\end{align}
Fourier transforming each of the matrix elements, we have
\begin{align}
&\frac{1}{Q}\langle 0 |  \cY_\bn^{ac} (x) \cY_n^{ad}(x) \bar \psi_{us(n)}(y)_\beta \delta(\tau_{us}-\hat \tau^{(0)}_{us})  \psi_{us(n)}(z)_\gamma \cY_\bn^{\dagger bc}(0)  \cY_n^{\dagger bd}(0) |0 \rangle  \nn \\
&\hspace{1cm}= \int \frac{d^4 r_1}{(2\pi)^4}  \frac{d^4 r_2}{(2\pi)^4}  \frac{d^4 r_3}{(2\pi)^4} e^{-ir_1 \cdot x} e^{-ir_2 \cdot y} e^{-ir_3 \cdot z} S_{\cB n\psi}^{(2)}(\tau_{us}, r_1, r_2, r_3,Q) \left(\frac{\Sl{\bar n}}{2} \right)_{\beta \gamma}\,,\nn \\
&\frac{1}{Q^2 N_c} \tr \langle 0 |  \cB^{d\mu}_{n\perp} (x)   (\bar \chi_n(z)  \Sl{\cB}_{n\perp} (z)  )_\alpha  \delta(\tau_n-\hat \tau^{(0)}_n)   (g\Sl{\cB}_{n\perp} (y) \chi_n(y)   )_\delta \cB^{d\nu}_{n\perp}(0) |0 \rangle \nn \\
&\hspace{1cm}=\int \frac{d^4 l_1}{(2\pi)^4}  \frac{d^4 l_2}{(2\pi)^4}  \frac{d^4 l_3}{(2\pi)^4}  e^{-il_1\cdot x} e^{-il_2\cdot y} e^{-il_3\cdot y} \cJ_{\cB n\psi}^{(2)}(\tau_n, l_1, l_2,l_3,Q)g^{\mu\nu}_\perp    \left(\frac{\Sl{\bar n}}{2} \right)_{\alpha\delta}\,,\nn\\
&\frac{1}{N_c} \tr \langle 0 |   \cB^{c\mu}_{\bn \perp}(x) \delta(\tau_{\bar n}-\hat \tau^{(0)}_{\bar n})   \cB^{c\nu}_{\bn \perp}(0)   |0 \rangle=\int \frac{d^4 k}{(2\pi)^4} e^{-ik\cdot x} \cJ_\bn(\tau_\bn, k,Q)g^{\mu \nu}_\perp  \,.
\end{align}
As before, this can be simplified to convolutions involving just the light cone positions
{\footnotesize
\begin{align}
&\frac{1}{\sigma_0}\frac{d\sigma^{(2),\text{rad.}}_{\psi, n,g} }{d\tau}= Q^5 H(Q^2) \int d \tau_n d\tau_\bn d\tau_{us} \delta(\tau -\tau_n -\tau_\bn -\tau_{us})   \left[  \int \frac{dk^-}{2\pi Q} \cJ_\bn(\tau_\bn, k^-,Q)  \right]  \int \frac{dr_2^+}{2\pi Q} \frac{dr_3^+}{2\pi Q}   \\
&\cdot \left[ Q \int \frac{d^4 r_1}{(2\pi)^4}   \int \frac{dr_2^-}{2\pi}  \frac{d^2r_2^\perp}{(2\pi)^2}   \int \frac{dr_3^-}{2\pi}  \frac{d^2r_3^\perp}{(2\pi)^2}       \frac{ S_{\cB n\psi}^{(2)}(\tau_{us}, r_1, r_2,r_3,Q)}  {r_2^+ r_3^+}  \right]   \cdot   \left[  Q^2 \int \frac{dl_1^+}{2\pi} \cJ_{\cB n \psi}(\tau_n, l_1^+, r_2^+,r_3^+,Q) \cdot r_2^+ r_3^+ \right]\nn\\
	&\equiv Q^5 H(Q^2)  \int d \tau_n d\tau_\bn d\tau_{us} \delta(\tau -\tau_n -\tau_\bn -\tau_{us}) J_\bn(Q^2\tau_\bn) 
	\cdot  \int \frac{dr_2^+}{2\pi Q} \frac{dr_3^+}{2\pi Q}  S_{\cB n\psi}^{(2)}(\tau_{us}, r_2^+,r_3^+,Q) J_{\cB n\psi}^{(2)}(\tau_n, r_2^+,r_3^+,Q)\nn\,.
\end{align}}%
Schematically, this can be illustrated as
\begin{align}
\fd{3cm}{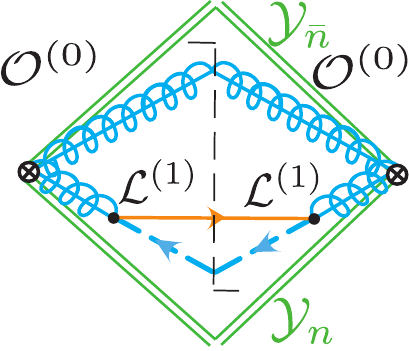} =  \int dr_2^+ dr_3^+ \fd{3cm}{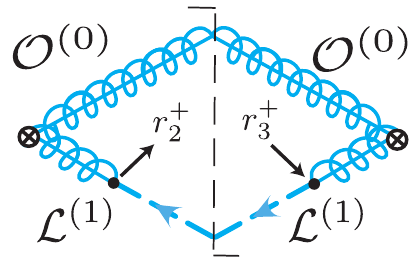} \otimes  \fd{3cm}{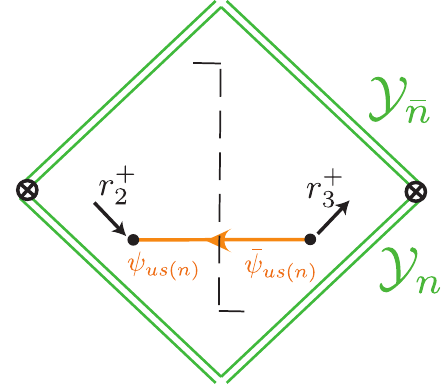}\,.
\end{align}
This takes an identical form to the case of the emission from a collinear quark, with the exception of the detailed structure of the Wilson lines and fields in the jet function.

\subsubsection{Soft Gluon Emission}\label{sec:sub_cross_gluon}

At lowest order in $\alpha_s$, we also have contributions from soft gluon emission. Due to the LBK theorem, these first arise through a single insertion of the $\cL^{(2)}$ Lagrangian. Here we will show how these can be factorized to all orders at the cross section level, for the insertion of the $\cL^{(2)}$ Lagrangian on either a collinear quark or gluon leg.

We begin by considering the case of the insertion on a collinear quark leg, corresponding to soft gluon emission from a collinear quark. We again take as a concrete process thrust for $e^+e^-\to$ dijets, as mediated by the hard scattering operator of \Eq{eq:LP_op_fact}. For convenience, we recall the form of the Lagrangian
\begin{align}
{\cal L}_{n \xi}^{(2) \text{BPS}} &=  \bar \chi_n  \left(  T^a \gamma^\mu_\perp \frac{1}{\bar \cP}  i \Sl \partial_{us\perp} - i  {\overleftarrow{\Sl \partial}}_{us\perp} \frac{1}{\bar \cP} T^a \gamma^\mu_\perp   \right)   \frac{\Sl \bn}{2} \chi_n  \,   g \cB_{us(n)}^{a\mu} \,.
\end{align}
This contributes to the cross section through the interference with the leading power hard scattering operator. As was the case at amplitude level, the factorization for the ultrasoft gluon emission is more complicated due to the presence of the ultrasoft derivative. We would like to arrange the ultrasoft derivative such that it acts only on the ultrasoft fields in the soft function.  This can be done using identical arguments as given in \Sec{sec:RF_amp_squark}. Flipping the derivative to act in the soft sector, and performing the Fierzing of the matrix element to obtain a factorized form, we find
\begin{align}
\frac{1}{\sigma_0}\frac{d\sigma^{(2),\text{rad}}_{\cB_{us}, n} }{d\tau}&= H(Q^2) \int d^4x d^4y \int d \tau_n d\tau_\bn d\tau_{us} \delta(\tau -\tau_n -\tau_\bn -\tau_{us}) \nn \\
&\cdot\frac{1}{QN_c} \tr \langle 0 | \chi_n(x)_\kappa \delta(\tau_n-\hat \tau_n)  \frac{1}{\bar \cP}\bar \chi_n(y)_\rho \chi_n(y)_\alpha \bar \chi_n(0)_\delta |0 \rangle  \left(\frac{\Sl{\bar n}}{2} \right)_{\kappa \rho}  \left(\frac{\Sl{\bar n}}{2} \right)_{\alpha \delta} \nn \\
& \cdot\frac{1}{QN_c} \tr \langle 0 |   \chi_\bn(x)_\beta \delta(\tau_n-\hat \tau_n)  \bar \chi_\bn(0)_\gamma   |0 \rangle \left(\frac{\Sl{ n}}{2} \right)_{\beta \gamma}  \nn \\
&\cdot \frac{1}{N_c} \tr \langle 0 |  Y_\bn^\dagger (x) Y_n^\dagger(x) \delta(\tau_{us}-\hat \tau_{us}) \partial_\perp \cdot \cB_{ us (n) \perp}(y) Y_n(0)  Y_\bn(0) |0 \rangle   \,.
\end{align}
The derivation of the Lorentz, Dirac, and color structure is left to \App{app:Fierzing}. 
Fourier transforming each of the functions
\begin{align}
&\frac{1}{Q N_c}\tr \langle 0 |  Y_\bn^\dagger (x) Y_n^\dagger(x) \delta(\tau_{us}-\hat \tau_{us}) \partial_\perp \cdot \cB_{us(n) \perp }(y) Y_n(0)  Y_\bn(0) |0 \rangle \nn \\
&\hspace{4cm}=\int \frac{d^4 r_1}{(2\pi)^4}  \frac{d^4 r_2}{(2\pi)^4} e^{-ir_1 \cdot x} e^{-ir_2 \cdot y} S_{n \cB_{us}}^{(2)}(\tau_{us}, r_1, r_2,Q)\,,\nn \\
&\frac{1}{Q N_c} \tr \langle 0 | \chi_n(x)_\beta \delta(\tau_n-\hat \tau_n) \frac{1}{\bar \cP}\bar \chi_n(y)_\gamma \chi_n(y)_\alpha \bar \chi_n(0)_\delta |0 \rangle \nn \\
&\hspace{3cm}=\int \frac{d^4 l_1}{(2\pi)^4}  \frac{d^4 l_2}{(2\pi)^4}  e^{-il_1\cdot x} e^{-il_2\cdot y} \cJ_{n \cB_{us}}^{(2)}(\tau_n, l_1, l_2,Q) \left(\frac{\Sl{\bar n}}{2} \right)_{\beta \gamma}  \left(\frac{\Sl{\bar n}}{2} \right)_{\alpha \delta} \,,\nn\\
&\frac{1}{Q N_c} \tr \langle 0 |   \chi_\bn(x)_\beta \delta(\tau_n-\hat \tau_n)  \bar \chi_\bn(0)_\gamma   |0 \rangle=\int \frac{d^4 k}{(2\pi)^4} e^{-ik\cdot x} \cJ_\bn(\tau_\bn, k,Q) \left(\frac{\Sl{ n}}{2} \right)_{\beta \gamma} \,,
\end{align}
and using the locality of the collinear functions to simplify the structure of the convolutions, we find
\begin{align}
	\frac{1}{\sigma_0}\frac{d\sigma^{(2),\text{rad}}_{\cB_{us}, n} }{d\tau}&= Q^5 H(Q^2)  \int d \tau_n d\tau_\bn d\tau_{us} \delta(\tau -\tau_n -\tau_\bn -\tau_{us})\cdot \left[  \int \frac{dk^-}{2\pi} \cJ_\bn(\tau_\bn, k^-,Q)  \right] \nn \\
	&\qquad\cdot   \int \frac{dr_2^+}{2\pi Q}  \cdot \left[  \int \frac{d^4 r_1}{(2\pi)^4}   \int \frac{dr_2^-}{2\pi}  \frac{d^2r_2^\perp}{(2\pi)^2}    \frac{ S_{n\cB_{us}}^{(2)}(\tau_{us}, r_1, r_2,Q)}  {r_2^+}  \right] \nn \\
	&\qquad \cdot   \left[   \int \frac{dl_1^+}{2\pi} \cJ_{n\cB_{us}}(\tau_n, l_1^+, r_2^+,Q) \cdot r_2^+ \right]
	\nn\\&\equiv
	Q^5 H(Q^2)  \int d \tau_n d\tau_\bn d\tau_{us} \delta(\tau -\tau_n -\tau_\bn -\tau_{us})  J_{\bar n}(Q^2\tau_\bn) \nn\\
	&\qquad\cdot \int \frac{dr_2^+}{2\pi Q}  S_{n\cB_{us}}^{(2)}(Q\tau_{us}, r_2^+)  J_{n\cB_{us}}^{(2)}(\tau_n, r_2^+,Q)\,.
\end{align}
Diagramatically, we have
\begin{align}
\fd{3cm}{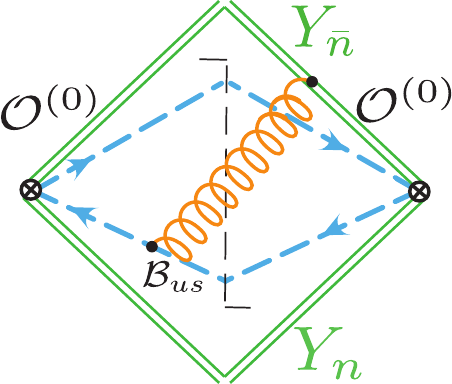}=\int d r_2^+ \fd{3cm}{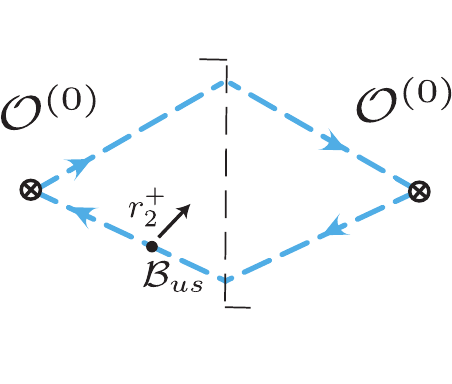} \otimes  \fd{3cm}{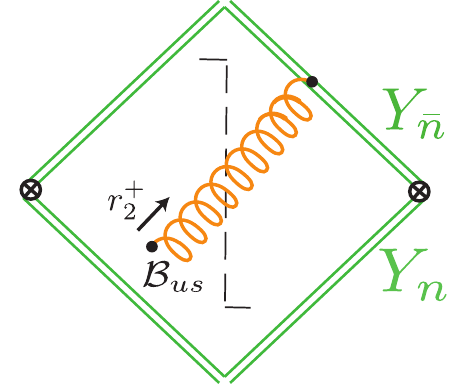}\,.
\end{align}
Here the green lines indicate Wilson lines, and the orange spring indicates the $\cB_{us(n)}$ field. As with the soft quark radiative function, this soft gluon radiative function is first non-vanishing with a single emission crossing the cut.
We have done some rearrangement, multiplying and dividing by a factor of $l_2^+$. This choice is to make the tree level expression for the jet function independent of $l_2^+$. In particular, we find a radiative jet function
\begin{align}
J_{n\cB_{us}}^{(2)}(\tau_n, r_2^+)=\left[   \int \frac{dl_1^+}{2\pi} \cJ_{n\cB_{us}}(\tau_n, l_1^+, r_2^+) \cdot r_2^+ \right]\,,
\end{align}
which incorporates all the collinear dynamics in the $n$ sector from which the $\cB_{us(n)}$ field is emitted. Since it is defined as a collinear matrix element, all loop corrections are collinear in nature, and are factorized from the soft loop corrections, which are described by the soft function.

We can also consider the insertion of an $\cL^{(2)}$ on a collinear gluon line, which corresponds at lowest order to the emission of a soft gluon from a collinear gluon leg. However, here we will find something interesting. If the collinear gluon field is taken to have no label perpendicular momentum, the $\cO(\lambda^1)$ Lagrangian insertion vanishes. However, at $\cO(\lambda^2)$, we have a contribution from the subsubleading gluon lagrangian $\cL_{ng}^{(2)}$, defined in \eq{subsubgluonlagr}. By dropping the terms proportional to the label perpendicular momentum, the relevant insertion comes from the three field operator
\be 
	\cL_{ng}^{(2)} \supset g \text{Tr} \left( \partial_\perp^{[\mu} \cB_{us \perp}^{\nu]}[ \cB^\perp_{n\mu}  , \cB^\perp_{n\nu}   ]\right)\,.
\ee
However, we note that
\begin{align}
	\partial_\perp^{[\mu} \cB_{us \perp}^{\nu]} &\equiv \partial_\perp^{[\mu} \big( S_n^\dagger \partial_\perp^{\nu]}S_n] + S_n^\dagger gA_{us \perp}^{\nu]}S_n \big) \nn \\
	&=[\partial_\perp^{[\mu} S_n^\dagger] [\partial_\perp^{\nu]}S_n] +  S_n^\dagger \partial_\perp^{[\mu} \partial_\perp^{\nu]} S_n+ \partial_\perp^{[\mu} S_n^\dagger gA_{us \perp}^{\nu]}S_n \nn \\
	&= \partial_\perp^{[\mu} gA_{us \perp}^{\nu]} + \cO(g^2)\,.
\end{align}
For the case of dijet production, where the two jets are back to back, all Wilson lines in the soft function are along the light cone directions $n$ or $\bar n$. Therefore, we see that the emission of a soft gluon from $\cL_{ng}^{(2)}$ gives a vanishing contribution to the cross section at lowest order in $\alpha_s$. This fact was used for the calculation of the leading log series in \Ref{Moult:2018jjd}. We will therefore not consider its factorization explicitly, since in this section we have only considered the factorization for contributions at lowest order in $\alpha_s$. Beyond $\cO(\alpha_s)$, there are other contributions that will be discussed in more detail in the next section.

\subsection{Discussion}\label{sec:discuss_squared}

In the section we have derived a factorized expression in terms of a convolution of gauge invariant soft and collinear matrix elements for the emission of a gauge invariant soft quark or gluon field at the cross section level. The main new feature at subleading power, are additional convolutions in the momentum passed between the soft and collinear sectors. Due to the multipole expansion in SCET this can be simplified to a single component $k^+_i$, which represents the position along the light cone. This also follows directly from the SCET power counting since only the $+$-momenta are the same size. To be able to separately renormalize the soft and collinear matrix elements, allowing for a resummation of subleading power logarithms, it is essential that both matrix elements are gauge invariant, and that no additional divergences appear in the single variable convolution. In a non-abelian gauge theory, the ability to formulate a soft emission in a gauge invariant manner is highly non-trivial since soft partons carry non-abelian charge, and it relies on our use of non-local gauge invariant soft quark and gluon fields, and an understanding of the all orders Lagrangian describing their interactions.

Some of the factorized expressions presented in this section are similar to those which have appeared in the $B$-physics literature \cite{Lee:2004ja}. In this case, instead of being vacuum matrix elements, the soft function is a matrix element of $B$ meson states. These were studied for the case of heavy to light decays where the heavy quark is treated using HQET. Nevertheless, their structure as non-local operators with insertions along the light cone is similar. This can be viewed as another advantage of using the operator based approach, namely since the SCET Lagrangians are universal, the same structures will appear in a variety of distinct physical processes allowing for a universal framework. As an example, although we have used for concreteness here the case of dijet production, since the insertions appear in a single collinear sector, this could be extended to N-jet production in a straightforward manner.

In this section we have explicitly worked out the structure, including the Lorentz and Dirac structure for  two examples, namely single soft quark and soft gluon emission, with the goal of showing in detail how the factorization with Lagrangian insertions can be performed, and how it gives rise to radiative functions at the cross section level. These two examples were chosen, since they contribute at lowest order. It should be clear that a similar factorization can be performed for an insertion of any term in the subleading power Lagrangians of \Eqs{eq:L1_fields}{eq:L4_fields}, or for multiple Lagrangian insertions. In \Sec{sec:RF_thrust} we will perform a systematic classification for the case of thrust in $e^+e^-\to$ dijets of the field structure of all possible radiative functions.

\section{Classification of Radiative Contributions for Dijet Event Shapes}\label{sec:RF_thrust}

Having shown how we can achieve a factorization of radiative contributions in terms of a convolution of gauge invariant soft and collinear matrix elements,  in this section we extend this to a complete classification of the radiative contributions for an SCET$_\text{I}$ event shape observable. For concreteness, we will consider thrust in $e^+e^-\to $ dijets, however thrust for $H\to gg$ would have a very similar form. Unlike in the previous section, where we worked out in detail the factorization for those contributions that contribute at lowest order in $\alpha_s$, here we will not explicitly perform the Fierzing and factorization. Instead, we will content ourselves with an understanding of the field structure, leaving the details of the factorization and resummation to future work. Nevertheless, a study of the field content provides valuable insight into the structure and complexity of the factorization at subleading power. Despite the large number of different field structures in the Lagrangians of \Eqs{eq:L1_fields}{eq:L4_fields}, we will see that due to the symmetries of the problem many terms can be shown not to contribute to all orders in $\alpha_s$. For example, we will show that there are no $\cO(\lambda)\sim \cO(\sqrt(\tau))$ power corrections, as expected from perturbative calculations. A much larger reduction in the number of jet and soft functions is achieved if one works to a fixed order in $\alpha_s$, for example $\cO(\alpha_s^2)$, and a study of the field content is sufficient to understand these aspects of the factorization.

In \Sec{sec:vanish}, we show that $\cO(\lambda)$ radiative contributions to the cross section vanish to all order in $\alpha_s$, and in \Sec{sec:novanish} we classify the $\cO(\lambda^2)$ contributions. For those operators that can contribute, we will explicitly give the field structure and contributing diagrams to $\cO(\alpha_s^2)$.

\subsection{Vanishing at $\cO(\lambda)$}\label{sec:vanish}

At $\cO(\lambda)$, the only possible radiative contribution is an $\cL^{(1)}$ insertion into the leading power hard scattering operators. If this $\cL^{(1)}$ insertion involves the conversion of a collinear quark to an ultrasoft quark, it will vanish, since it will give rise to a vacuum matrix element involving an odd fermion number. To show that the remaining possible $\cL^{(1)}$ insertions also vanish, we note that by power counting they all involve either an ultrasoft gluon field, or an ultrasoft derivative operator, but not both. 

First consider the case of an ultrasoft derivative operator. With a single Lagrangian insertion into the leading power operators,  after performing the BPS field redefinition, the ultrasoft derivative operator must act on collinear fields, and is therefore factorized into the jet function. We now use the dimensional regularization rule for residual momenta 
\begin{align}
\sum\limits_{q_l} \int d^d q_r (q_r)^j F(q_l^-, q_l^\perp, q_r^+)=0\,,
\end{align}
where $(q_r)^j$ denotes positive powers of the $q_r^-$ and $q_r^\perp$ momenta, which are the only residual momenta which appear in the subleading power Lagrangians, to show that this vanishes. After eliminating these terms, we can simplify the structure of the  $\cL^{(1)}$ Lagrangian to
\begin{align}
\cL_n^{(1)\text{BPS}}&\sim    \cB_{n\perp} \cB_{n\perp} \cB_{us(n) \perp} \cP_\perp +\bar \chi_n \chi_n \cB_{us(n) \perp} \cP_\perp+  \cB_{n\perp} \cB_{n\perp} \cB_{n\perp} \cB_{us(n) \perp} +\bar \chi_n \chi_n \cB_{n\perp} \cB_{us(n) \perp}\,,
\end{align}
each term of which involves a $\cB_{us(n)}$ field. After performing the factorization, the $\cB_{us(n)}$ field can be factorized into the soft function. However, the soft function must be rotationally invariant about the $n-\bar n$ axes. Since the only other objects which appear in the soft functions at this power are Wilson lines, this implies that it is only possible to form a rotationally invariant soft function from the $\bar n\cdot \cB_{us(n)}$, or $n\cdot \cB_{us(n)}$ components of the field. However, only the  $\cB_{us(n)\perp}$ component appears at this power. Therefore, there are no $\cO(\lambda)$ power corrections from Lagrangian insertions. In previous papers \cite{Feige:2017zci,Moult:2017rpl}, we have shown that $\cO(\lambda)$ contributions from hard scattering operators vanish for both a $q\bar q$ and $gg$ current, as do subleading power corrections to the measurement operator. Therefore, the present analysis provides a complete proof that there are no $\cO(\lambda)$ power corrections to all orders in $\alpha_s$.

\subsection{Contributions at $\cO(\lambda^2)$}\label{sec:novanish}

The $\cO(\lambda^2)$ radiative contributions to the cross section do not vanish. As discussed in \Sec{sec:sub-fact}, in addition to an insertion of the $\cO(\lambda^2)$ Lagrangian, one must also consider two insertions of the $\cO(\lambda)$ Lagrangian, as well as an insertion of the $\cO(\lambda)$ Lagrangian into an $\cO(\lambda)$ hard scattering operator.\footnote{One must also consider subleading corrections to the measurement function. However, as discussed in \Sec{sec:obs_fact}, the subleading power corrections to the measurement function first enter at $\cO(\lambda^2)$, and therefore if we are working only to $\cO(\lambda^2)$, we do not need to consider the interference of the subleading power measurement function with Lagrangian insertions. At higher powers, this contribution could be treated in a similar way to the contributions discussed in this paper. } The complete expression for the radiative contribution is therefore
{\begin{small}
\begin{align}\label{eq:xsec_lam2_rad}
&\frac{d\sigma}{d\tau}^{(2),\text{rad.}} =  N \sum_{X,i}  \tilde \delta^{(4)}_q   \int d^4x  \bra{0} C_i^{(1)*} \tO_i^{(1)\dagger}(0) \ket{X}\bra{X} \TO( i \cL^{(1)}(x)) C^{(0)} \tO^{(0)}(0) \ket{0}   \delta\big( \tau - \tau^{(0)}(X) \big)+\text{h.c.}    \nn\\
&+ N \sum_{X,i}  \tilde \delta^{(4)}_q   \int d^4x  \bra{0}  C^{(0)*} \tO^{(0)\dagger}(0)\ket{X}\bra{X}\TO ( i \cL^{(1)}(x)) C_i^{(1)} \tO_i^{(1)}(0)  \ket{0}   \delta\big( \tau - \tau^{(0)}(X) \big)+\text{h.c.}    \nn\\
&+ N \sum_X  \tilde \delta^{(4)}_q \int d^4x  \bra{0}  C^{(0)*} \tO^{(0)\dagger}(0) \ket{X}\bra{X} \TO ( i \cL^{(2)}(x)) C^{(0)} \tO^{(0)}(0) \ket{0}   \delta\big( \tau - \tau^{(0)}(X) \big)  +\text{h.c.} \nn\\
&-\frac{N}{2} \sum_X    \tilde \delta^{(4)}_q  \int d^4x \int d^4y \bra{0}\ATO \cL^{(1)}(x) \cL^{(1)}(y)  C^{(0)*} \tO^{(0)\dagger}(0) \ket{X}\bra{X}  C^{(0)} \tO^{(0)}0) \ket{0}   \delta\big( \tau - \tau^{(0)}(X) \big)  +\text{h.c.} \nn\\
&+\frac{N}{2} \sum_X   \tilde \delta^{(4)}_q  \int d^4x  \int d^4y \bra{0}\ATO\cL^{(1)}(x)  C^{(0)*} \tO^{(0)\dagger}(0) \ket{X}\bra{X} \cL^{(1)}(y) C^{(0)} \tO^{(0)}0) \ket{0}  \delta\big( \tau - \tau^{(0)}(X) \big) +\text{h.c.} \,,
\end{align}
\end{small}}%
which is a subset of the complete set of $\cO(\lambda^2)$ contributions given in \Eq{eq:xsec_lam2}.
We will consider the different contributions from each of these cases in turn. For simplicity, we will not separately discuss the different possible positions of the final state cut, since the operator structure is the same in all cases. 

\subsubsection{Contributions from $\cO(\lambda)$ Hard Scattering Operators }\label{sec:lam_hard}
We first consider term the first two lines of \eq{xsec_lam2_rad} involving a single $\cO(\lambda)$ hard scattering operators, $\cO^{(1)}$, and a single $\cO(\lambda)$ Lagrangian insertion, $\cL^{(1)}$.
A detailed derivation of the $\cO(\lambda)$ hard scattering operators for $e^+e^-\to$ dijets was given in \cite{Feige:2017zci}. The operators are of two types, and either involve a collinear quark and a collinear gluon field in the same collinear sector, or two collinear quark fields in the same collinear sector. We can use arguments identical to those presented in \Sec{sec:vanish} to show that these terms do not contribute to all orders in $\alpha_s$. In particular, since the hard scattering operators at $\cO(\lambda)$ do not involve additional ultrasoft fields, the rotational invariance of the soft function, combined with the structure of the $\cO(\lambda)$ Lagrangian, which involves only $\cB_{us(n)\perp}$ fields  guarantees that all contributions involving ultrasoft field insertions vanish. Similarly, all contributions involving insertions of an ultrasoft derivative from the Lagrangian vanish for the same reasons as described in \Sec{sec:vanish}.

\subsubsection{Contributions from $\cO(\lambda^2)$ Lagrangian Insertions}\label{sec:thrust_single_insert}

Next we consider the third line of \eq{xsec_lam2_rad}, involving the T-products of leading power hard scattering operators and $\cL^{(2)}$.
When considering contributions to the cross section,  the structure of the  $\cL^{(2)}$ Lagrangian can be simplified, since one cannot have a single insertion involving an ultrasoft fermion, or involving just ultrasoft derivatives acting in the collinear sector. This reduces the structure down to $10$ distinct field structures
\begin{align}
\cL_n^{(2)\text{BPS}}&\sim  \cB_{n\perp} \cB_{n\perp} \cB_{us(n)} \cB_{us(n)} + \cB_{n\perp} \cB_{n\perp} \partial_{us} \cB_{us(n)}  + \cB_{n\perp} \cB_{n\perp} \cB_{n\perp} \cB_{n\perp} \cB_{us(n)} \nn \\
&+ \cB_{n\perp} \cB_{n\perp} \cB_{us(n)} \cP_\perp^2 +\cB_{n\perp} \cB_{n\perp} \cP_\perp \cB_{n} \cB_{us(n)}+\bar \chi_n \chi_n \cB_{us(n)} \cB_{us(n)} \nn \\
& +\bar \chi_n \chi_n \partial_{us} \cB_{us(n)}  +\bar \chi_n \chi_n \cB_{n\perp} \cB_{n\perp} \cB_{us(n)}+\bar \chi \chi \cB_{us(n)} \cP_\perp^2 +\bar \chi_n \chi_n \cP_\perp \cB_{n} \cB_{us(n)}\,,
\end{align}
the explicit form of which was given in \Eq{eq:lam2_BPS}.
Unlike at $\cO(\lambda)$, these contributions do not vanish.

For each of these different contributions, it is easy to understand at which order they can first contribute. Those with a $\cP_\perp$ insertion will first contribute at one-higher order, as they require an additional collinear emission to provide a non-zero $\perp$ momentum. In particular, we see that while there are a large number of different radiative functions that are required for an all orders descriptions, many first contribute at high loops. Considering only those terms which contribute to $\cO(\alpha_s^2)$, we have only $6$ terms
\begin{align}
\cL_n^{(2)\text{BPS}}&\sim \bar \chi_n \chi_n \partial_{us} \cB_{us(n)}+\bar \chi_n \chi_n \cB_{us(n)} \cB_{us(n)}  + \cB_{n\perp} \cB_{n\perp} \partial_{us} \cB_{us(n)}  +  \cB_{n\perp} \cB_{n\perp} \cB_{us(n)} \cP_\perp^2 \nn \\
& +\bar \chi \chi \cB_{us(n)} \cP_\perp^2 +\bar \chi_n \chi_n \cP_\perp \cB_{n} \cB_{us(n)}\,.
\end{align}
The first term contributes at $\cO(\alpha_s)$. The second term contributes at $\cO(\alpha_s^2)$, and  has been analyzed in detail in \Sec{sec:sub_cross_gluon} where we worked out the complete structure of the factorization, and its tree level diagram is shown in \Tab{tab:LO}. The remaining terms first contribute at $\cO(\alpha_s^2)$, and representative diagrams, along with the field structure of the factorization are given in \Tab{tab:NLO_1} for those with a single $\cB_{us(n)}$ field, and in \Tab{tab:NLO_2} for those with two $\cB_{us(n)}$ fields.

\subsubsection{Contributions from $[\cO(\lambda)]^2$ Lagrangian Insertions }\label{sec:thrust_double_insert}

At $\cO(\lambda^2)$ one receives contributions from two $\cL^{(1)}$ insertions as given by the last two lines of \eq{xsec_lam2_rad}. The non-vanishing contributions can have either one, or two soft fields. Since the $\cL^{(1)}$ has the field structure 
\begin{align}\label{eq:L1_doubleinsert}
\cL_n^{(1)\text{BPS}}&\sim \cB_{n\perp} \cB_{n\perp} \partial_{us} \cP_\perp+\bar \chi_n \chi_n \partial_{us} \cP_\perp  +  \cB_{n\perp} \cB_{n\perp} \cB_{us(n)} \cP_\perp\nn \\
& +\bar \chi_n \chi_n \cB_{us(n)} \cP_\perp+  \cB_{n\perp} \cB_{n\perp} \cB_{n\perp} \cB_{us(n)}  +\bar \chi_n \chi_n \cB_{n\perp} \cB_{us(n)} +    \cB_{n\perp} \cB_{n\perp} \cB_{n\perp} \partial_{us} \nn \\
&+\bar \chi_n \chi_n \cB_{n\perp} \partial_{us} +\bar \chi_n \cB_{n\perp} \psi_{us(n)}\,,
\end{align}
this gives rise to a large number of distinct possibilities. While it is possible to write the radiative functions for all the different possible cross terms of \Eq{eq:L1_doubleinsert}, here we focus only on those that contribute up to $\cO(\alpha_s^2)$. Note that the two insertions do not need to be in the same collinear sector. However, those in distinct sectors first contribute at 3 loops. We therefore do not draw them explicitly.

The terms in the $\cL^{(1)}$ Lagrangian contain at most one ultrasoft field. Since radiative contributions where both $\cL^{(1)}$ insertions involve no collinear fields vanish,  the double $\cL^{(1)}$ insertions can contain either one or two ultrasoft fields. Except for those terms involving the ultrasoft quarks, all terms in the $\cL^{(1)}$ Lagrangian involve either $>2$ collinear fields, or a $\cP_\perp$ operator. In both cases, these first appear at $\cO(\alpha_s^2)$. The only tree level contribution from $\cL^{(1)}$ insertions is the that of the soft quarks, which was discussed in detail in \Sec{sec:sub_cross_quark} where the complete factorization structure was worked out, and whose leading order diagram is shown in \Tab{tab:LO}. In \Tab{tab:NLO_1} we summarize the field structures of the different radiative terms with a single emission arising from two  $\cL^{(1)}$ insertions, as well as a representative diagram at $\cO(\alpha_s^2)$. In all cases, these can be viewed as a propagator correction, marked with a cross, and a $\cB_{us(n)}$ emission from  $\cL^{(1)}$. The contributions with two $\cB_{us(n)}$ fields are shown in \Tab{tab:NLO_2}, along with representative diagrams at $\cO(\alpha_s^2)$. Note that here, and throughout this section, we will explicitly draw soft fields in the radiative jet functions to illustrate where they are attached. These fields will of course be factorized into the soft functions.

It is interesting to briefly consider the simplification of the $\cL^{(1)} \cdot \cL^{(1)}$ terms in the threshold limit. The threshold limit has received much attention in the power corrections literature \cite{Dokshitzer:2005bf,Grunberg:2007nc,Laenen:2008gt,Laenen:2008ux,Grunberg:2009yi,Laenen:2010uz,Almasy:2010wn,Bonocore:2014wua,White:2014qia,deFlorian:2014vta,Bonocore:2015esa,Bonocore:2016awd} due to the large amount of available perturbative data  \cite{Matsuura:1987wt,Matsuura:1988sm,Hamberg:1990np,Anastasiou:2014lda,Anastasiou:2014vaa,Anastasiou:2015ema,Anastasiou:2015yha,Dulat:2017prg}. Up to crossing, the production of a color singlet Drell-Yan pair in a proton proton collision is identical to the case of $e^+e^-\to$ dijets considered here, and therefore power corrections in $(1-z)$ can be considered using the same formalism and operators. In the case of threshold production, collinear particles are kinematically forbidden from crossing the cut into the final state. This greatly reduces the number of possible contributions at $\cO(\alpha_s^2)$. In particular, since for the $\cL^{(1)}\cdot \cL^{(1)}$ to contribute at $\cO(\alpha_s^2)$, the collinear gluon must cross the cut, these terms will not contribute in the threshold limit to $\cO(\alpha_s^3)$.

{
\renewcommand{\arraystretch}{1.4}
\begin{table}[t!]
\scalebox{0.842}{
\hspace{0.2cm}\begin{tabular}{| l | c | c |c |c|c| r| }
  \hline                       
  $T$-Product & Example Diagram & Soft Function& Jet Function \\
  \hline
  $\cL^{(2)}$ & $\fd{3cm}{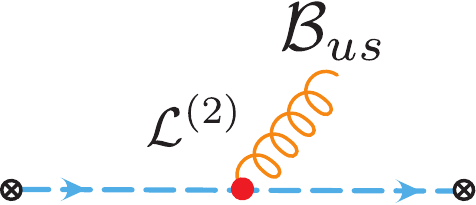}$ &  $\langle0| Y_n  Y_{\bar n}(0) \partial \cB_{us(n)}(x)Y_{\bar n} Y_n(0)   |0\rangle$& $ \langle0| \bar \chi_n(y) \bar \chi_n \chi_n(x)  \chi_n (0)|0\rangle$ \\
  \hline
   $\cL^{(1)}\cdot \cL^{(1)}$ & $\fd{3cm}{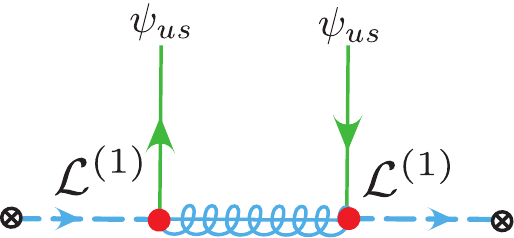}$ & {\begin{small}$ \langle0|Y_n  Y_{\bar n}(0) \bar \psi_{us}(x) \psi_{us}(z)Y_{\bar n} Y_n(0)  |0\rangle$\end{small}} &  {\begin{small}$ \langle0| \bar \chi_n(y) \bar \chi_n \cB_{n\perp} (z)\cdot  \bar \chi_n \cB_{n\perp} (x) \chi_n (0)|0\rangle$  \end{small}} \\
  \hline  
\end{tabular}}
\caption{
Leading order contributions to the $\cO(\lambda^2)$ thrust cross section from radiative functions. The soft gluon emission arises from a single $\cL^{(2)}$ insertions, in accord with the LBK theorem, while the soft quark contribution arises from two $\cL^{(1)}$ insertions. We do not explicitly write the corresponding functions when the soft gluon or quark is emitted from the $\bar n$ collinear sector.
}
\label{tab:LO}
\end{table}
}

{
\renewcommand{\arraystretch}{1.4}
\begin{table}[t!]
\scalebox{0.842}{
\hspace{-0.5cm}\begin{tabular}{| l | c | c |c |c|c| r| }
  \hline                       
  $T$-Product & Example Diagram & Soft Function & Jet Function \\
  \hline
  $\cL^{(2)}$ & $\fd{2.5cm}{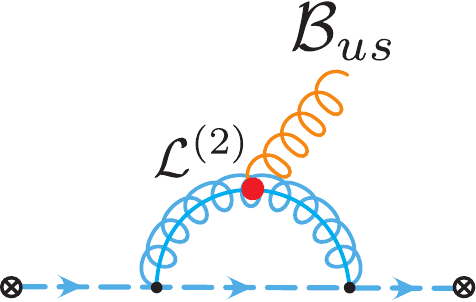}$ &  $\langle0| Y_n  Y_{\bar n}(0)  \partial \cB_{us(n)} (x) Y_{\bar n} Y_n(0)  |0\rangle$& $  \langle0| \bar \chi_n(y) \cB_{n\perp} \cB_{n\perp}(x) \chi_n (0)|0\rangle $ \\
  \hline 
    $\cL^{(2)}$ & $\fd{2.5cm}{figures_classify_all/one_loop_quark_gluon_insertion_low.pdf}$ &  $\langle0| Y_n  Y_{\bar n}(0)\cB_{us(n)}(x)Y_{\bar n} Y_n(0)  |0\rangle$& $ \langle0| \bar \chi_n(y) \cB_{n\perp} \cB_{n\perp}  \cP_\perp^2 (x) \chi_n (0)|0\rangle$ \\
  \hline 
   $\cL^{(2)}$  & $\fd{2.5cm}{figures_classify_all/1collinear_1soft_L2_low.pdf} $ &  $\langle0| Y_n  Y_{\bar n}(0)\partial \cB_{us(n)}(x) Y_{\bar n} Y_n(0) |0\rangle$& $ \langle0| \bar \chi_n(y) \bar \chi \chi  \cP_\perp^2 (x) \chi_n(0)|0\rangle$ \\
  \hline  
     $\cL^{(2)}$  & $\fd{2.5cm}{figures_classify_all/1collinear_1soft_low.pdf}$ &  $\langle0|Y_n  Y_{\bar n}(0) \cB_{us(n)}(x) Y_{\bar n} Y_n(0) |0\rangle$& $ \langle0| \bar \chi_n(y) \bar \chi_n \chi_n \cP_\perp \cB_{n}  (x) \chi_n(0)|0\rangle$ \\
    \hline  
     $\cL^{(1)}\cdot \cL^{(1)}$  & $\fd{2.5cm}{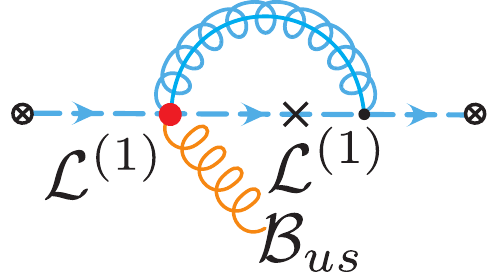}$ &  $\langle0|Y_n  Y_{\bar n}(0) \partial \cB_{us(n)}(z)Y_{\bar n} Y_n(0) |0\rangle$& $\langle0| \bar \chi_n(y) \bar \chi_n \chi_n \cB_{n\perp}(z)\cdot \bar \chi_n \chi_n  \cP_\perp (x) \chi_n (0)|0\rangle$ \\
      \hline  
     $\cL^{(1)}\cdot \cL^{(1)}$  & $\fd{2.5cm}{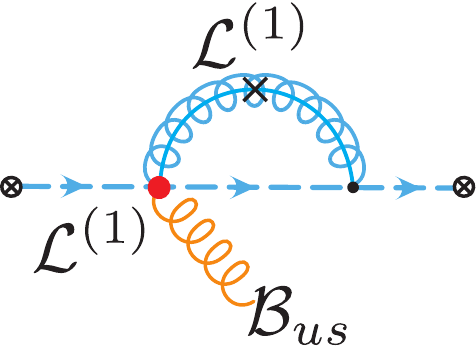}$ &  $\langle0|Y_n  Y_{\bar n}(0) \partial \cB_{us(n)}(z)Y_{\bar n} Y_n(0) |0\rangle$& $ \langle0| \bar \chi_n(y) \bar \chi_n \chi_n \cB_{n\perp} (z)\cdot \cB_{n\perp} \cB_{n\perp}  \cP_\perp (x) \chi_n(0)|0\rangle$ \\
       \hline  
     $\cL^{(1)}\cdot \cL^{(1)}$  & $\fd{2.5cm}{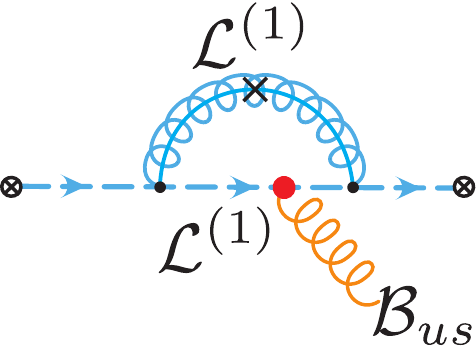}$ &  $\langle0|Y_n  Y_{\bar n}(0) \partial \cB_{us(n)}(z)Y_{\bar n} Y_n(0) |0\rangle$& $ \langle0| \bar \chi_n(y) \bar \chi_n \chi_n \cP_\perp(z)\cdot \cB_{n\perp} \cB_{n\perp}  \cP_\perp (x) \chi_n (0)|0\rangle$ \\
       \hline  
     $\cL^{(1)}\cdot \cL^{(1)}$  & $\fd{2.5cm}{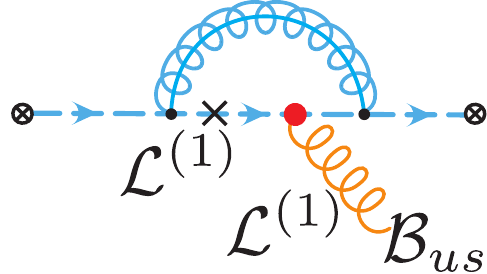}$ &  $\langle0|Y_n  Y_{\bar n}(0) \partial \cB_{us(n)}(z)Y_{\bar n} Y_n(0) |0\rangle$& $ \langle0| \bar \chi_n(y) \bar \chi_n \chi_n \cP_\perp(z)\cdot \bar \chi_n \chi_n  \cP_\perp (x) \chi_n (0)|0\rangle$ \\
       \hline  
     $\cL^{(1)}\cdot \cL^{(1)}$  & $\fd{2.5cm}{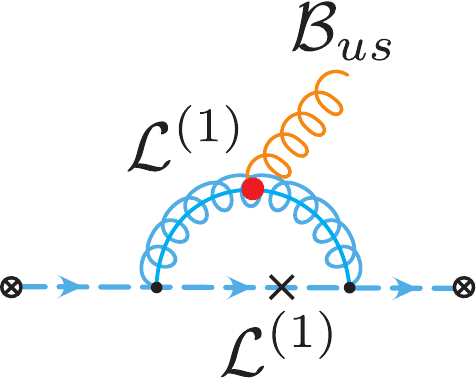}$ &  $\langle0|Y_n  Y_{\bar n}(0)  \partial \cB_{us(n)}(z)Y_{\bar n} Y_n(0) |0\rangle$& $ \langle0|\bar \chi_n(y)  \cB_{n\perp} \cB_{n\perp}  \cP_\perp (z)\cdot \bar \chi_n \chi_n  \cP_\perp (x) \chi_n (0)|0\rangle$ \\
       \hline  
     $\cL^{(1)}\cdot \cL^{(1)}$  & $\fd{2.5cm}{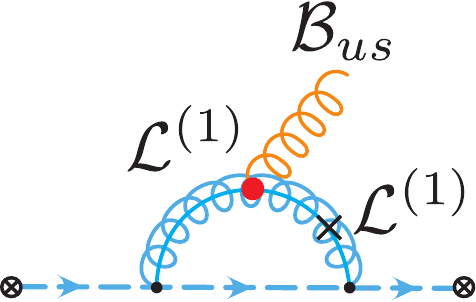}$ &  $\langle0|Y_n  Y_{\bar n}(0) \partial \cB_{us(n)}(z)Y_{\bar n} Y_n(0) |0\rangle$& $ \langle0|\bar \chi_n(y)  \cB_{n\perp} \cB_{n\perp} \cP_\perp (z)\cdot \cB_{n\perp} \cB_{n\perp}  \cP_\perp (x) \chi_n (0)|0\rangle$ \\    
  \hline 
\end{tabular}}
\caption{
Radiative functions which contribute at $\cO(\lambda^2)$ to the thrust cross section starting at 2 loops with a single soft emission. This includes both terms arising from a single $\cL^{(2)}$ insertion, and from two $\cL^{(1)}$ insertions. We have not explicitly written the functions where the emission occurs in the $\bar n$ collinear sector, since they can easily be obtained from those given. We have also not explicitly written the measurement function.
}
\label{tab:NLO_1}
\end{table}
}
{
\renewcommand{\arraystretch}{1.4}
\begin{table}[t!]
\scalebox{0.842}{
\hspace{-0.9cm}\begin{tabular}{| l | c | c |c |c|c| r| }
  \hline                       
  $T$-Product & Example Diagram & Soft Function&Jet Function \\
  \hline
  $\cL^{(2)}$ & $\fd{3cm}{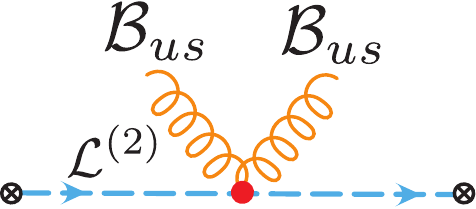}$ &  $\langle0|Y_n  Y_{\bar n}(0) \cB_{us(n)}  \cB_{us(n)} (x) Y_{\bar n} Y_n(0) |0\rangle$& $ \langle0| \bar \chi_n(y) \bar \chi_n \chi_n  (x) \chi_n (0)|0\rangle$ \\
  \hline 
   $\cL^{(1)} \cdot \cL^{(1)} $ & $\fd{3cm}{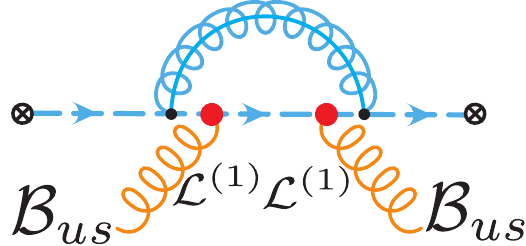}$ &  $\langle0|Y_n  Y_{\bar n}(0) \cB_{us(n)} (x) \cB_{us(n)} (z)Y_{\bar n} Y_n(0)  |0\rangle$& $\langle0| \bar \chi_n(y) \bar \chi_n \chi_n \cP_\perp (z)\cdot  \bar \chi_n \chi_n  \cP_\perp  (x) \chi_n(0)|0\rangle$ \\
  \hline  
     $\cL^{(1)} \cdot \cL^{(1)} $ & $\fd{3cm}{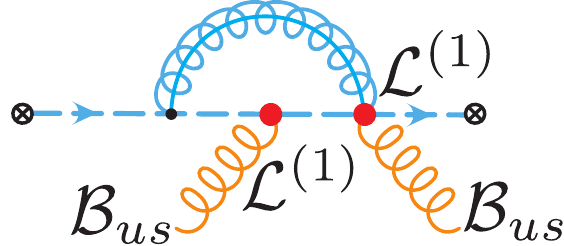}$ &  $\langle0|Y_n  Y_{\bar n}(0) \cB_{us(n)} (x) \cB_{us(n)}(z) Y_{\bar n} Y_n(0) |0\rangle$& $ \langle0| \bar \chi_n(y) \bar \chi_n \chi_n  \cP_\perp(z) \cdot  \bar \chi_n \chi_n \cB_{n\perp}  (x) \chi_n (0)|0\rangle$ \\
  \hline  
       $\cL^{(1)} \cdot \cL^{(1)} $ & $\fd{3cm}{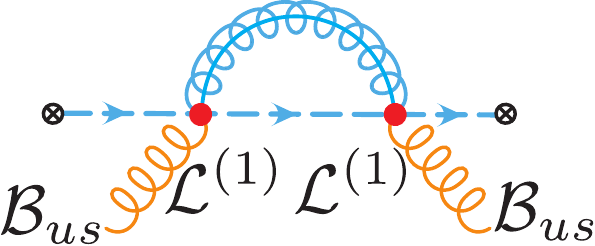}$ &  $\langle0|Y_n  Y_{\bar n}(0) \cB_{us(n)}(x)  \cB_{us(n)}(z) Y_{\bar n} Y_n(0) |0\rangle$& $ \langle0| \bar \chi_n(y) \bar \chi_n \chi_n \cB_{n\perp} (z) \cdot \bar \chi_n \chi_n \cB_{n\perp}  (x) \chi_n (0)|0\rangle$ \\
  \hline  
\end{tabular}}
\caption{
Radiative functions which contribute at $\cO(\lambda^2)$ to the thrust cross section starting at 2 loops with two soft emissions. In the restriction to the threshold limit, only the first $T$-product contributes, since collinear partons cannot cross the cut, making the other terms first contribute at $\cO(\alpha_s^3)$. We have not explicitly written the functions where the emission occurs in the $\bar n$ collinear sector, since they can easily be obtained from those given.
}
\label{tab:NLO_2}
\end{table}
}

\subsection{Discussion}\label{sec:discuss}

The analysis of this section illustrates that the general structure of the Lagrangian contributions is quite complicated. In particular, while there are only several contributions at lowest order in $\alpha_s$, there are a large number of contributions which first appear at higher orders, coming from Lagrangian insertions which involve multiple collinear fields. Indeed, there are terms whose first non-vanishing contribution is at $4$-loops. These contributions do not seem in general to be related to the renormalization group evolution of those operators that contribute at lowest order, and thus appear to complicate the general structure.

 We believe that the leading logarithmic contributions are captured by the renormalization group evolution of the two radiative functions that contribute at lowest order, which are given in \Tab{tab:LO}, and that the Lagrangian insertions involving higher numbers of collinear fields are only required to reproduce subleading logarithms. This has been explicitly verified in the subleading power leading logarithmic resummation \cite{Moult:2018jjd} for thrust in $H\to gg$ in pure Yang-Mills theory (i.e. without quarks), and here a proof of this statement was given under the assumption that the convolutions appearing in the factorization formula converge. However, we do not currently have a more general proof of this statement. This expectation comes from the fact that collinear loops from these operators involving multiple collinear fields are less singular than the identical collinear loops involving collinear Wilson line emissions. If this is true, then the leading logarithmic resummation may take a simple form, but beyond leading logarithmic order an increasing number of operators will be required. We plan to develop the leading logarithmic renormalization group evolution of these leading power operators in a future paper, and by comparing with explicit calculations it should be possible to understand the role of the operators involving additional collinear fields.

\section{Conclusions}\label{sec:conclusions}

In this paper we have derived an all orders factorization for the emission of soft partons from a jet, expressed as a convolution of gauge invariant soft and collinear matrix elements. Unlike for leading power soft emissions, which are only sensitive to the color charge and direction of the collinear particles, radiative functions couple the momentum of the soft and collinear sectors in a non-trivial manner, and are also sensitive to the spin structure of the collinear particles.  Using SCET, we have shown that the radiative functions are given by matrix elements of universal subleading power Lagrangians describing the interactions of non-local gauge invariant quark and gluon fields. The use of non-local gauge invariant fields is crucial to achieve a gauge invariant factorization for soft parton emission, since in a non-abelian gauge theory the emitted parton carries a charge. The multipole expansion in the effective theory allows the factorization to be expressed as a single variable convolution describing the position along the light cone of the operator insertion, which dresses the leading power Wilson lines, and describes the breakdown of eikonalization. 

A key advantage of our approach is that the radiative functions are derived from factorized matrix elements of the SCET Lagrangians, which describe the interactions of the gauge invariant quark and gluon fields to all orders. This allows us to provide a complete classification of all radiative functions that will contribute at a given power in the expansion. The number of such operators at each given power is finite, and soft and collinear emissions to all orders in $\alpha_s$ are described by the intricate Wilson line structure of the operators, dictated by the symmetries of the effective theory. As one particular example, we introduced a radiative function describing the emission of a soft quark. As was shown in the calculation of subleading power thrust \cite{Moult:2016fqy,Boughezal:2016zws,Moult:2017jsg} contributions from soft quarks give a leading logarithmic contribution at subleading power. Soft quarks are also required to compute subleading power corrections to the threshold limit for Drell-Yan production in the $qg$ channel. We performed a classification of all the radiative functions which contribute for the event shape thrust in $e^+e^-\to$ dijets, and derived in detail the structure of the radiative functions which contribute at tree level. This provides the final missing ingredient in achieving factorization and resummation of event shapes at subleading power. 

Combined with the subleading power operator bases for $\bar q \Gamma q$ \cite{Feige:2017zci,Chang:2017atu} and $gg$ \cite{Moult:2017rpl}, and the analysis of the subleading power measurement function \cite{Feige:2017zci}, all the sources of subleading power corrections in SCET (as were summarized in \Sec{sec:sum_fact}) have now been characterized in detail. It will now be of significant interest to derive a complete subleading power factorization theorem for an event shape observable, and to study the renormalization group evolution of the different components, a direction which we intend to pursue in future work. The renormalization of higher twist matrix elements has been studied  \cite{Balitsky:1987bk,Ratcliffe:1985mp,Ji:1990br,Ali:1991em,Kodaira:1996md,Balitsky:1996uh,Mueller:1997yk,Belitsky:1997zw,Belitsky:1999ru,Belitsky:1999bf,Vogelsang:2009pj,Braun:2009mi}, as has that of subleading power operators in $B$ physics \cite{Hill:2004if,Beneke:2005gs}, and of several power suppressed operators for $e^+e^-\to$ dijets \cite{Freedman:2014uta,Goerke:2017lei} and $N$-jet production \cite{Beneke:2017ztn,Beneke:2018rbh}.  We anticipate that combining all these ingredients will allow for the resummation of subleading power corrections for event shape observables.

\begin{acknowledgments}
We thank Daniel Kolodrubetz for collaboration at initial stages of this work, and Cyuan-Han Chang, Robert Szafron, Sebastian Jaskiewicz, and HuaXing Zhu for discussions.
We thank the participants of the 2016 Edinburgh workshop ``Threshold Logarithms Beyond Leading Power", as well as those of the 2018 Nikhef ``Next to leading power corrections in particle physics'' workshop for stimulating discussions. 
We thank the ``Challenges and Concepts for Field Theory and Applications in the Era of LHC Run-2''  workshop for hospitality and support while portions of this work were completed.
This work was supported in part by the Office of Nuclear Physics of the U.S. Department of Energy under the Grant No. DE-SCD011090, by the Office of High Energy Physics of the U.S. Department of Energy under Contract No. DE-AC02-05CH11231, and the LDRD Program of LBNL. I.S. was also supported by the Simons Foundation through the Investigator grant 327942.

\end{acknowledgments}

\appendix

\section{BPS Identities}\label{app:BPS_identities}

In this appendix we collect a number of basic identities related to the BPS transformation. Given an operator $\hat{O}$ we denote the BPS transformed operator by $\BPS[\hat{O}]$. Using the fundamental representation for the ultrasoft Wilson lines $Y_n$ we have
\begin{align}
	&\BPS[W_n] = Y_n W_n^{(0)} Y_n^\dagger \,, &&\BPS[W_n^\dagger] = Y_n W_n^{(0)\dagger} Y_n^\dagger\,, \nn\\
	&\BPS[\chi_n] = Y_n \chi^{(0)}_n \,,	&&\BPS[\hat{O}_{us}] = \hat{O}_{us}  \,,\nn
\end{align}
\be
	\BPS[\cD_i] = \BPS[ W_n^\dagger D_i W_n] = Y_n W_n^{(0)\dagger} Y_n^\dagger\, \BPS[D_i]\, Y_n W_n^{(0)} Y_n^\dagger \,.
\ee
Ghost particles $c_n$, transform under BPS field redefinitions according to
\begin{align}
\BPS[c_n] = Y_n c_n Y_n^\dagger\,.
\end{align}
Other useful relations ($A,B$ are generic operators) are the following:
\begin{align}
	&[Y_n,\cP_\perp^\mu] = 0 \implies \cP_\perp^\mu = Y_n \cP_\perp^\mu Y_n^\dagger\,, \nn\\
	&Y_n^\dagger in\cdot D_{us} Y_n = i n \cdot \partial_{us}\,, \qquad &&Y_n^\dagger gn \cdot A_{us} Y_n = i n \cdot \partial_{us} - Y_n^\dagger in\cdot \partial_{us} Y_n \,,\nn\\
    &[ Y_n  A \,Y_n^\dagger\,,\, B] = Y_n\,[A\,,\,Y^\dagger_n B \,Y_n]\, Y_n^\dagger\,,\qquad
    &&[ Y_n A \, Y_n^\dagger, \, Y_n \,B\, Y_n^\dagger] = Y_n \,[A, B] \,Y^\dagger_n \label{eq:commextr}\,.
\end{align}
Using these relations we can compute the BPS field redefinition of the derivative operators appearing in the Lagrangian
\begin{align}
	\BPS[iD^\mu_{n \perp}] &\equiv \BPS[P_\perp^\mu +g A^\mu_{n \perp}] =  Y_n iD^{(0)\mu}_{n \perp} Y_n^\dagger\,,\nn \\
	\BPS[i\cD^\mu_{n\perp}] &\equiv \BPS[W_n^\dagger iD^\mu_{n \perp} W_n] = Y_n i\cD^{(0) \mu}_{n \perp} Y_n^\dagger\,,\nn \\
	\BPS[i\cD^\mu_{n}] &= Y_n \left( i\cD^{(0) \mu}_{n \perp} + \frac{n^\mu}{2} \cPbar \right) Y_n^\dagger + \frac{\bar n^\mu}{2}Y_n W_n^{(0)\dagger} \, ( Y_n^\dagger in\cdot \partial_{us}Y_n + g n\cdot A^{(0)}_n)\,  W_n^{(0)} Y_n^\dagger \nn \\
	&= Y_n \left( i\cD^{(0) \mu}_{n \perp} + \frac{n^\mu}{2} \cPbar \right) Y_n^\dagger + \frac{\bar n^\mu}{2}Y_n W_n^{(0)\dagger} \, ( Y_n^\dagger [in\cdot \partial_{us}Y_n] + in\cdot \partial_{us} + g n\cdot A^{(0)}_n)\,  W_n^{(0)} Y_n^\dagger \nn \\
	&= Y_n i\cD^{(0) \mu}_{n } Y_n^\dagger + \frac{\bar n^\mu}{2}Y_n W_n^{(0)\dagger} \, Y_n^\dagger [in\cdot \partial_{us}Y_n]  W_n^{(0)} Y_n^\dagger\,.
\end{align}
A less trivial calculation is how $i\cD^\mu_{ns}$ transforms under BPS field redefinition. We have
\begin{align}
	\BPS[i\cD^\mu_{ns}] &\equiv \BPS[W_n^\dagger iD^\mu_{ns} W_n] = \BPS[i\cD_n^\mu] + \frac{\bn^\mu}{2} Y_n W_n^{(0)\dagger} Y_n^\dagger\, gn \cdot A_{us} \, Y_n W_n^{(0)} Y_n^\dagger  \nn\\ 
	&= \BPS[i\cD_n^\mu] - \frac{\bn^\mu}{2} Y_n W_n^{(0)\dagger} Y_n^\dagger\, in \cdot \partial_{us} \, Y_n W_n^{(0)} Y_n^\dagger + \frac{\bn^\mu}{2} Y_n W_n^{(0)\dagger} in\cdot \partial_{us} W_n^{(0)} Y_n^\dagger \nn\\
	&= Y_n \left( i\cD^{(0) \mu}_{n \perp} + \frac{n^\mu}{2} \cPbar  + \frac{\bn^\mu}{2} in\cdot \cD^{0}_n \right) Y_n^\dagger \equiv Y_n i\cD^{(0) \mu}_{n } Y_n^\dagger \,.
\end{align}
The most important thing to note here is that 
\be
	\BPS[i\cD^\mu_{n \perp}] = Y_n i\cD^{(0)\mu}_{n \perp} Y_n^\dagger\,,\label{eq:Dperptransf}
\ee
but
\be
	\BPS[i\cD^\mu_{n}] \neq Y_n i\cD^{(0)\mu}_{n} Y_n^\dagger \,. \label{eq:Dstransf}
\ee

\section{Fierzing for Radiative Jet Functions}\label{app:Fierzing}

In this appendix we collect some details related to the color and Dirac structure of the radiative jet functions. To obtain scalar jet and soft functions, one must factorize in Dirac and color space.  Factorization in Dirac and color space can be achieved using the SCET Fierz relation \cite{Lee:2004ja}
\begin{align}\label{eq:SCET_fierz}
\big(\delta^{\a'\a}\delta^{i'i} \big)\big( \delta^{\b\b'}\delta^{jj'}\big) 
&= \frac{1}{2}\sum\limits_{k=1}^6 (F_k^{\bar n})^{\b\a}_{ji} \otimes (F_k^n)^{\a'\b'}_{i'j'} \nn\\
&=\frac{1}{2}\Big[
\frac{\Sl{\bar n}}{2N_C} \otimes \frac{\Sl{n}}{2} 
- \frac{\Sl{\bar n}\gamma^5}{2N_C} \otimes \frac{\Sl{n}\gamma^5}{2} 
- \frac{\Sl{\bar n}\gamma^\alpha_\perp}{2N_C} \otimes \frac{\Sl{n}\gamma^\perp_\alpha}{2} \nn \\
&\hspace{1cm}
+ \Sl{\bar n}T^a \otimes \frac{\Sl{n}T^a}{2} 
- \Sl{\bar n}\gamma^5T^a \otimes \frac{\Sl{n}\gamma^5T^a}{2} 
- \Sl{\bar n}\gamma^\alpha_\perp T^a \otimes \frac{\Sl{n}\gamma^\perp_\alpha T^a}{2} 
\Big]\,.
\end{align}
In the text, we primarily focused on the convolution structure. Here we consider the derivation of the Dirac and color structure. 

Consider first the soft quark radiative jet function. Since we are only interested in the Dirac and color structure, it is notationally simplest to ignore all the measurement functions, and just consider the vacuum matrix element of the fields. For the radiative jet function involving a soft quark, we have from \Sec{sec:sub_cross_quark}
\begin{align}
\cM^{(2)}_{\psi_{us}}= \int d^4x d^4y~ &\langle 0 | \left[ \bar \chi_n(x) Y_n^\dagger(x) \gamma^\mu_\perp Y_{\bar n}(x) \chi_{\bar n}(x)   \right]    \left[ \bar \psi_{us(n)}(y) g \Sl{\cB}_{n\perp}(y) \chi_n(y)   \right] \nn \\
&\hspace{1cm}\cdot \left[  \bar \chi_n(z) g\Sl{\cB}_{n\perp}(z)   \psi_{us(n)}(z) \right] 
\left[  \bar \chi_{\bar n} (0) Y_{\bar n}^\dagger(0) \gamma^\mu_\perp Y_n(0) \chi_n(0)  \right] |0\rangle\,,
\end{align}
which will enter the expression for the cross section.
Applying the Fierz relation of \Eq{eq:SCET_fierz} three times, we obtain the factorized form of the matrix element
\begin{align}
\cM^{(2)}_{\psi_{us}}=\int d^4x d^4y~&\langle 0 | \left[  \bar \chi_{\bar n} (0) F^n_k \chi_{\bar n}(x)  \right] |0\rangle \langle 0 | \left[  \bar \chi_{n} (x) F^{\bar n}_{k'} \chi_{n}(0)  \right]  \left[ \bar \chi_n(z) g \Sl{\cB}_{n\perp}(z) F^{\bar n}_l g  \Sl{\cB}_{n\perp}(y) \chi_n(y)  \right]   |0\rangle  \nn \\
&\hspace{-0.9cm}\cdot\langle 0 |   \tr \left[ F^n_{k'} Y_n^\dagger (x) \gamma^\mu_\perp Y_{\bar n}(x) F^{\bar n}_k Y_{\bar n}^\dagger(0) \gamma^\mu_\perp Y_n(0)   \right]  \left[ \bar \psi_{us(n)}(y) F^n_l \psi_{us(n)}(z)   \right] |0\rangle\,.
\end{align}
The symmetries of the soft and collinear sectors can then be used to set $k=k'=l=1$. We can then further simplify the structure of the $n$-collinear matrix element by applying another Fierz relation. We then obtain
\begin{align}
\cM^{(2)}_{\psi_{us}}=\int d^4x d^4y&\langle 0 | \left[  \bar \chi_{\bar n} (0) \frac{\Sl n}{2} \chi_{\bar n}(x)  \right] |0\rangle  \langle 0 |   \tr \left[ Y_n^\dagger (x) Y_{\bar n}(x)  Y_{\bar n}^\dagger(0)  Y_n(0)   \right]  \left[ \bar \psi_{us(n)}(y) \frac{\Sl n}{2} \psi_{us(n)}(z)   \right] |0\rangle \nn \\
&\cdot \langle 0 | \left[  \bar \chi_{n} (x) \frac{\Sl {\bar n}}{2} \chi_n(y) \right]     \tr \left[ g \cB_{n\perp}(z)  \cdot g  \cB_{n\perp}(y)     \right]     \left[  \bar \chi_n(z)   \frac{\Sl {\bar n}}{2} \chi_n(0) \right]   |0\rangle 
\,.
\end{align}
Reinstating the time ordering and measurement function, this corresponds to the result given in \Sec{sec:sub_cross_quark}.

For the case of the radiative jet function involving a gluon emission, from \Sec{sec:sub_cross_gluon}, we must consider the factorization of the matrix element
\begin{align}
\cM^{(2)}_{\cB_{us}}=\langle 0 |    \left[ \bar \chi_n(x) Y_n^\dagger(x) \gamma^\mu_\perp Y_{\bar n}(x) \chi_{\bar n}(x)   \right] &   \bar \chi_n  \left[  T^a \gamma_{\perp \mu} \frac{1}{\bar \cP}  i \Sl \partial_{us\perp} - i  {\overleftarrow{\Sl \partial}}_{us\perp} \frac{1}{\bar \cP} T^a \gamma_{\perp \mu}   \right]   \frac{\Sl \bn}{2} \chi_n (y) \nn \\
&\cdot\left[  \bar \chi_{\bar n} (0) Y_{\bar n}^\dagger(0) \gamma^\mu_\perp Y_n(0) \chi_n(0)  \right]             |0 \rangle\,,
\end{align}
we can flip the action of the ultrasoft derivative onto the $\cB_{us}$ field, and apply twice the Fierz relation of \Eq{eq:SCET_fierz} to obtain
\begin{align}
\cM^{(2)}_{\cB_{us}}=&\langle 0 | \left[ \bar \chi_{\bar n}(0) \frac{\Sl n}{2}  \chi_{\bar n}(x)   \right] |0 \rangle \langle 0 | \left[ \bar \chi_n(x) F_{\bar n}^k \chi_n(y) \right] \frac{1}{\bar \cP}   \left[ \bar \chi_n(y) F_{\bar n}^{k'} \chi_n(0) \right]|0 \rangle \nn \\
&\hspace{2cm}\cdot\langle 0 | \tr \left[   (-i \Sl{\partial}_\perp g\Sl{\cB}_{us(n) \perp} (y))  \frac{\Sl{\bar n}}{2} F_n^k Y_n^\dagger (x) Y_{\bar n}(x) \frac{\Sl{\bar n}}{2} Y_{\bar n}^\dagger (0) Y_n(0) F_n^{k'} \right]\,.
\end{align}
The tree level contribution is with $k=k'=1$. Without having a closed quark loop from which the soft gluon field is emitted, the only possibilities are $k, k'= \frac{\Sl n}{2},  \frac{\Sl n}{2} T^a$. The color neutrality of the vacuum in the collinear sector implies that one would have to contract the color indices between $k$ and $k'$. It seems that one could potentially get this configuration from a fermion bubble type diagram, but this is first at two loops, and should not contribute to LL. (i.e. we know it doesn't from the explicit result, which doesn't have an $n_f$). Simplifying the Dirac structure, we then obtain
\begin{align}
\cM^{(2)}_{\cB_{us}}=&\langle 0 | \left[ \bar \chi_{\bar n}(0) \frac{\Sl n}{2}  \chi_{\bar n}(x)   \right] |0 \rangle \langle 0 | \left[ \bar \chi_n(x)  \frac{\Sl{\bar n}}{2} \chi_n(y) \right] \frac{1}{\bar \cP}   \left[ \bar \chi_n(y)  \frac{\Sl{\bar n}}{2} \chi_n(0) \right]|0 \rangle \nn \\
&\hspace{2cm}\cdot\langle 0 | \tr \left[   (-i \partial_\perp \cdot g\cB_{us(n) \perp} (y))  Y_n^\dagger (x) Y_{\bar n}(x)  Y_{\bar n}^\dagger (0) Y_n(0) \right]|0\rangle\,.
\end{align}
Reinstating the time ordering and measurement function, this corresponds to the result given in \Sec{sec:sub_cross_gluon}.

\bibliography{../../overallbib}{}
\bibliographystyle{JHEP}

\end{document}